%% file: Paper_MultDepSpectra.tex
\documentclass[ALICE,manyauthors]{Includes/cernphprep}
\input{Includes/Header.tex}
\usepackage{orcidlink}
\begin{document}
\begin{titlepage}
	\PHyear{2022}
	\PHnumber{266}      
	\PHdate{28 November}  
	\title{Multiplicity dependence of charged-particle production\\in \pp, \ppb, \xexe and \pbpb collisions at the LHC}
	\ShortTitle{Multiplicity dependence of charged-particle production at the LHC}   
	\Collaboration{ALICE Collaboration\thanks{See Appendix~\ref{app:collab} for the list of collaboration members}}
	\ShortAuthor{ALICE Collaboration} 
\begin{abstract}
\input{Content/0--Abstract.tex}
\end{abstract}
\end{titlepage}

\setcounter{page}{2}
\input{Content/1--Introduction.tex}
\input{Content/2--Experiment.tex}
\input{Content/3--Analysis.tex}
\input{Content/4--Results.tex}
\input{Content/5--Summary.tex}

\newenvironment{acknowledgement}{\relax}{\relax}
\begin{acknowledgement}
\section*{Acknowledgements}
We thank Klaus Werner for providing the EPOS3 predictions and Bj\"orn Schenke, Prithwish Tribedy and Chun Shen for the hydrodynamical calculations.
We are grateful to Larry McLerran and Michal Praszalowicz for discussions on the geometrical scaling. 
\input{fa_2022-11-03_Opt_C.tex}
\end{acknowledgement}
\clearpage
\bibliographystyle{Bibliography/utphys}
\bibliography{Bibliography/bibliography}

\newpage
\appendix
\clearpage

\clearpage
\section{The ALICE Collaboration}
\label{app:collab}
\input{2022-11-03-Alice_Authorlist_2022-11-03_0_Opt_C.tex}  
\end{document}

%% file: Includes/Header.tex
\usepackage[utf8]{inputenc}
\usepackage[comma,square,numbers,sort&compress]{natbib}
\usepackage{hyperref}
\usepackage{lineno}
\usepackage{multirow}
\usepackage{xspace}
\usepackage[shortcuts]{extdash}
\DeclareGraphicsExtensions{.pdf, .png, .jpg, .gif, .eps}
\usepackage{siunitx}

\newcommand{\pp}{pp\xspace}
\newcommand{\pbpb}{Pb\==Pb\xspace}
\newcommand{\ppb}{p\==Pb\xspace}
\newcommand{\xexe}{Xe\==Xe\xspace}
\newcommand{\gevc}{\ensuremath{{~\mathrm{GeV}}/c}\xspace}

\newcommand{\tev}{\ensuremath{~\mathrm{TeV}}\xspace}

\newcommand{\vzrange}{\ensuremath{{\abs{V_\mathrm {z}} < 10~\text{cm}}}\xspace}
\newcommand{\vzmeasrange}{\ensuremath{{\abs{V^\mathrm{meas}_\mathrm {z}} < 10~\text{cm}}}\xspace}
\newcommand{\pt}{\ensuremath{p_{\mathrm T}}\xspace}
\newcommand{\ptmeas}{\ensuremath{p_{\mathrm T}^{\mathrm {meas}}}\xspace}

\newcommand{\ptRange}{\ensuremath{0.15\gevc < \pt < 10\gevc}\xspace}

\newcommand{\mpt}{\ensuremath{\langle p_\mathrm T\rangle}\xspace}
\newcommand{\sigmapt}{\ensuremath{\sigma(p_\mathrm T)}\xspace}
\newcommand{\nch}{\ensuremath{N_{\mathrm {ch}}}\xspace}
\newcommand{\nchmeas}{\ensuremath{N_{\mathrm {ch}}^{\mathrm {meas}}}\xspace}

\newcommand{\mnch}{\ensuremath{\langle N_{\mathrm {ch}}\rangle}\xspace}
\newcommand{\sigmanch}{\ensuremath{\sigma(N_{\mathrm {ch}})}\xspace}

\newcommand{\sqrsn}{\ensuremath{\sqrt{s_\mathrm{NN}\,}}\xspace}
\newcommand{\sqrs}{\ensuremath{\sqrt{s\,}}\xspace}

\renewcommand{\v}[1]{\ensuremath{\mathbf{#1}}} 						
\newcommand{\abs}[1]{\left| #1 \right|} 								
\let\baraccent=\= 													
\renewcommand{\=}[1]{\stackrel{#1}{=}} 								

\usepackage{todonotes}
\usepackage{silence}
\newcommand\plotWidth{0.45\textwidth}
\newcommand\plotWidthTwoD{0.585\textwidth} 
\newcommand\plotDistance{0.05\textwidth}

%% file: Content/0--Abstract.tex
Multiplicity (\nch) distributions and transverse momentum (\pt) spectra of inclusive primary charged particles in the kinematic range of $|\eta| < 0.8$ and \ptRange are reported for \pp, \ppb, \xexe and \pbpb collisions at centre-of-mass energies per nucleon pair ranging from
$\sqrsn = 2.76\tev$ up to $13\tev$. 
A sequential two-dimensional unfolding procedure is used to extract the correlation between the transverse momentum of primary charged particles and the charged-particle multiplicity of the corresponding collision.
This correlation sharply characterises important features of the final state of a collision and, therefore, can be used as a stringent test of theoretical models.
The multiplicity distributions as well as the mean and standard deviation derived from the \pt spectra are compared to state-of-the-art model predictions.
Providing these fundamental observables of bulk particle production consistently across a wide range of collision energies and system sizes can serve as an important input for tuning Monte Carlo event generators.

%% file: Content/1--Introduction.tex
\section{Introduction}
In high-energy nucleus--nucleus~(AA) collisions, a hot and dense state of deconfined strongly-interacting matter, commonly denoted as the quark--gluon plasma, is formed~\cite{Busza:2018rrf}. 
This system undergoes hydrodynamic evolution~\cite{Heinz:2013th,Bernhard:2016tnd,Gardim:2019xjs} and exhibits hadron yields indicating chemical equilibrium~\cite{Andronic:2017pug, BraunMunzinger:2003zd}.  
At low to intermediate transverse momentum (\pt), up to about a few $\mathrm{GeV}/c$, charged-particle production is influenced by the collective expansion (flow) of the system, reflected in the shapes of single-particle transverse-momentum spectra~\cite{Adam:2016tre,ALICE:2018vuu} and multi-particle correlations~\cite{Heinz:2013th}. 
It came as a major surprise that even small collision systems such as proton--proton (\pp) and proton--nucleus (pA) collisions at Large Hadron Collider (LHC) energies exhibit features that can be attributed to collective expansion~\cite{Adam:2016dau,ABELEV:2013wsa,Khachatryan:2014jra,Chatrchyan:2013nka,Khachatryan:2015waa,Aad:2013fja,Dusling:2013oia,Blok:2017pui,Schenke:2021mxx}. 

Multi-particle correlations measured in all collision systems, contain an imprint of the initial state of the colliding partners, characterised via their quark and gluon degrees of freedom, allowing the extraction of fundamental properties of the quark--gluon plasma~\cite{Bernhard:2016tnd,Bernhard:2019bmu,Gardim:2019xjs}.
Hydrodynamic-like (final-state) collective flow is  increasingly part of the modelling of small collision systems~\cite{Moreland:2018gsh,Schenke:2020mbo} in an interplay with initial-state phenomena (see reviews in Refs.~\cite{Dusling:2015gta,Schenke:2021mxx}).
Such collision systems are usually modelled in the framework of the colour glass condensate (CGC)~\cite{McLerran:2013una}, where multi-particle production proceeds via the decay of colour flux tubes that stretch between two colliding hadrons in the longitudinal direction and are coherent in the transverse direction over a range that is inversely proportional to a saturation scale~$Q_{\textrm{s}}$, which appears due to the non-linearity of parton evolution at high energies.

In the PYTHIA8 event generator~\cite{Sjostrand:2014zea}, which describes a broad range of observables in pp collisions, the initial state is determined by parton distribution functions extracted from measurements.
With increasing collision energy, the role of multi-parton interactions, i.e., when two or more distinct (hard) parton interactions occur within a pp collision, becomes more and more important~\cite{Skands:2010ak}. 
The respective colour strings may cut each other or reconnect, which leads to a redistribution of energy from particle production to transverse momentum, and, therefore, a smaller number of particles but with higher average \pt.
A phenomenon recently implemented in PYTHIA is the interaction between transversely-extended colour strings, exerting pressure on each other~\cite{Bierlich:2017vhg}. This produces effects similar to those resulting from final-state collective dynamics, akin to that of a long-lived quark--gluon medium. 
Recently, PYTHIA8 was extended with the Angantyr model for heavy-ion collisions~\cite{Bierlich:2018xfw}, which uses a Glauber-based initial-state modelling with Gribov colour fluctuations to determine the number of \pp sub-collisions. 
In current PYTHIA8 implementations with the Angantyr model, there is no colour reconnection between individual \pp sub-collisions but between the multiple partonic interactions of the \pp sub-collisions.

A broad range of observables is also described successfully in the EPOS family Monte Carlo (MC) event generators~\cite{Drescher:2000ha}. 
The initial state is realised in EPOS through a parton-based Gribov-Regge theory~\cite{Drescher:2000ha} which is a multiple scattering framework, recently augmented with  the treatment of saturation effects~\cite{Werner:2019aqa}.
While in EPOS3~\cite{Werner:2013tya}, a full hydrodynamic evolution is included for the final state, in EPOS~LHC~\cite{Pierog:2013ria} a parameterised hydrodynamic component of a small volume with high density of thermalised matter is used together with a dilute region, i.e. a core plus corona implementation.
In both PYTHIA8 and EPOS~LHC models, the respective parameters are tuned using the Run 1 data at the LHC~\cite{Skands:2014pea,Pierog:2013ria}. 

The mean transverse momentum, \mpt, of the charged-particle \pt spectrum
and its correlation with the charged-particle multiplicity \nch 
carry essential information on the underlying particle production mechanism.
This has been studied by many experiments at hadron colliders in 
pp($\overline{\mathrm{p}}$) covering centre-of-mass energies from ${\sqrs = 31~\textrm{GeV}}$ up to 
13\tev~\cite{Arnison:1982ed,ABCDHW,Albajar:1989an, PhysRevLett.60.1622,Adams:2006xb,Aaltonen:2009ne,Aamodt:2010my,Khachatryan:2010nk, Aad:2010ac,Adam:2015pza} as well as in \xexe~\cite{Acharya:2018eaq} and \pbpb~\cite{ALICE:2018vuu} collisions at ${\sqrsn = 5.44\tev}$ and 5.02\tev, respectively. 
All experiments observed an increase of \mpt with \nch in the central rapidity region, explained in the PYTHIA approach as a consequence of non-trivial colour reconnections (see discussion in Ref.~\cite{Christiansen_2015}).
While in the CGC approach \mpt is a universal function of the ratio of the charged-particle multiplicity to the transverse area of the collision~\cite{McLerran:2014apa}, in the EPOS~LHC model it is determined by the collective expansion~\cite{Pierog:2013ria}.
For all collision systems, the $\mpt\==\nch$ correlation is a basic observable for tuning or calibrating the theoretical models~\cite{Moreland:2018gsh}, a simple observable which allows extracting fundamental properties of a deconfined quark--gluon~medium~\cite{Gardim:2019xjs}.

As bulk production at the LHC is driven by a complex interplay of soft and hard QCD processes, the endeavour to find a consistent model description for particle production in all collision systems has not been concluded yet.
At the LHC, a large amount of data was recorded in Run~1 and Run~2, covering different collision systems from \pp to heavy-ion collisions at various centre-of-mass energies, which allows a detailed study of particle production over a wide range of charged-particle multiplicity.
This Letter presents a measurement of the charged-particle multiplicity distributions and the corresponding transverse momentum spectra as a function of \nch in \pp, \ppb, \xexe and \pbpb collisions at centre-of-mass energies per nucleon pair ranging from
$\sqrsn = 2.76\tev$ up to 13\tev.
From these spectra, the mean \mpt and standard deviation ${\sigmapt = \sqrt{\langle (p_\mathrm{T} - \langle p_\mathrm {T}\rangle)^{2} \rangle}}$ within \ptRange are extracted.
The comprehensive set of measurements presented in this Letter can serve as a high-precision input for tuning phenomenological models towards the goal of understanding particle production in the non-perturbative regime of QCD and its transition to the perturbative regime.

This Letter is structured as follows. Section~\ref{sec:experiment} briefly describes the experimental setup. In Section~\ref{sec:analysis}, the data used for this analysis and a detailed description of the analysis procedure are presented. Results are discussed in Section~\ref{sec:results}, and a summary is given in Section~\ref{sec:summary}.

%% file: Content/2--Experiment.tex
\section{Experimental setup}\label{sec:experiment}

The measurements reported in this Letter were obtained with ALICE at the Large Hadron Collider. 
In the following, the detectors relevant for this work are discussed briefly.
A more detailed description of the ALICE apparatus in its configurations of LHC Run 1 and Run 2 can be found in Refs.~\cite{ALICE:2008ngc,AlicePerf2014}.

The present analysis is based on tracking information from the Inner Tracking System (ITS)~\cite{Aamodt:2010aa} and the Time Projection Chamber (TPC)~\cite{Alme:2010ke}.
Both detectors are located within a large solenoidal magnet which provides a nominal field strength of 0.5~T for all of the data taking periods analysed in this work, except for the \xexe data taking, where the magnetic field was reduced to 0.2~T in order to extend the kinematic acceptance of the detector to lower transverse momentum.
The ITS is comprised of six cylindrical layers of silicon detectors with radii between 3.9~cm and 43~cm.
Its two innermost layers are equipped with silicon pixel detectors (SPD), the two intermediate layers consist of silicon drift detectors (SDD), and the two outermost layers are made of double-sided silicon strip detectors (SSD).
The large cylindrical TPC has an active radial range from about 85~cm to 250~cm and an overall length along the beam direction of about 500~cm. 
It covers the full azimuth in the pseudorapidity range ${|\eta| < 0.9}$ and provides track reconstruction with up to 159 space points along the trajectory of a charged particle as well as particle identification via the measurement of specific energy loss ${\rm d}E/{\rm d}x$.

The V0 detector, which consists of two scintillator arrays covering the pseudorapidity ranges of $2.8<\eta<5.1$~(V0A) and ${-3.7<\eta<-1.7}$~(V0C), is used for triggering on hadronic collisions and multiplicity measurements at forward rapidity~\cite{ALICE:2018cpu,ALICE:2018vuu}.
Contamination from electromagnetic interactions in \pbpb and \xexe collisions was strongly suppressed using signals from two neutron-Zero-Degree Calorimeters (ZDC), positioned on both sides of the interaction point at 114.0 m distance, see~\cite{AlicePerf2014} for details.

%% file: Content/3--Analysis.tex
\section{Analysis procedure}\label{sec:analysis}

\begin{figure}[t]
	\center
	\includegraphics[width=\plotWidthTwoD]{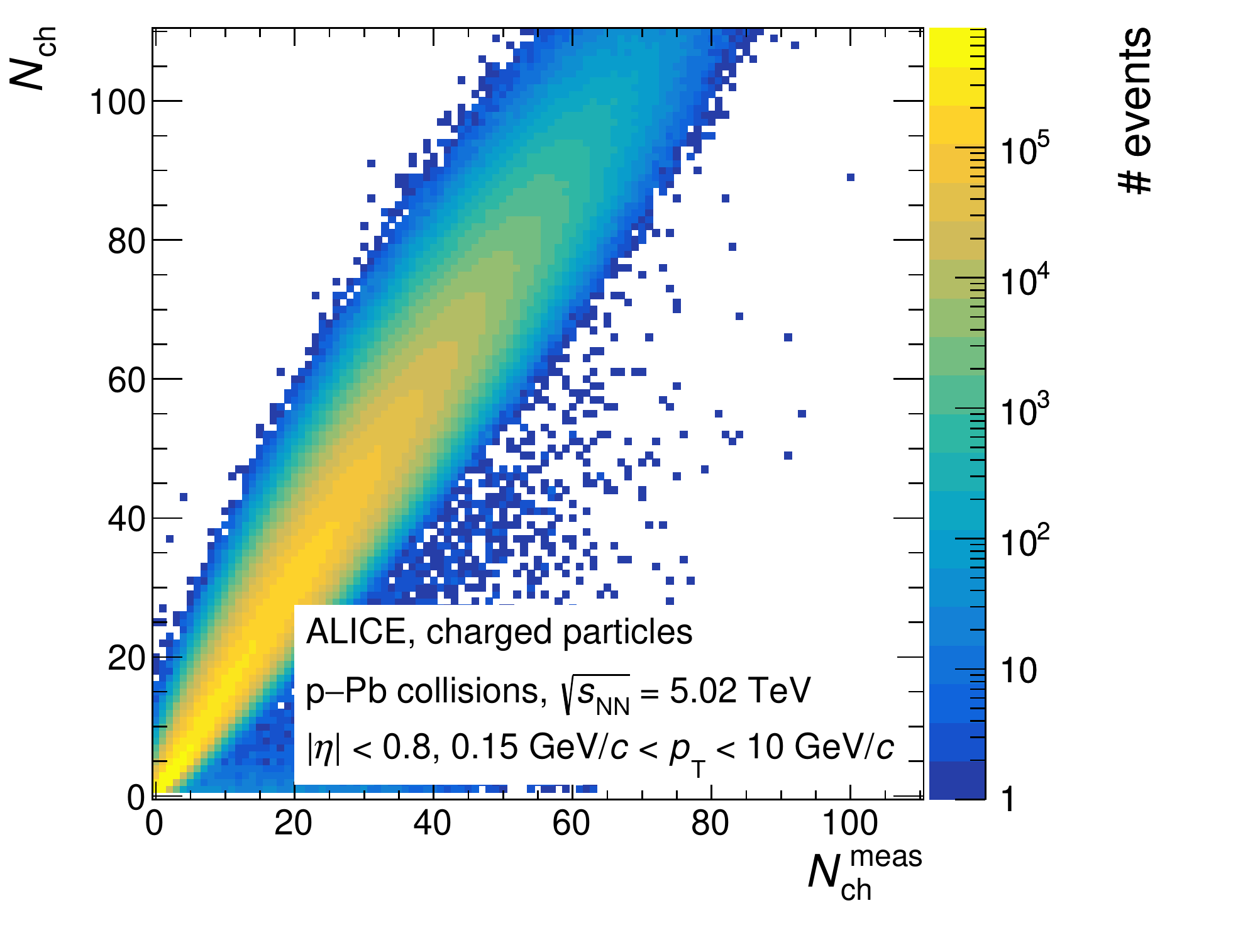}
	\caption{Multiplicity correlation matrix for \ppb collisions at $\sqrsn = 5.02\tev$.}
	\label{fig:multCorrel}
\end{figure}

This analysis aims to obtain the correlation between primary charged-particle \pt spectra and their corresponding event multiplicities \nch, both defined consistently in the kinematic range ${|\eta|~<~0.8}$ and \ptRange. 
This kinematic selection ensures optimal momentum resolution and homogeneous tracking efficiency over the entire range.
In addition, the multiplicity distributions for ${\nch > 0}$ events are reported. 
For each collision, the number of reconstructed tracks selected for the analysis provides an experimental measure (\nchmeas) for its multiplicity.
Due to detector acceptance and reconstruction efficiency, this measured track multiplicity \nchmeas is different from the actual number of primary charged particles (\nch) produced in the kinematic region under study. 
Secondary particles from weak decays or from interactions with the detector material as well as tracks that are smeared into the acceptance (i.e. from $|\eta| \geq 0.8$ and $\pt \leq 0.15\gevc,\ \pt \geq 10\gevc$) remaining in the sample of selected tracks further contribute to the measured \pt spectra and consequently to the measured track multiplicity.
The event-by-event fluctuations of reconstruction efficiency and contamination effects lead to a rather broad correlation between \nch and \nchmeas that is shown in Fig.~\ref{fig:multCorrel} for an example data set.
As a result, the transverse momentum spectrum associated to a given \nchmeas value carries contributions on particle production from a range of different \nch values.
In addition, the finite resolution of the transverse momentum measurement itself results in a (small) smearing of the measured transverse momentum \ptmeas.
The correct correlation between the collision-characteristic \nch value and its corresponding \pt distribution can be extracted by unfolding the measured quantities.

\paragraph{Data sets and event selection}
The data analysed in this work were collected between 2010 and 2018 during the LHC Run~1 and Run~2 data-taking periods.
They comprise \pp, \ppb, \xexe, and \pbpb collisions at centre-of-mass energies per nucleon pair ranging from ${\sqrsn = 2.76}$ up to 13\tev.
Hadronic collisions were selected with two different minimum-bias (MB) interaction triggers.
In Run~1, a single hit in either of V0A or V0C detectors or in the SPD was required (denoted as V0OR) in coincidence with a crossing of two particle bunches in the LHC at the nominal ALICE interaction point.
For the Run~2 data taking period, a signal in both V0A and V0C was necessary to fulfil the MB trigger condition (denoted V0AND).
Therefore, the former is also sensitive to single diffractive \pp events while the latter almost exclusively selects non-single diffractive interactions.
The offline event selection (for details see Ref.~\cite{ALICE:2018cpu,ALICE:2018vuu}) is optimised to reject beam-induced background and pileup collisions.
Events with no reconstructed vertex and those with poor vertex resolution are rejected.
Both the trigger efficiency and the vertex requirements affect mostly low multiplicity events.
To ensure full pseudorapidity coverage of the tracking detectors (in particular by the inner ITS layers) and therefore avoid a possible asymmetry in the kinematic selection of the tracks, all collisions are required to have their reconstructed primary vertex located within \vzmeasrange along the beam direction with respect to the nominal interaction point.
Table~\ref{tab:datasets} shows an overview of the data sets and their corresponding number of events selected for this analysis.
In order to have comparable results regardless of the trigger setup and event selections, all measurements presented in this Letter are corrected using Monte Carlo simulations such that they represent collisions with at least one charged particle produced in the kinematic range ${|\eta|~<~0.8}$ and \ptRange. 
For this event class, the possible bias originating from Monte Carlo event generators for very low \nch values (partially originating from diffractive or electromagnetic events) is minimal.

\begin{table}[t]
\center
\caption{\label{tab:datasets} Overview of the analysed data sets. The definitions of the two MB triggers are explained in the text.}
\begin{tabular}{@{}lccccc@{}} 
\hline\hline
 & \sqrsn (TeV) & year & MB trigger & no. events analysed (M)\\ 
\hline
\multirow{5}{*}{\pp} & 2.76 & 2011 & V0OR & 48 \\
   & $5.02$ & 2017 & V0AND & 316 \\
   & $7$    & 2010 & V0OR & 129 \\
   & $8$    & 2012 & V0AND & 26 \\
   & $13$   & 2016 & V0AND & 180 \\
  \hline
\multirow{2}{*}{\ppb} & $5.02$ & 2016 & V0AND & 309 \\
 & $8.16$ & 2016 & V0AND & 15 \\
\hline
\xexe & $5.44$ & 2017 & V0AND & 1 \\
\hline
\multirow{2}{*}{\pbpb} & $2.76$ & 2010 & V0OR & 19 \\
 & $5.02$ & 2015, 2018 & V0AND & 239 \\
 \hline\hline
\end{tabular}
\end{table}

\paragraph{Track selection}
A primary charged particle~\cite{ALICE-PUBLIC-2017-005} is defined as a charged particle with a mean proper lifetime $\tau$ larger than ${1~\textrm{cm}/c}$, which is either produced directly in the interaction or from decays of particles with $\tau$ smaller than ${1~\textrm{cm}/c}$, excluding particles produced in interactions with the detector material.
Charged-particle trajectories are reconstructed using both the ITS and TPC detectors.
In order to select only tracks with good \pt resolution for analysis, a minimal length in the active detector volume as well as a good agreement of the final track parameterisation with its comprising space points are required.
The contamination of the track sample with weakly decaying particles, secondary particles from interactions of primary particles with the detector material and from pileup events is significantly reduced by selecting only tracks originating from a location close to the primary interaction vertex.
In previous ALICE publications~\cite{ALICE:2018vuu,Acharya:2018eaq} those selection criteria were studied in great detail and optimised for best track quality and minimal contamination from secondary particles.

\paragraph{Particle-composition correction}
The measured data are complemented by Monte Carlo simulations implementing a realistic GEANT3~\cite{GEANT3} model of the ALICE detector and mimicking the experimental conditions present during the data taking.
From these simulations, information about efficiency, secondary contamination, and smearing of \nch as well as \pt is obtained.
However, it was found in previous measurements~\cite{ALICE:strangeEnhance,ALICE:strangeEnhanceMult} that the current state-of-the-art MC event generators do not perfectly reproduce the relative particle abundances, in particular for hyperons.
Since the detector response, as well as the effect of the track selection, vary for the different particle species (e.g. due to different lifetimes), a purely MC-based correction for efficiency and feed-down contamination of inclusive charged particles would depend on the accuracy of the relative hadron abundances produced by the respective underlying event generator.
To take this effect into account, the particle abundances from the event generator are re-weighted by means of a data-driven approach that was already employed in other ALICE analyses~\cite{ALICE:2018vuu,Acharya:2018eaq}.
This particle-composition correction utilises several ALICE measurements of identified ($\pi, \textrm{K}, \textrm{p}, \Lambda$) particle \pt spectra as a function of multiplicity (in coarse intervals) for a range of collision systems~\cite{ALICE:strangeEnhanceMult,ALICE:2019avo,Adam:2016dau,ALICE:2013wgn,ALICE:2013mez} as input and offers \nch- and \pt-dependent correction factors for particle abundances.
These data-driven adjustments for the generator bias result in a more accurate description of the detector performance and are applied prior to the unfolding corrections described in the following.

\paragraph{Event-level unfolding}
The measurement of the raw charged-particle multiplicity distribution is influenced by several effects. 
In the experiment, some collisions that occurred within \vzrange with respect to the nominal interaction vertex are not detected by the minimum-bias trigger or discarded by the subsequent event selection.
Due to the experimental vertex-position resolution, a valid collision event may also be reconstructed outside of \vzmeasrange and therefore rejected in the analysis.
On the other hand, the measured and selected sample of events may contain collisions without any primary charged particle produced within the kinematic range of interest (i.e. events with $\nch = 0$) or collisions with a true vertex position located outside \vzrange.
In addition, as discussed before, the value of the measured track multiplicity \nchmeas itself is affected by 
tracking efficiency and track selection as well as contamination with secondaries and particles smeared into the kinematic acceptance, resulting in a broad correlation between the actual number of primary charged particles \nch and the measured track multiplicity \nchmeas.

Using the particle-composition corrected MC simulation, the measured track multiplicity distribution can be corrected for the efficiency, contamination, and smearing effects by means of an iterative unfolding procedure proposed by D'Agostini~\cite{DAgostini:1994fjx} and implemented in the RooUnfold~\cite{rooUnfold} software package.

Starting with an initial assumption (prior) for the desired multiplicity distribution, which in this case is taken from the MC simulation, unfolding weights (posterior probabilities) are calculated by combining the prior with the detector response and the measured track multiplicity distribution according to Bayes’ theorem. By again applying these posterior probabilities to the measured track multiplicity distribution, an updated and more accurate guess for the prior is calculated.
This procedure is repeated at least three times and, in order to avoid overfitting, immediately stopped once the $\chi^2/\textrm{ndf}$ between the resulting multiplicity distributions of two consecutive iterations drops below unity. In this context, the number of degrees of freedom refers to the number of data points in the respective distribution.
The procedure is found to be very stable and the resulting unfolded spectrum after convergence is not sensitive to the choice of a prior.

\paragraph{Track-level unfolding}
While the one-dimensional multiplicity distributions are straightforward to unfold with the iterative D'Agostini method described above, extracting the correlation between \pt spectra and the corresponding multiplicity poses a greater challenge.
In principle, this two-dimensional (2D) deconvolution could be done in the same manner, i.e., by unfolding the distribution of $(\nchmeas, \ptmeas)$-pairs and thereby extracting the corrected $(\nch, \pt)$-distribution of primary charged particles.
However, for the highly-granular measurement carried out here, this would imply a huge number of possible combinations and therefore, in practice, require an unreasonably large Monte Carlo event sample to sufficiently populate the smearing matrix that represents the detector response.
Hence, in this analysis, a new approach was developed aiming to effectively achieve the 2D~unfolding in a simpler way, by splitting it into multiple one-dimensional unfolding problems.
Starting from the raw yield of charged-particle tracks as a function of measured track transverse momentum \ptmeas and measured track multiplicity \nchmeas, this technique yields the fully corrected transverse momentum spectra of primary charged particles as a function of their corresponding primary charged particle multiplicity \nch.
These fully corrected \nch-dependent \pt spectra are then normalised to the number of ${\nch > 0}$ events obtained from the unfolded multiplicity distributions.
In addition, the spectra are divided by the widths $\Delta \pt$ and $\Delta \nch$ of the respective intervals chosen for analysis.
In the present work, for \pp, \ppb and AA collisions with $\nch \leq 100$, multiplicity intervals of $\Delta \nch = 1$ are used, while for the rest of the range in AA collisions intervals of $\Delta \nch = 9$ are chosen.
This choice is driven by optimising the resolution of the unfolding procedure versus computing time.
As the bulk of particles are produced at low transverse momentum, the \pt intervals are set to have decreasing granularity from low to high \pt.

In the experiment, event-level as well as track-level effects influence whether a charged particle with transverse momentum \pt from an event of multiplicity \nch is detected with the measured properties \ptmeas and \nchmeas.
For the reasons described above, entire collision events, and in consequence all of their corresponding particles, can either be lost if the event is rejected or incorrectly selected and as a result contribute to the background contained in the measured track sample.
Further, for an event which is selected for analysis, the \ptmeas spectrum as well as the measured track multiplicity \nchmeas are affected by tracking efficiency, transverse momentum resolution, and contamination from secondaries or particles smeared into the kinematic acceptance of the measurement.
While the event-level effects change the \pt-integrated detector response and are intrinsically multiplicity dependent, the track-level detector response mainly depends on the transverse momentum of the respective particle.
However, there is still a small \pt dependence of contamination and efficiency related to the event selection as the trigger may bias toward specific types of events 
(e.g., by selectively rejecting diffractive collisions which may have a transverse momentum shape different from that of non-diffractive events with the same multiplicity).
On the other hand, also the tracking efficiency and contamination are (slightly) multiplicity dependent, as for instance the particle composition of the event changes with multiplicity, which is relevant in particular for AA collisions.

The basic idea behind the novel procedure employed in this analysis is that the multiplicity dimension (which is mostly affected by event-level effects) and the \pt dimension (which is dominated by track-level effects) can be treated in two separate, sequential unfolding stages.
In a first step, the \nchmeas dependent raw track yield in each \ptmeas interval is unfolded separately using the event-level efficiencies and contamination as well as the (\pt integrated) multiplicity smearing matrix of primary charged particles, which contains the probability for a primary charged particle from a collision with true multiplicity \nch to be found in an event with measured track multiplicity \nchmeas.
As a result of this deconvolution, the measured tracks are reassigned to the true multiplicities \nch and corrected for track losses related to the event selection and contamination from background events.
In a second step, the \ptmeas dependent track yield is unfolded individually in each \nch interval using the corresponding \pt dependent tracking efficiencies and \ptmeas dependent contamination, as well as the (multiplicity-integrated) transverse momentum smearing matrix of primary charged particles.
Note that since the measured track distributions in each of the individually unfolded \ptmeas or \nch intervals are different, unique posterior probabilities (i.e., unfolding weights) are obtained in each of these cases.
For all of the \ptmeas intervals, the \ptmeas integrated \nch distribution of measured primary charged particles taken from the MC simulation is used as the initial prior for the unfolding, while for all the \nch intervals the \nch integrated \pt distribution of measured primary charged particles is used.

\begin{figure}[htb]
	\center
	\includegraphics[width=\plotWidth]{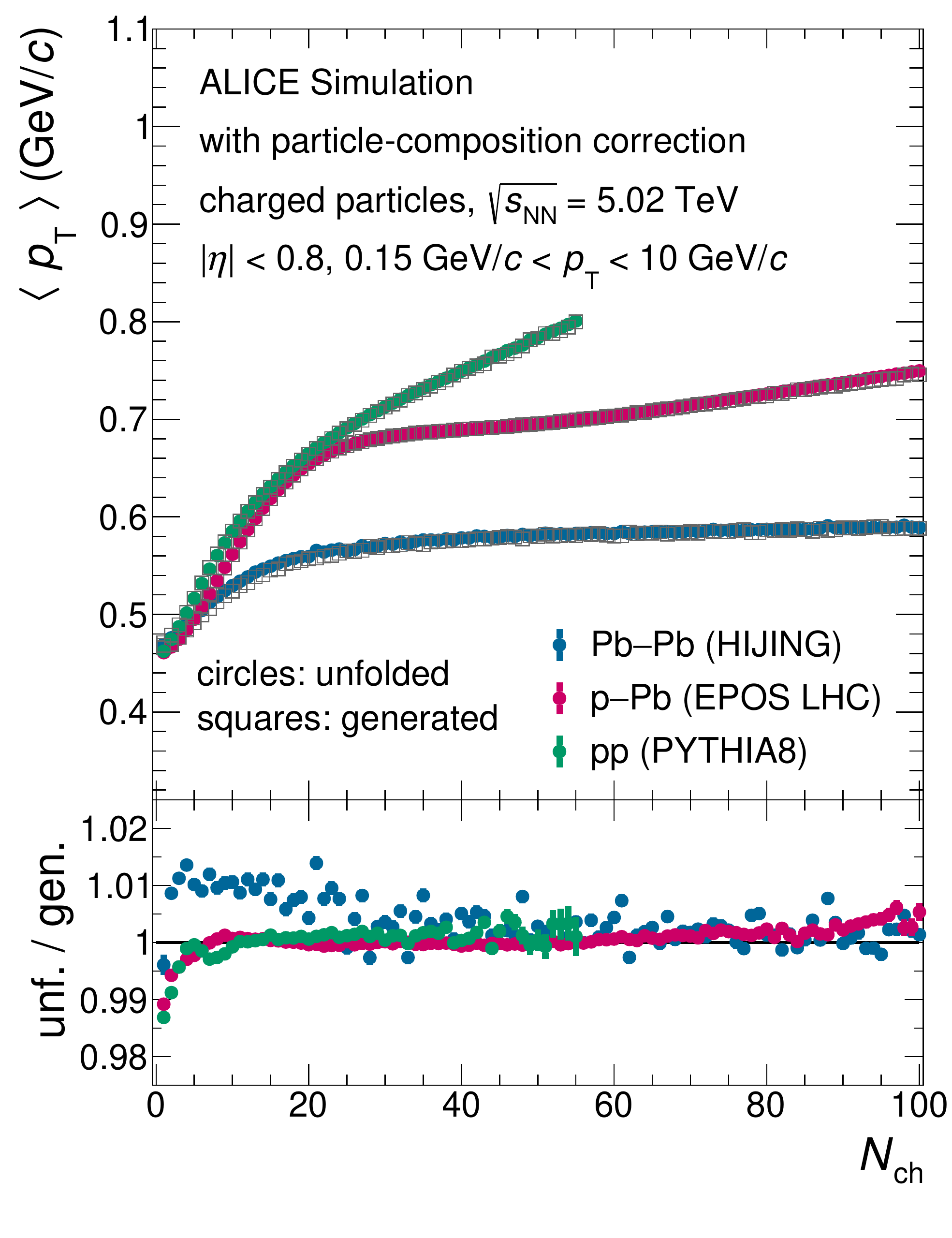}
	\hspace{\plotDistance}
	\includegraphics[width=\plotWidth]{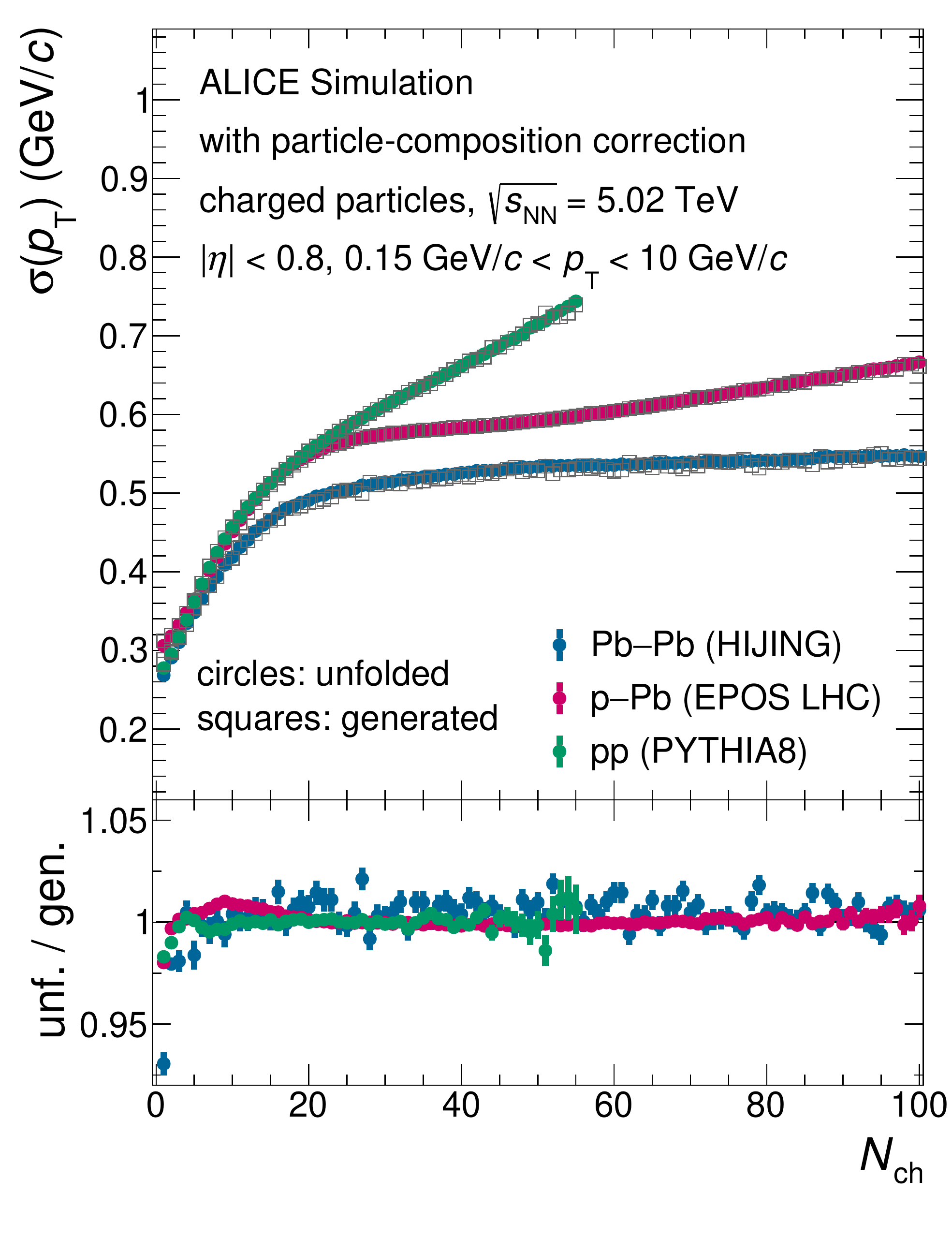}
	\caption{The \nch dependence of the mean (left panel) and standard deviation (right panel) of the \pt distributions for (particle-composition corrected) Monte Carlo events in \pp, \ppb and \pbpb collisions at $\sqrsn = 5.02\tev$. 
	Results propagated through a full GEANT model of ALICE and corrected with the sequential 2D unfolding (closed circles) procedure described in the text are compared with the generator-level (open squares) distributions and their ratios are shown in the bottom panels.}
	\label{fig:closure}
\end{figure}

To validate the self-consistency of this sequential unfolding approach, it is applied to a MC sample which includes the transport of particles through the detector and the results are then compared with its underlying generator-level expectation.
In Fig.~\ref{fig:closure} this closure test is shown for the mean and standard deviation of the unfolded \nch dependent transverse momentum spectra simulated for \pp, \ppb and \pbpb collisions at $\sqrsn = 5.02\tev$ using the PYTHIA8, EPOS LHC and HIJING~\cite{Wang:1991hta} event generators, respectively. Note that these are the particle-composition corrected simulations.
The ratios in the bottom panels show a very good agreement between the unfolded and generated distributions over the whole range of multiplicities.
The non-closure is mostly well below 1\% and the remaining relative differences are used as an estimate for the systematic uncertainty of the method.
The closure test was validated by cross-checking the largest data set, i.e. pp collisions at $\sqrs = 13\tev$, with EPOS~LHC as an alternative MC generator.

\begin{figure}[htb]
	\center
	\includegraphics[width=\plotWidthTwoD]{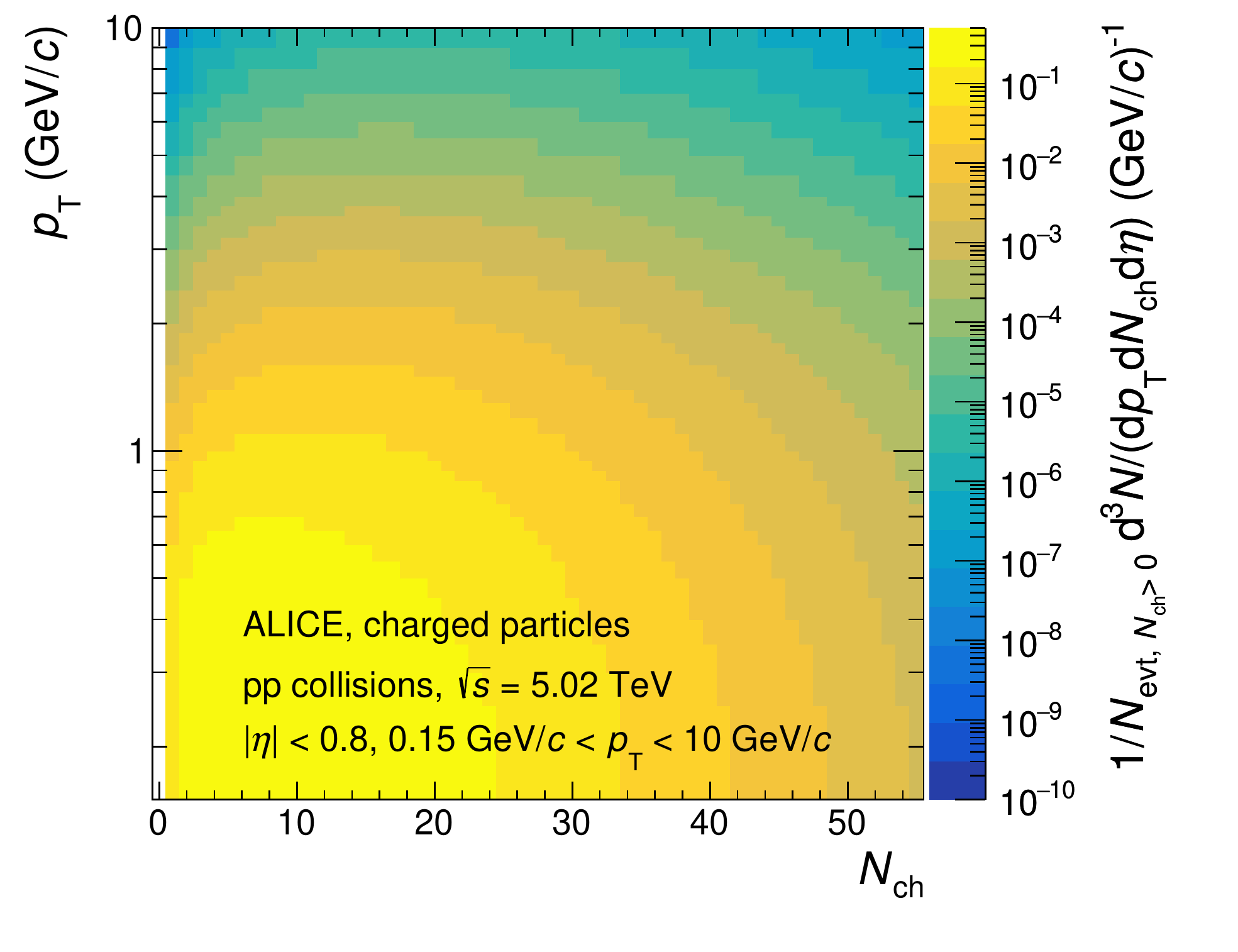}
	\\
	\includegraphics[width=\plotWidth]{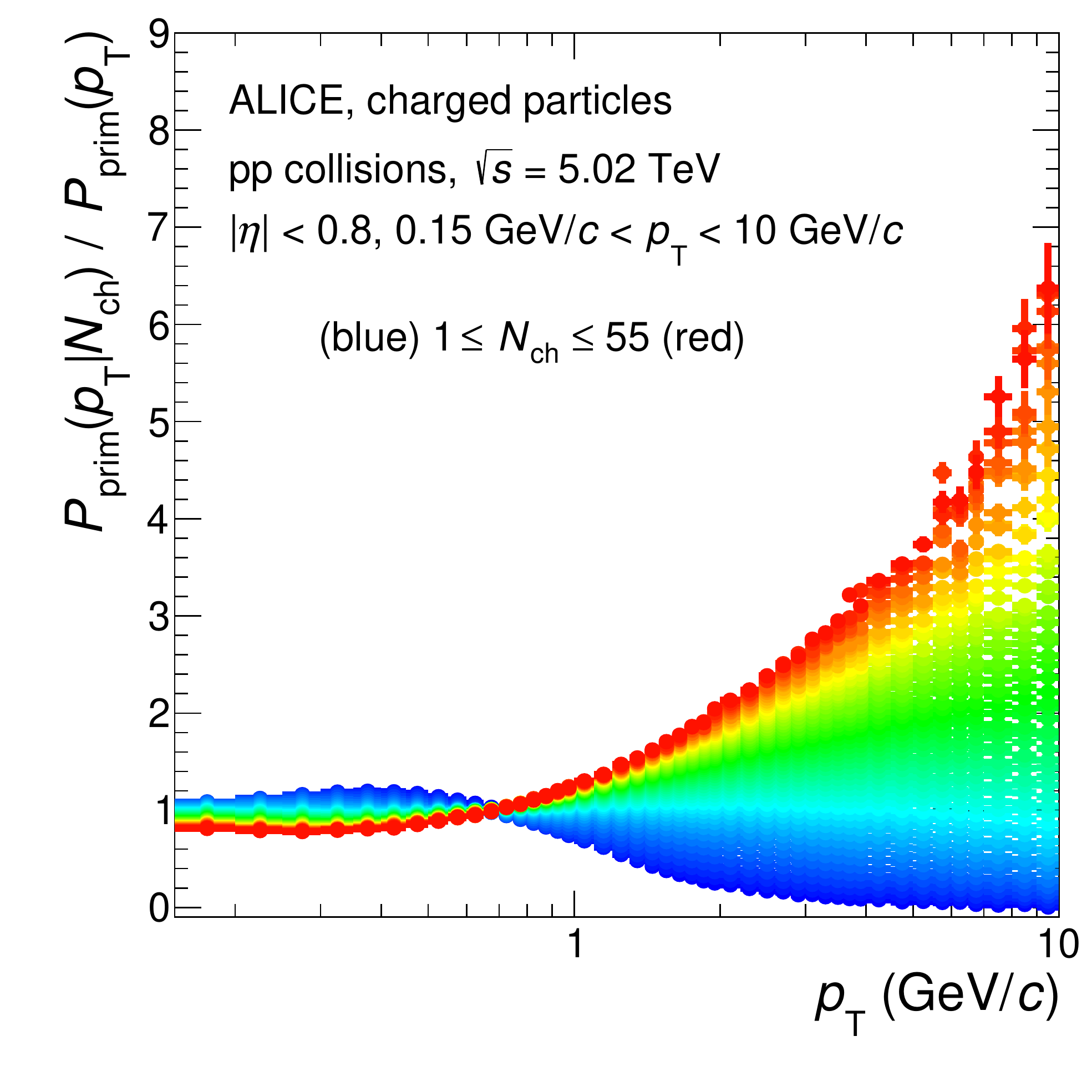}
	\hspace{\plotDistance}
	\includegraphics[width=\plotWidth]{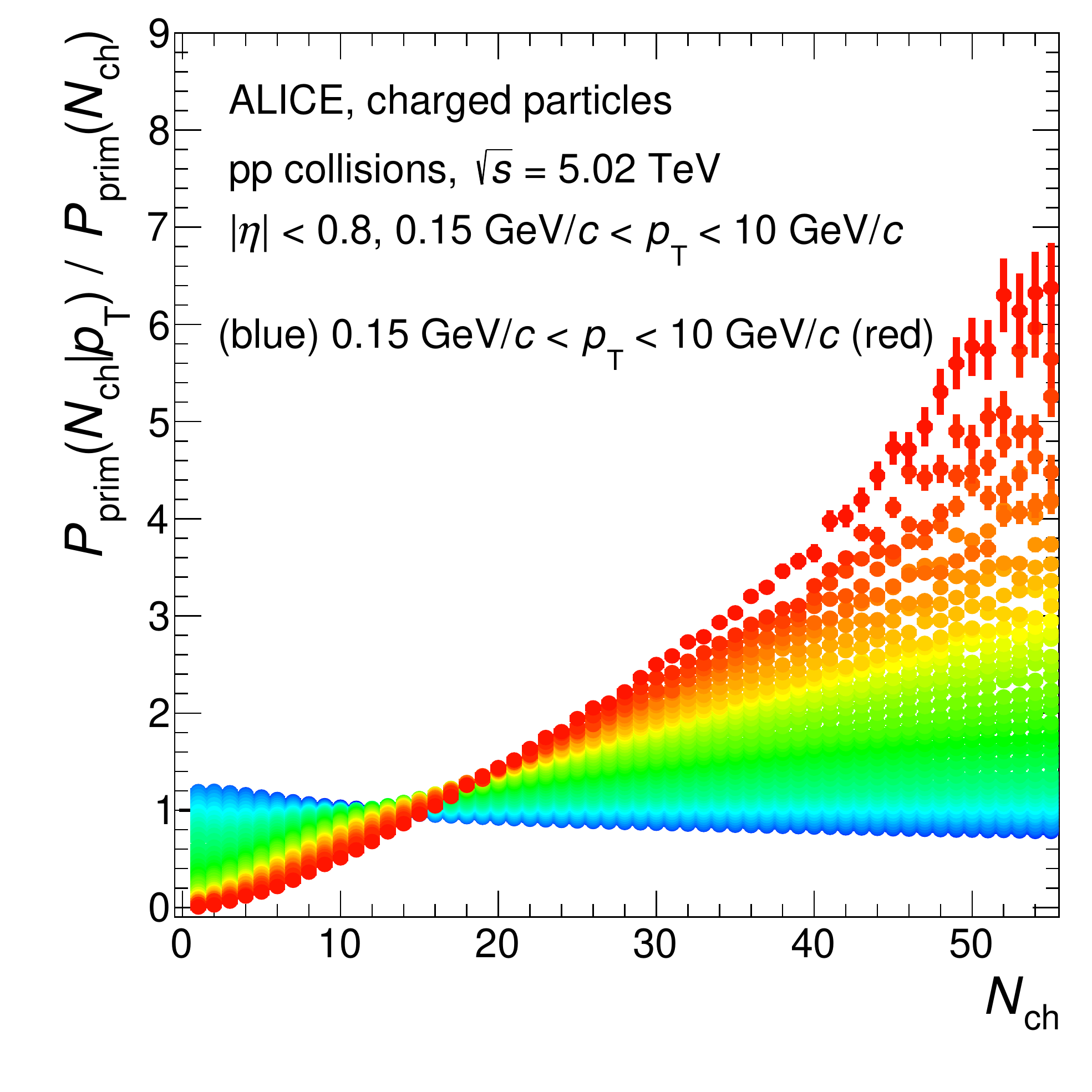}
	\caption{Top panel: the correlation of primary charged particle \pt spectra with multiplicity per $\nch > 0$ event for \pp collisions at $\sqrt{s} = 5.02\tev$.
	Bottom panels: the corresponding relative change of \pt (left) and \nch (right) distributions with respect to the inclusive ones. In the left panel, each of the curves represents a single \nch value, ranging from $\nch = 1$ (blue) to $\nch = 55$ (red). In the right panel, the colours represent the \pt intervals used in this analysis from the lowest in blue to the highest one in red. 
	}
	\label{fig:spectral-shapes}
\end{figure}

In the top panel of Fig.~\ref{fig:spectral-shapes}, the evolution of spectral shapes with multiplicity obtained with this unfolding procedure is shown for pp collisions at $\sqrt{s} = 5.02\tev$.
This double-differential measurement not only allows characteristic properties of the spectra, e.g. \mpt, and \sigmapt, to be extracted but also allows a direct comparison of the spectral shape produced in collisions with different multiplicities.
The bottom left panel of Fig.~\ref{fig:spectral-shapes} shows the ratio of multiplicity-dependent self normalised \pt distributions $P_{\textrm{prim}}(\pt|\nch)$ to the multiplicity-integrated self normalised \pt distribution $P_{\textrm{prim}}(\pt)$ of primary charged particles. 
A hardening of the spectra with multiplicity is apparent, which continuously increases with \nch, a trend observed earlier in coarser multiplicity intervals~\cite{Adam:2015pza} and with different event selection methods~\cite{ALICE:2019dfi}.
In the bottom right panel of Fig.~\ref{fig:spectral-shapes} the self normalised multiplicity distribution of primary charged particles for fixed transverse momentum $P_{\textrm{prim}}(\nch|\pt)$ divided by the \pt-integrated \nch distribution $P_{\textrm{prim}}(\nch)$ is shown as a function of \nch. 
As expected, it is evident that high \pt particles are produced mostly in high-multiplicity events. 

\paragraph{Systematic uncertainties}
The accuracy of the corrections applied in this work depends on how well the measured track properties are reproduced by the MC simulations.
The systematic uncertainties related to a possible disagreement were estimated by varying the track-selection criteria in reasonable ranges.
A detailed list of those track quality criteria and their variations can be found in previous ALICE publications~\cite{ALICE:2018vuu,Acharya:2018eaq}.
In order to estimate the systematic uncertainty related to the particle-composition correction, the yields of identified particles are varied within their respective systematic uncertainties and the extrapolation of those spectra to $\pt = 0.15\gevc$ is performed with different functions.
In addition, an uncertainty for the accuracy of the unfolding procedure is assigned that is quantified by the level of agreement between the unfolded simulated measurement and the expected distributions produced by the generator (MC closure test).
For each variation, the fully corrected results are calculated and their deviation to the nominal result is considered an uncertainty. Therefore, any effect of the variations on the performance of the analysis procedure is included in the respective systematic uncertainty.
To obtain the overall systematic uncertainties, all individual contributions are assumed to be fully uncorrelated and are added in quadrature.

The systematic uncertainties of multiplicity distributions are around $2\==5\%$ at low \nch, but increase towards higher multiplicities up to $10\==20\%$, depending on the collision system and energy.
While for \pp and \ppb collisions the track quality requirements are the most relevant contributions, in AA collisions the systematic uncertainty related to the particle-composition correction is the most prominent one.
The systematic uncertainty of the mean transverse momentum of the unfolded spectra is largely dominated by the contributions from track-selection variations, yet the total systematic uncertainties in most of the reported \nch range are around $1\%$.
At the lowest and highest multiplicities the contribution from the closure tests increases to up to $2\%$.
The contribution from the particle-composition correction is around $0.5\%$.
The systematic uncertainties of the standard deviation of the spectra \sigmapt are dominated by the track quality requirements and are below $2\%$ on average.
At very low and very high \nch, the systematic uncertainties go up to $3\==5\%$ due to the MC closure test contribution.

%% file: Content/4--Results.tex
\section{Results and discussion}\label{sec:results}

\begin{table}[t]
\center
\caption{Global characteristics of the analysed data sets with corresponding systematic uncertainties. Both the \nch and the \pt spectra are consistently defined in the kinematic range ${|\eta|~<~0.8}$ and \ptRange and only events with $\nch > 0$ are considered.}
\begin{tabular}{@{}lccccc@{}} 
\hline\hline
 & \sqrsn (TeV) & \mnch & \sigmanch & $\mpt_\mathrm{incl}$ ($\mathrm{MeV}/c$) & $\sigmapt_\mathrm{incl}$ ($\mathrm{MeV}/c$)\\
\hline
\multirow{5}{*}{\pp} & $2.76$ & $7.18 \pm 0.24$ & $6.05 \pm 0.17$ & $589.7 \pm 2.6$ & $483 \pm 4$ \\
   & $5.02$ & $8.21 \pm 0.10$ & $7.20 \pm 0.08$ & $612.2 \pm 2.7$ & $520.2 \pm 1.0$ \\
   & $7$    & $8.86 \pm 0.12$ & $7.88 \pm 0.11$ & $627.1 \pm 1.6$ & $541.3 \pm 2.1$ \\
   & $8$    & $9.05 \pm 0.22$ & $8.11 \pm 0.18$ & $631 \pm 5$ & $547 \pm 4$ \\
   & $13$   & $10.31 \pm 0.09$~~ & $9.48 \pm 0.07$ & $654.0 \pm 1.0$ & $582.4 \pm 0.9$ \\
  \hline
\multirow{2}{*}{\ppb} & $5.02$ & $25.51 \pm 0.25$~~ & $19.79 \pm 0.20$~~ & $711.9 \pm 1.3$ & $619.8 \pm 1.1$ \\
      & $8.16$ & $29.56 \pm 0.26$~~ & $23.13 \pm 0.23$~~ & $741.5 \pm 1.4$ & $657.0 \pm 1.3$ \\
\hline
\xexe & $5.44$ & $458 \pm 10$~~ & $514 \pm 13$~~ & $717.4 \pm 1.8$ & $568.4 \pm 1.4$ \\
\hline
\multirow{2}{*}{\pbpb} & $2.76$ & $573 \pm 9$~~~~ & $667 \pm 12$~~ & $687.3 \pm 1.3$ & $528.0 \pm 1.7$ \\
       & $5.02$ & $682 \pm 13$~~ & $819 \pm 16$~~ & $724.1 \pm 1.1$ & $564.9 \pm 1.0$ \\
 \hline\hline
\end{tabular}
\label{tab:globalProp}
\end{table}

Multiplicity distributions as well as the mean and standard deviation of charged-particle \pt spectra as a function of \nch are presented in comparison with model predictions.
Table~\ref{tab:globalProp} summarises the mean and standard deviation of both the multiplicity distributions and \nch-integrated \pt spectra for \pp, \ppb, \xexe, and \pbpb collisions at the various centre-of-mass energies per nucleon pair.

\begin{figure}[htb]
	\center
	\includegraphics[width=\plotWidth]{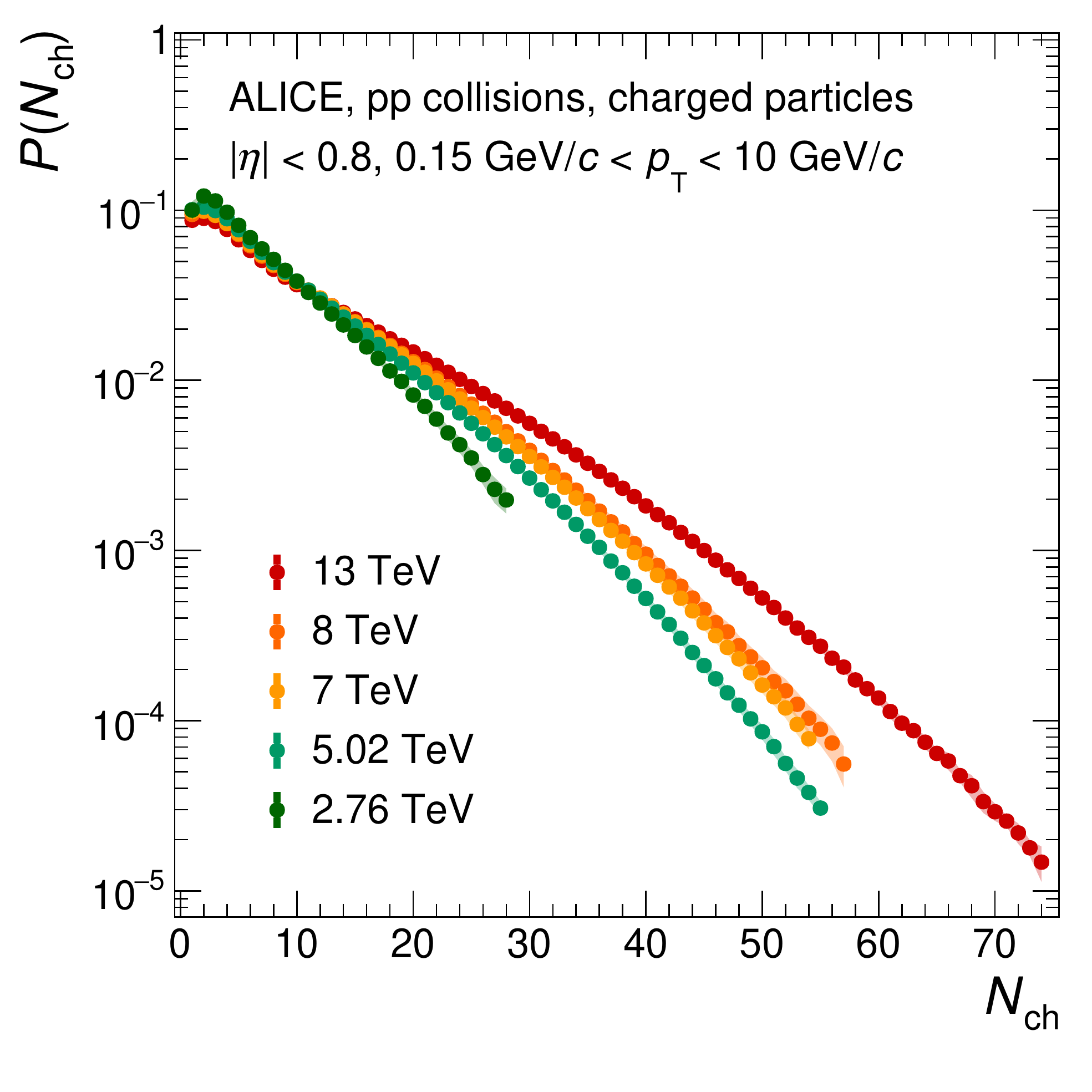}
	\hspace{\plotDistance}
	\includegraphics[width=\plotWidth]{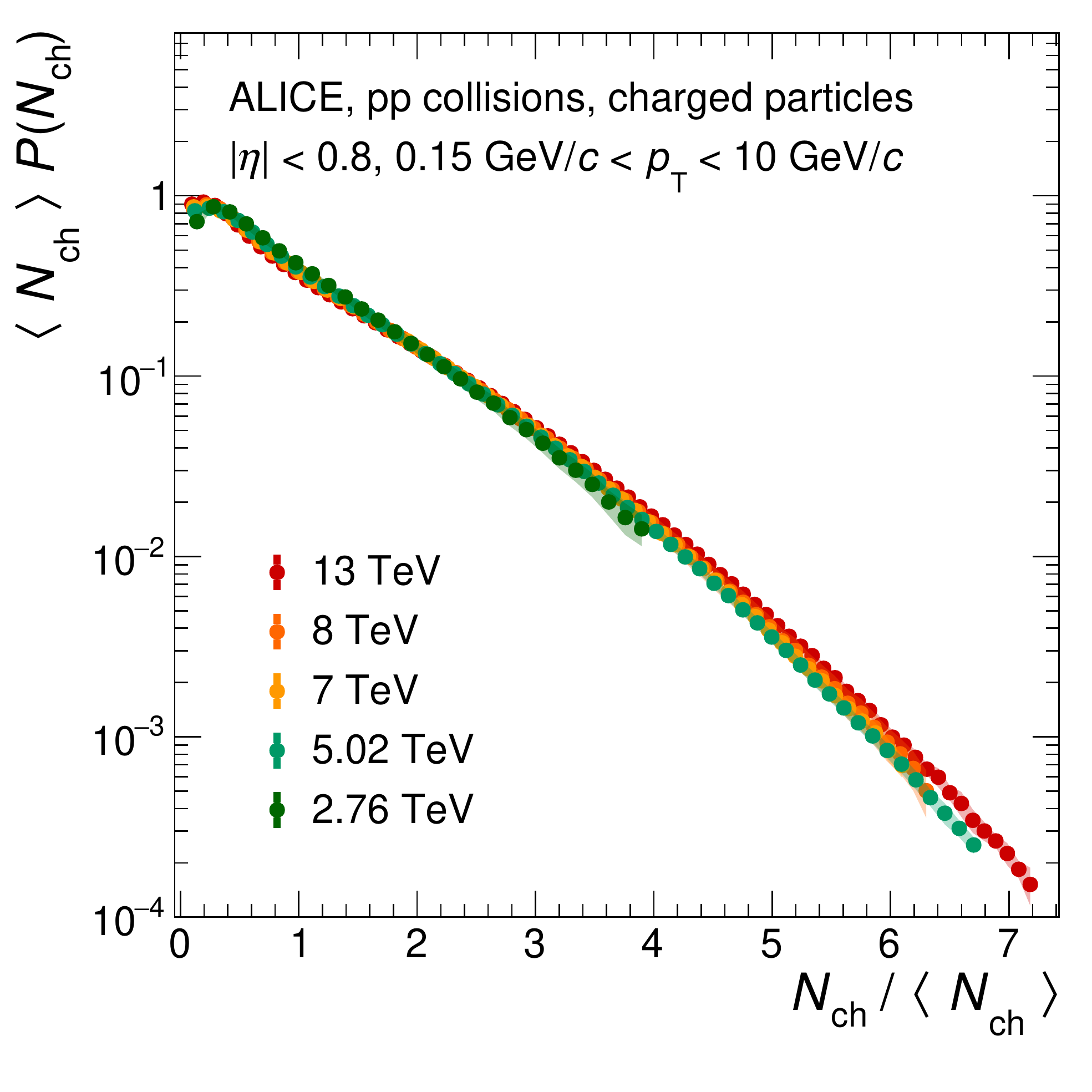}
	\\
	\includegraphics[width=\plotWidth]{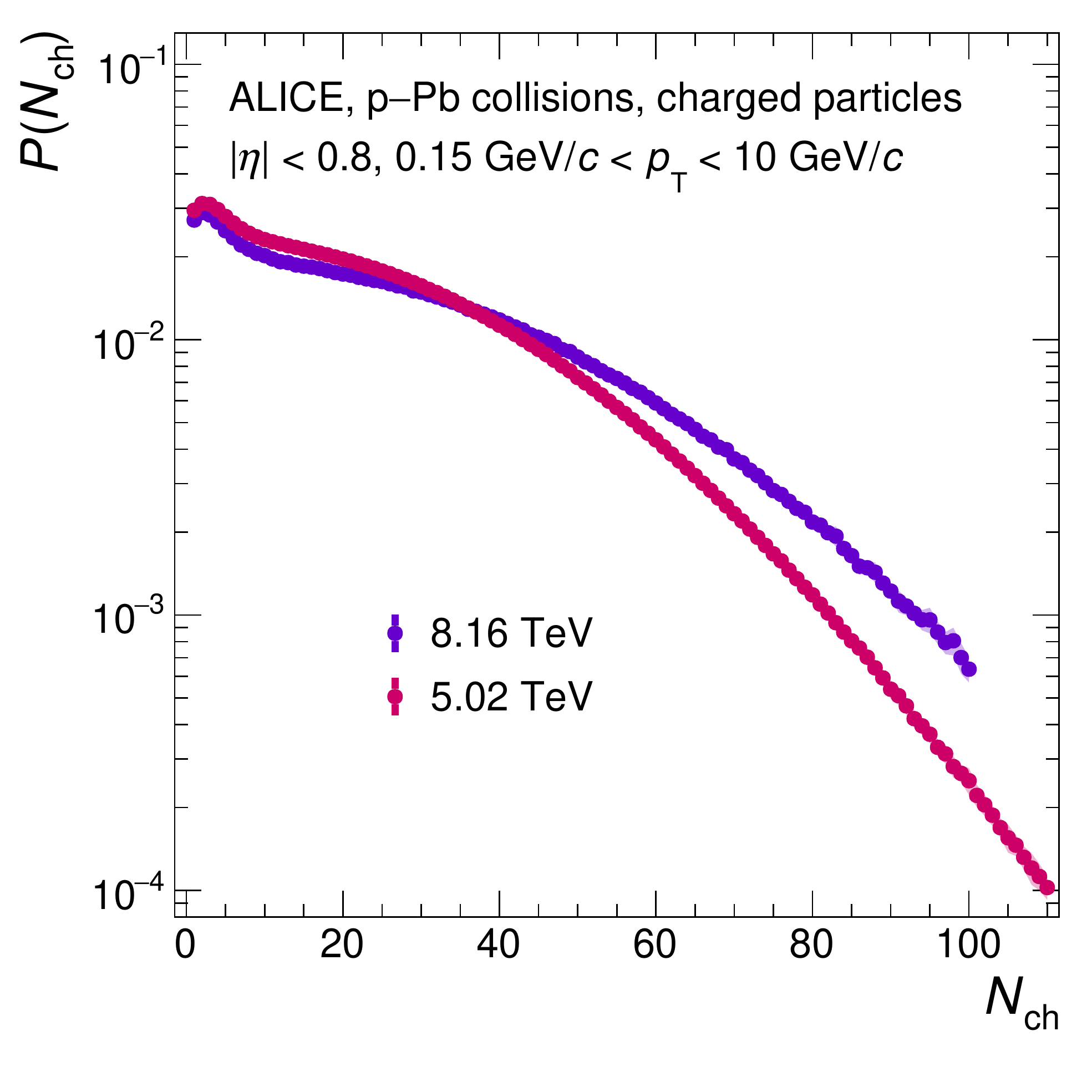}
	\hspace{\plotDistance}
	\includegraphics[width=\plotWidth]{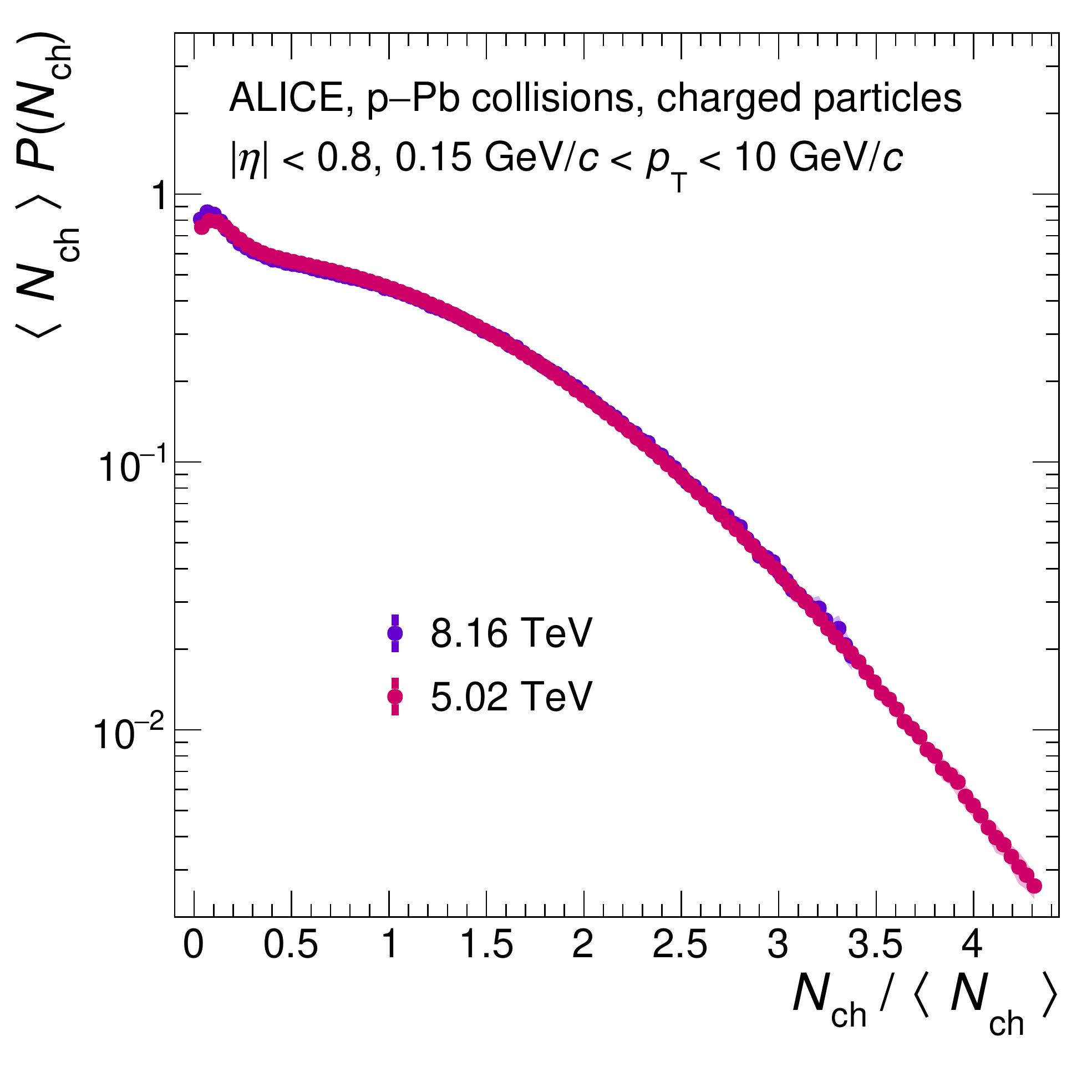}
	\\
	\includegraphics[width=\plotWidth]{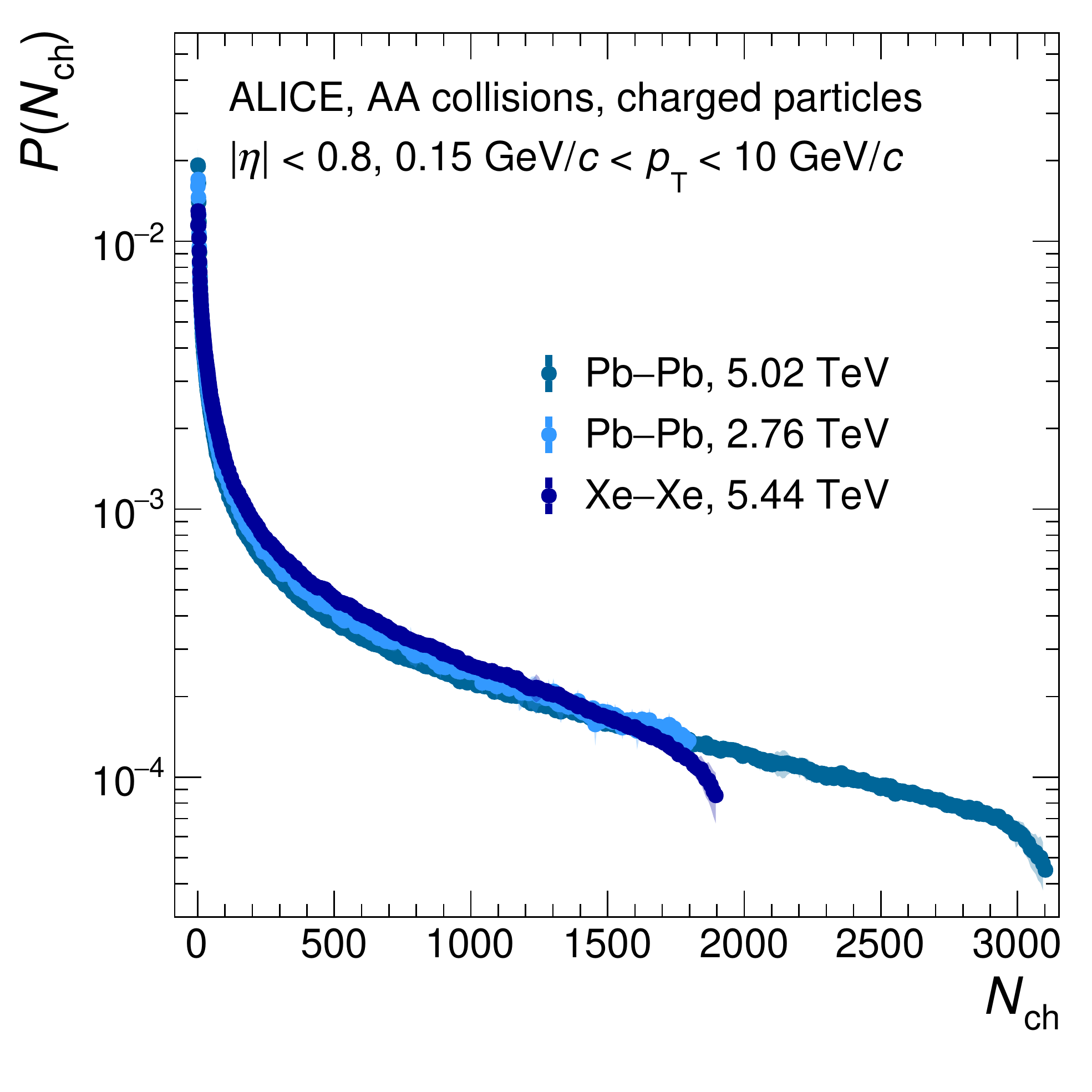}
	\hspace{\plotDistance}
	\includegraphics[width=\plotWidth]{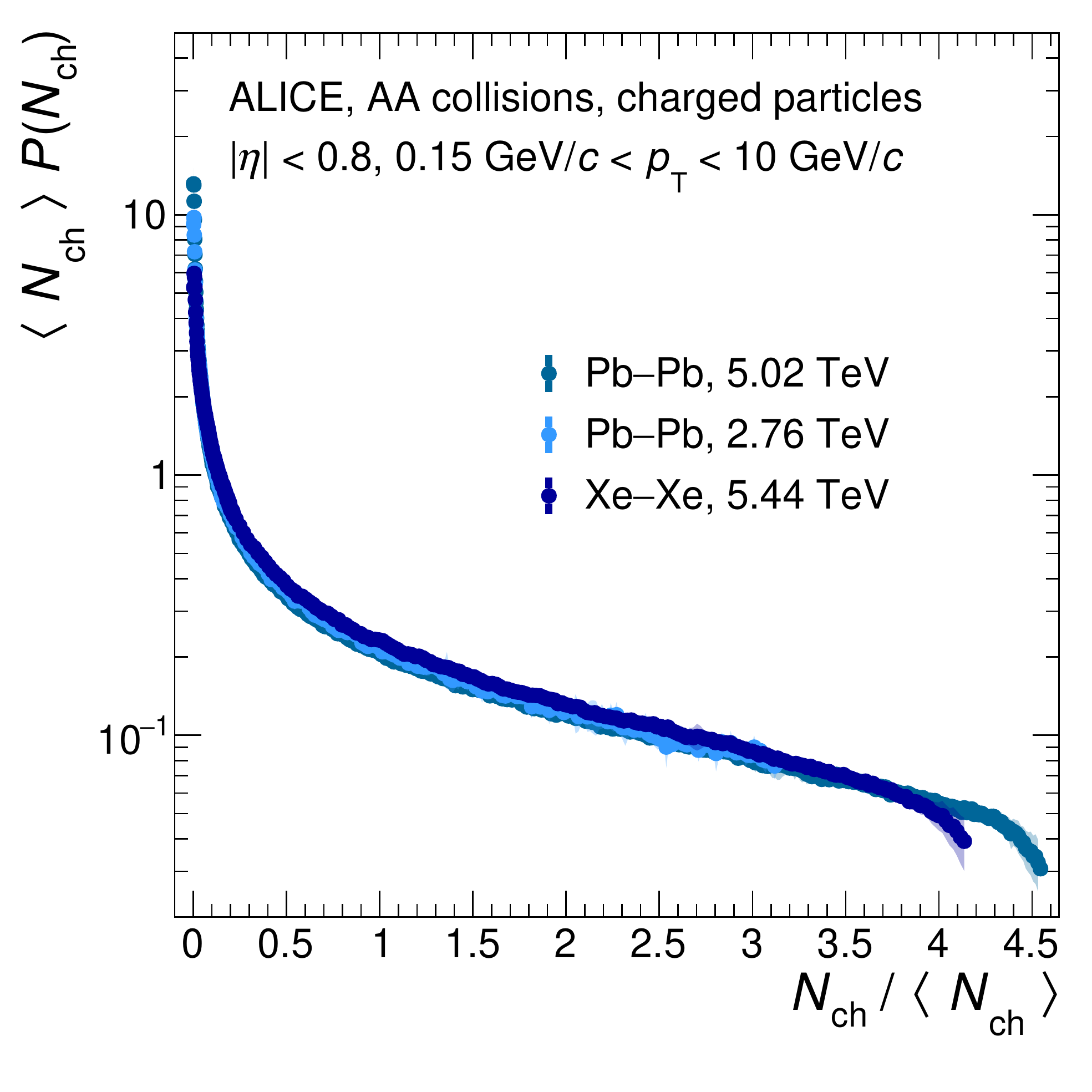}
	\caption{Probability density of charged-particle multiplicity \nch (left) and the corresponding KNO-scaled distributions (right) for \pp (top), \ppb (middle), and AA (bottom) collisions at different centre-of-mass energies per nucleon pair. Statistical and systematic uncertainties are shown as bars and semi-transparent bands, respectively.}
	\label{fig:multDist}
\end{figure}

\begin{figure}[htb]
	\center
	\includegraphics[width=\plotWidth]{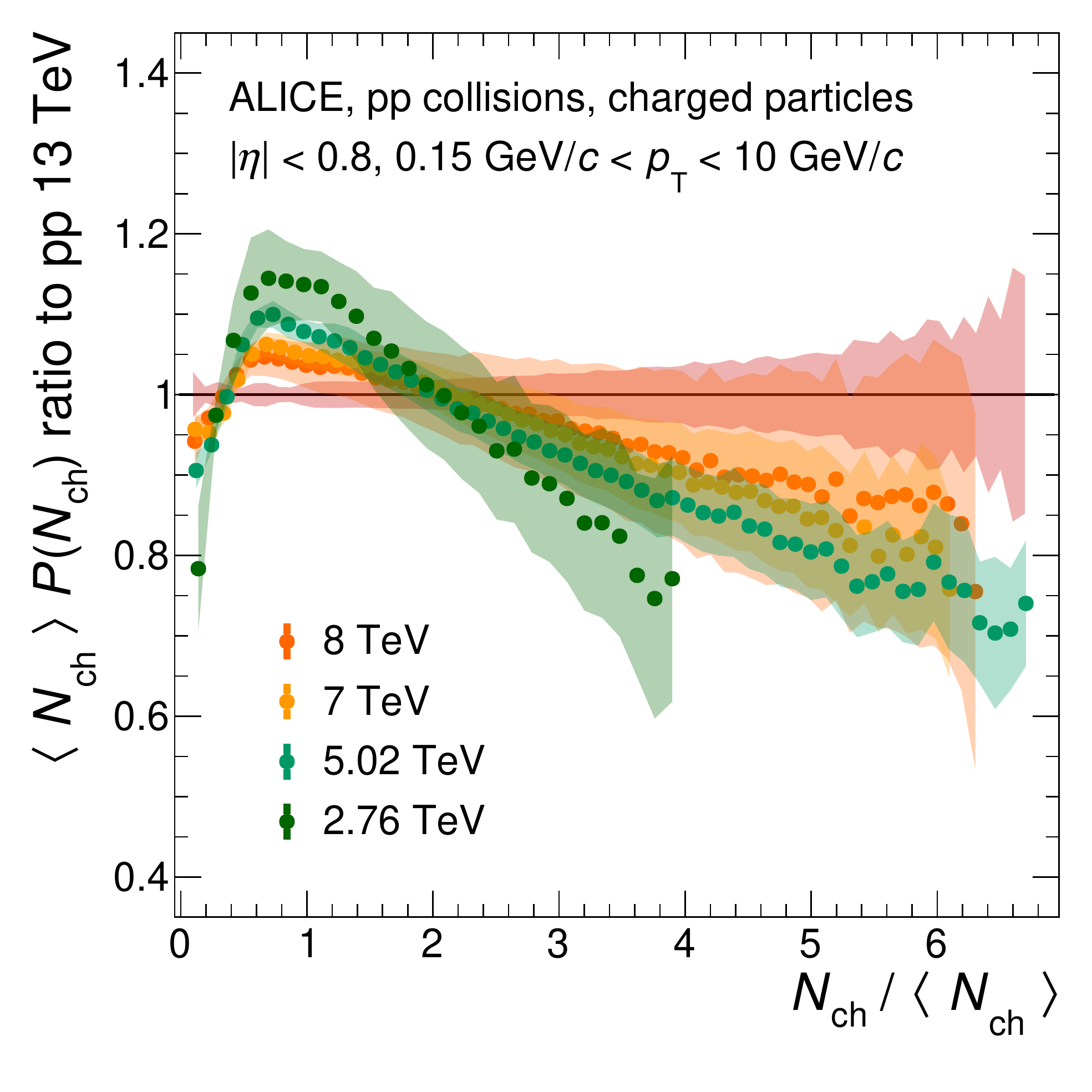}
	\\
	\includegraphics[width=\plotWidth]{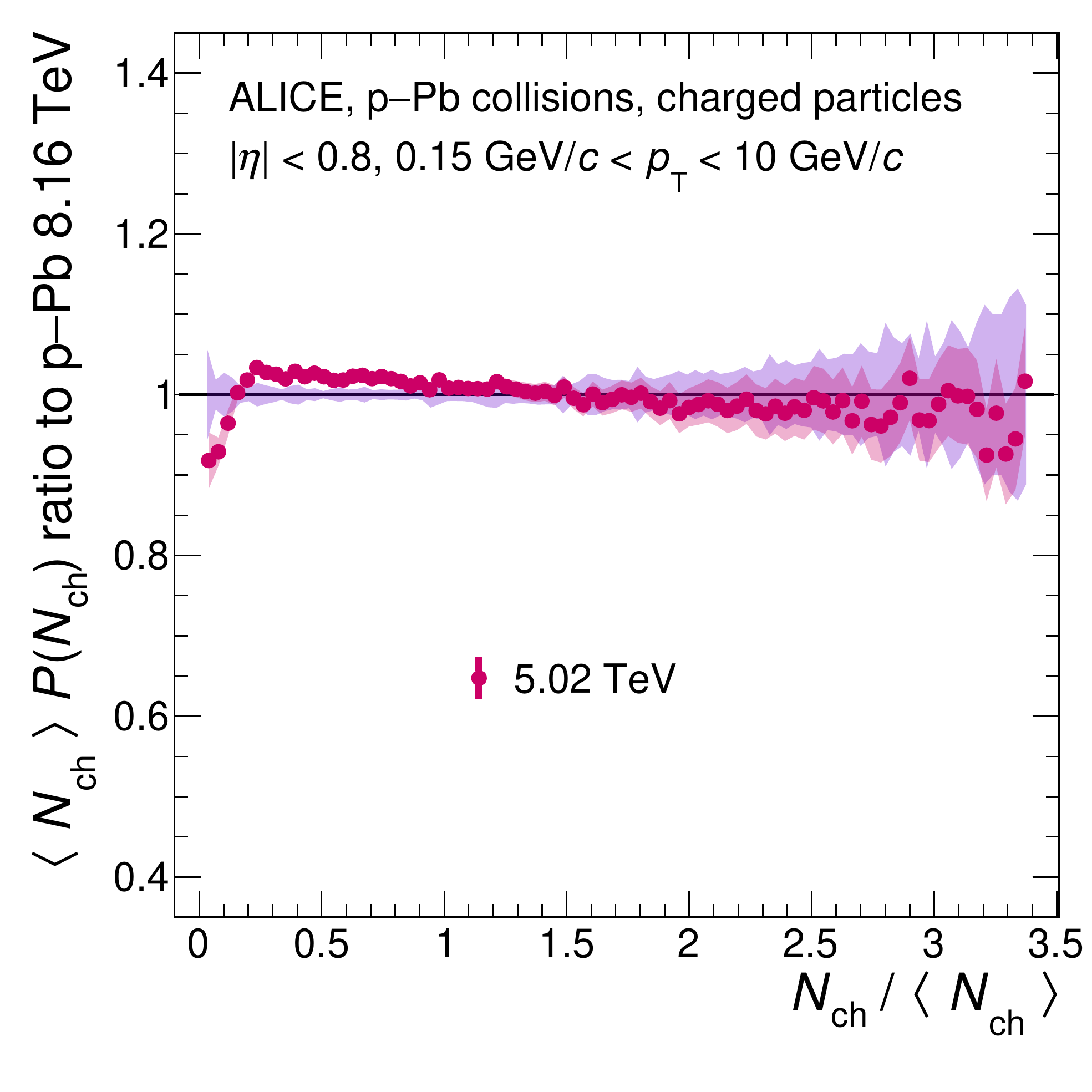}
	\hspace{\plotDistance}
    \includegraphics[width=\plotWidth]{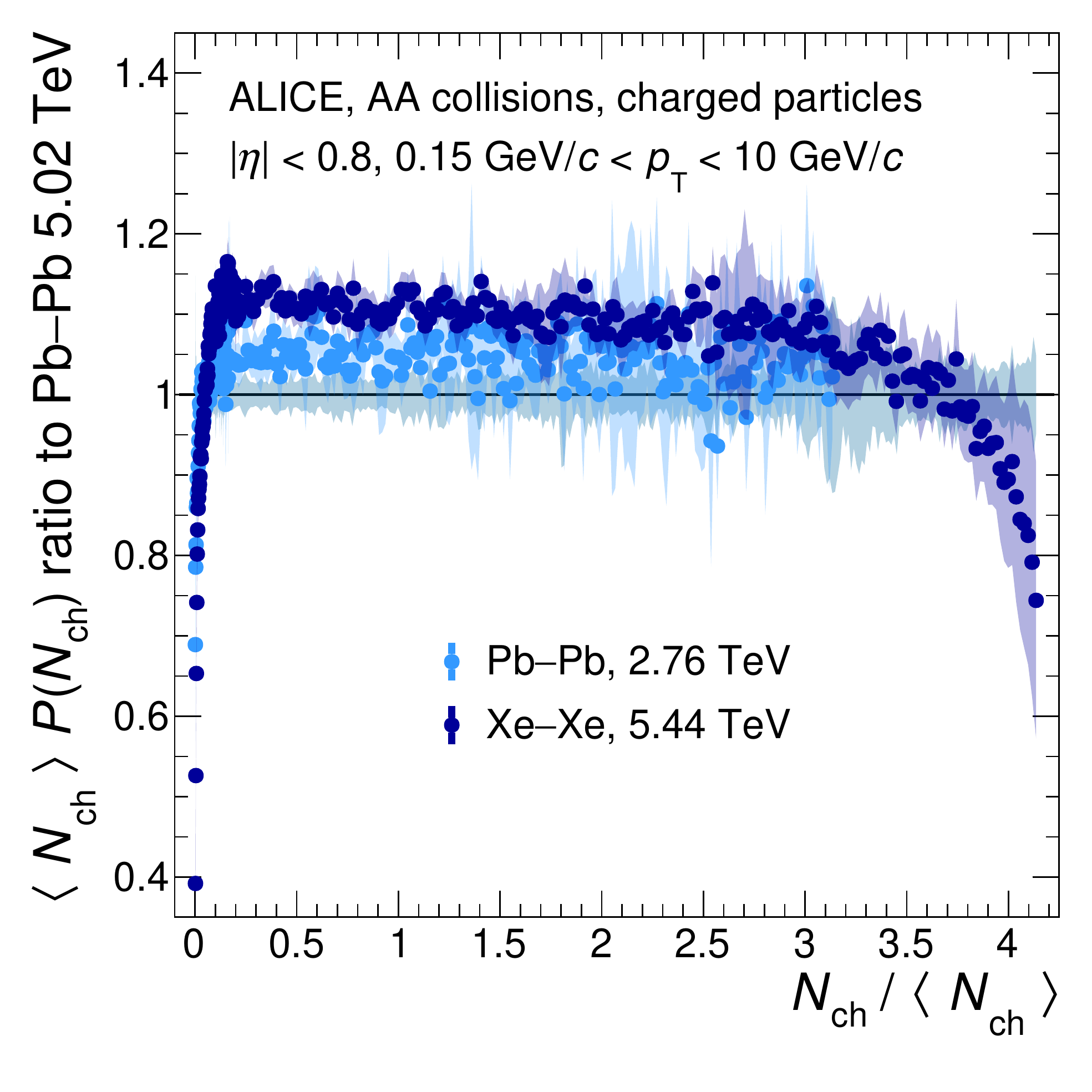}
	\caption{Ratios of the KNO-scaled multiplicity distributions at various centre-of-mass energies per nucleon pair relative to $\sqrs = 13\tev$ for \pp collisions (top panel) and relative to $\sqrsn = 8.16\tev$ and $\sqrsn = 5.02\tev$ for \ppb and \pbpb collisions, respectively (left and right bottom panels). Statistical and systematic uncertainties are shown as bars and semi-transparent bands, respectively.}
	\label{fig:multDistKNO_ratioData}
\end{figure}

The left panel of Fig.~\ref{fig:multDist} shows the probability density of charged-particle multiplicity \nch for \pp (top), \ppb (middle), and AA (bottom) collisions at different energies per nucleon pair.
For all collision systems, the distributions reach a maximum around $\nch \approx$~2 and then fall steeply off over several orders of magnitude. 
In the \pp and \ppb systems, the slope of the decay with \nch decreases with increasing collision energy. This can be attributed to the larger momentum transfer in the initial hard scattering which results in larger multiplicities.
The right panel of Fig.~\ref{fig:multDist} presents the data after scaling the probability density and the charged-particle multiplicity with the average number of charged particles $\langle \nch \rangle$ according to the Koba--Nielsen--Olesen (KNO)~\cite{KOBA1972317} scaling.
Figure~\ref{fig:multDistKNO_ratioData} shows the corresponding ratios of the KNO-scaled multiplicity distributions at various centre-of-mass energies per nucleon pair relative to $\sqrs = 13\tev$, $\sqrsn = 8.16\tev$ and $\sqrsn = 5.02\tev$ for pp, \ppb and \pbpb collisions, respectively.
KNO scaling apparently holds in \pp and AA collisions within $20\%$, and in \ppb collisions within $10\%$.

\begin{figure}[htb]
	\center
	\includegraphics[width=\plotWidth]{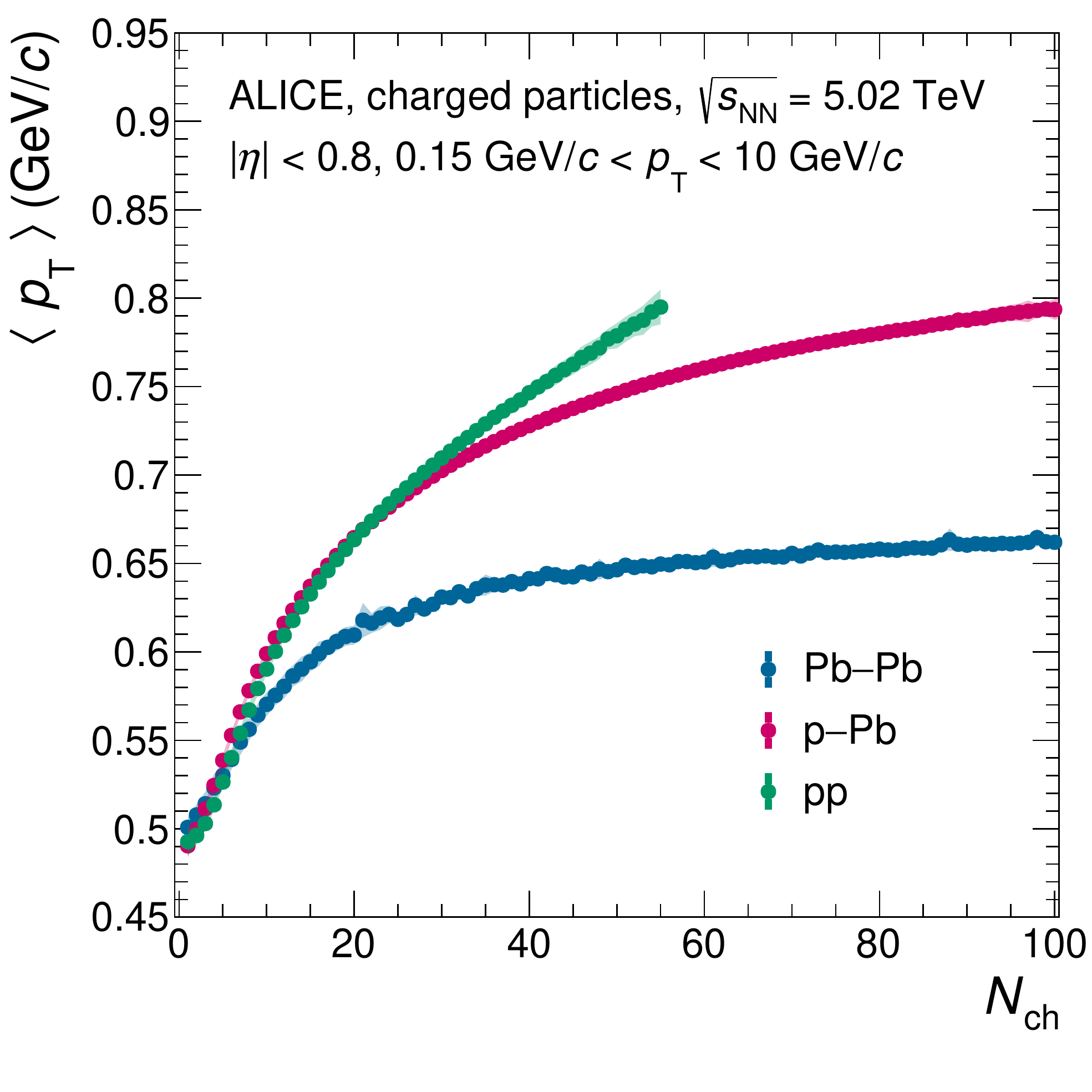}
	\hspace{\plotDistance}
	\includegraphics[width=\plotWidth]{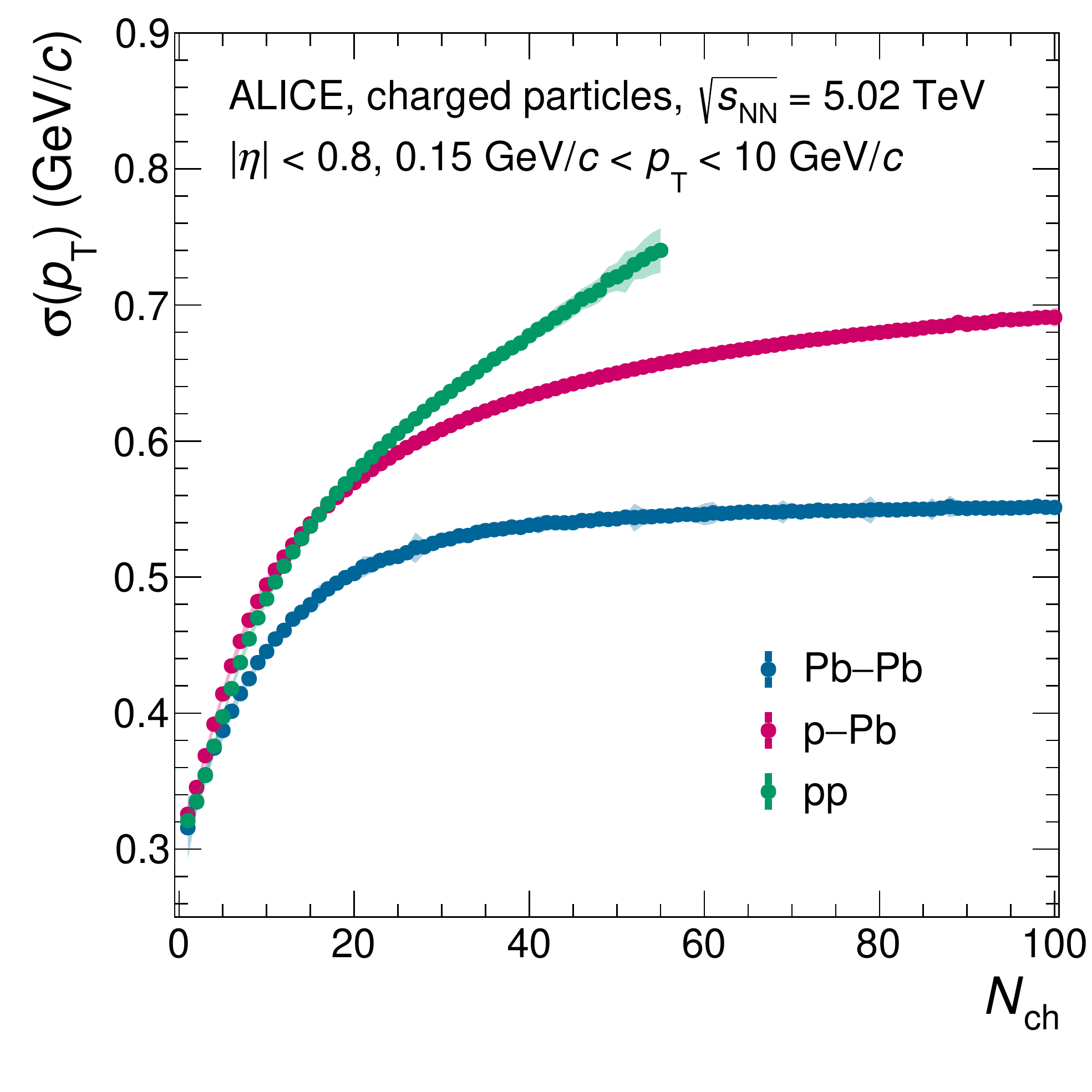}
	\caption{Mean (left) and standard deviation (right) of the charged-particle transverse momentum spectra as a function of the charged-particle multiplicity for \pp, \ppb, and \pbpb collisions at a centre-of-mass energy per nucleon pair of $\sqrsn = 5.02\tev$. Statistical and systematic uncertainties are shown as bars and semi-transparent bands, respectively.}
	\label{fig:mpt-sigmapt-5TeV}
\end{figure}

\begin{figure}[htb]
	\center
	\includegraphics[width=\plotWidth]{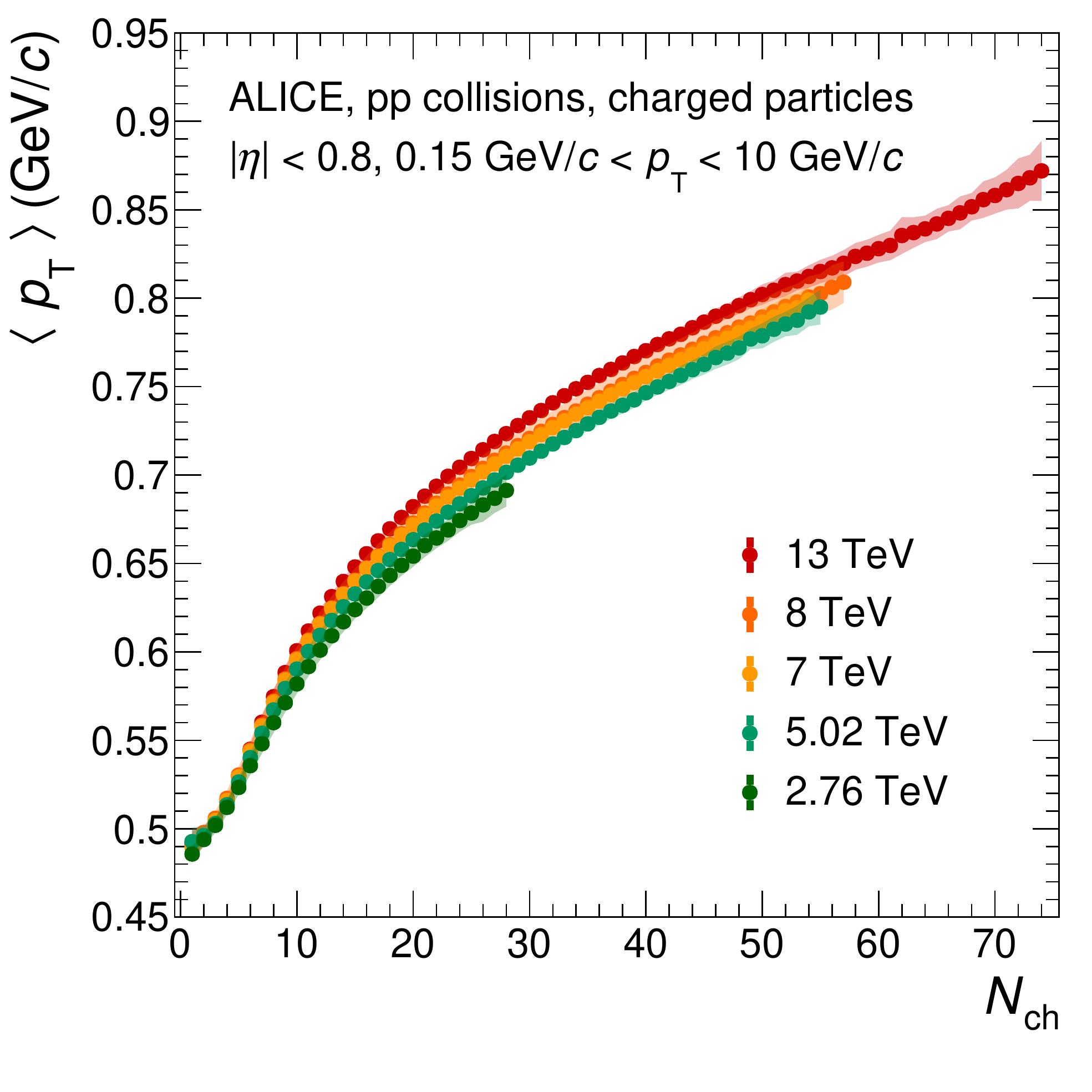}
	\hspace{\plotDistance}
    \includegraphics[width=\plotWidth]{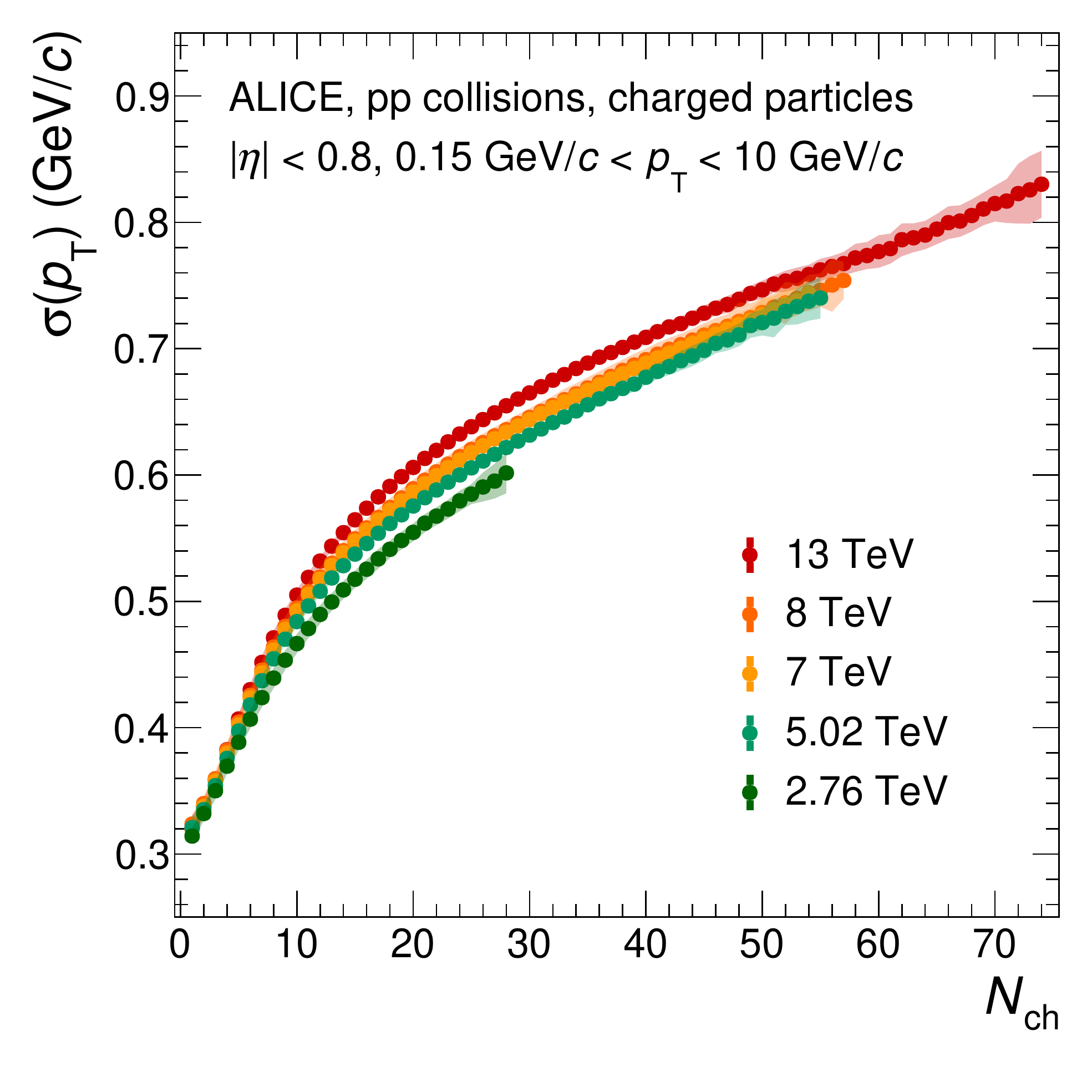}
	\\
	\includegraphics[width=\plotWidth]{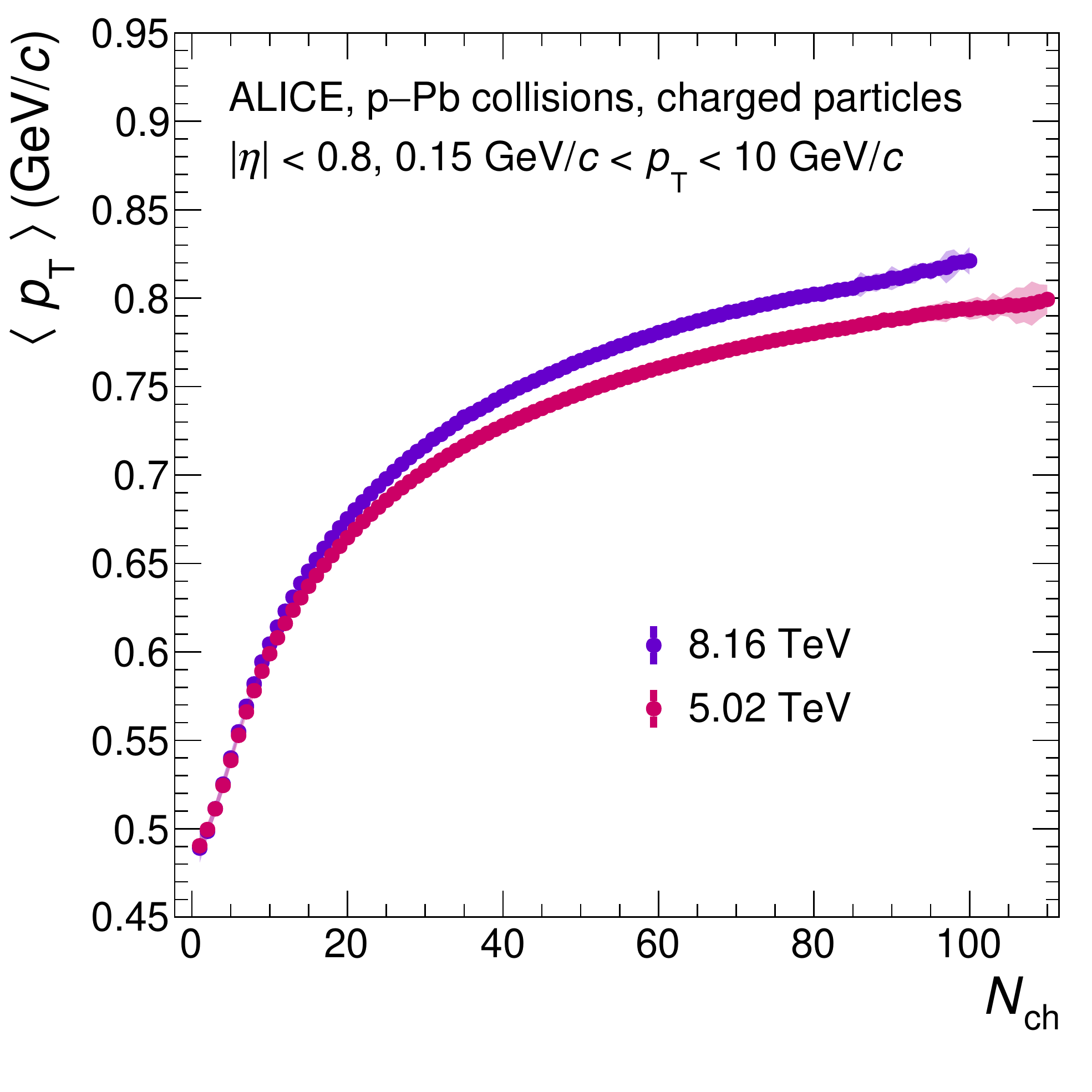}
	\hspace{\plotDistance}
	\includegraphics[width=\plotWidth]{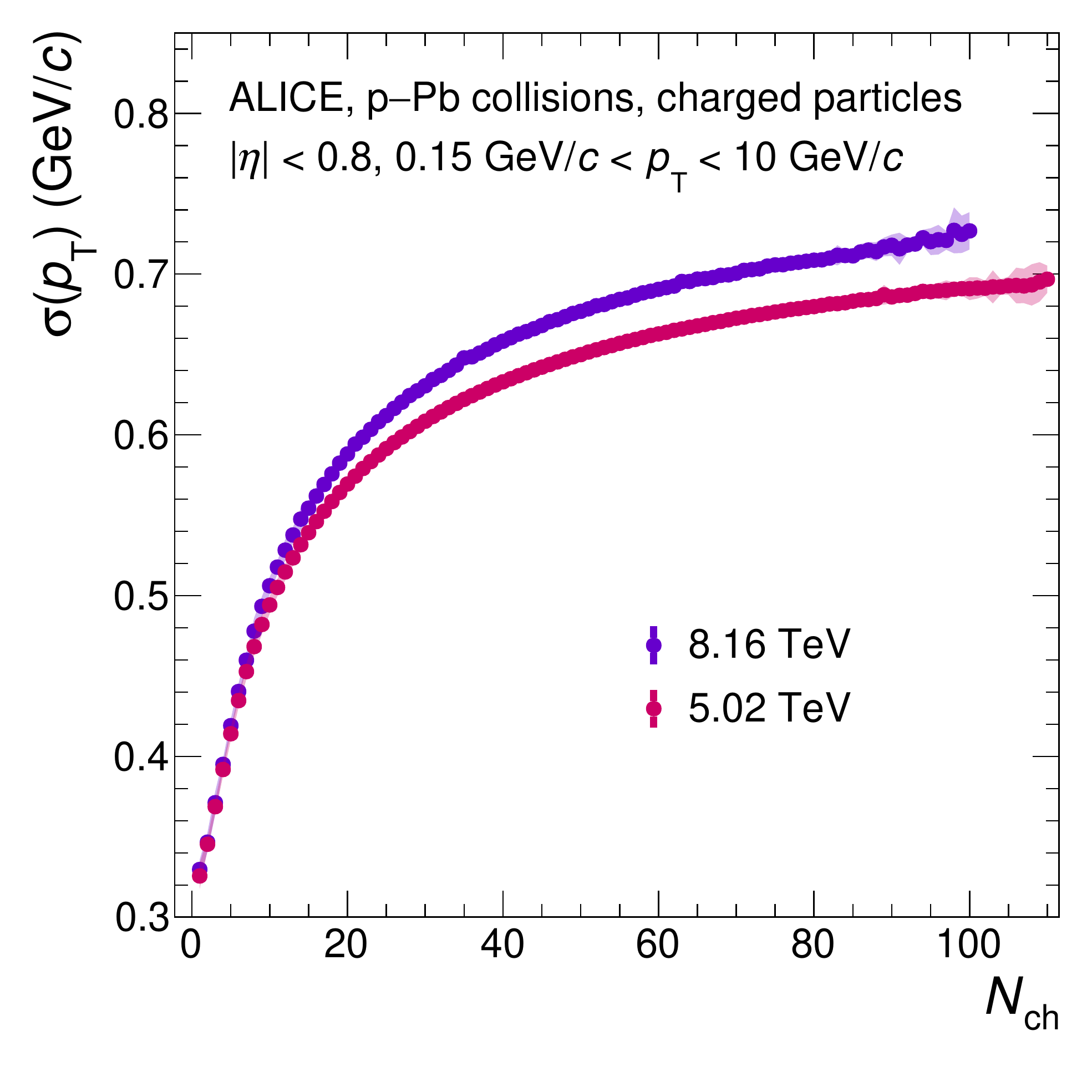}
	\\
	\includegraphics[width=\plotWidth]{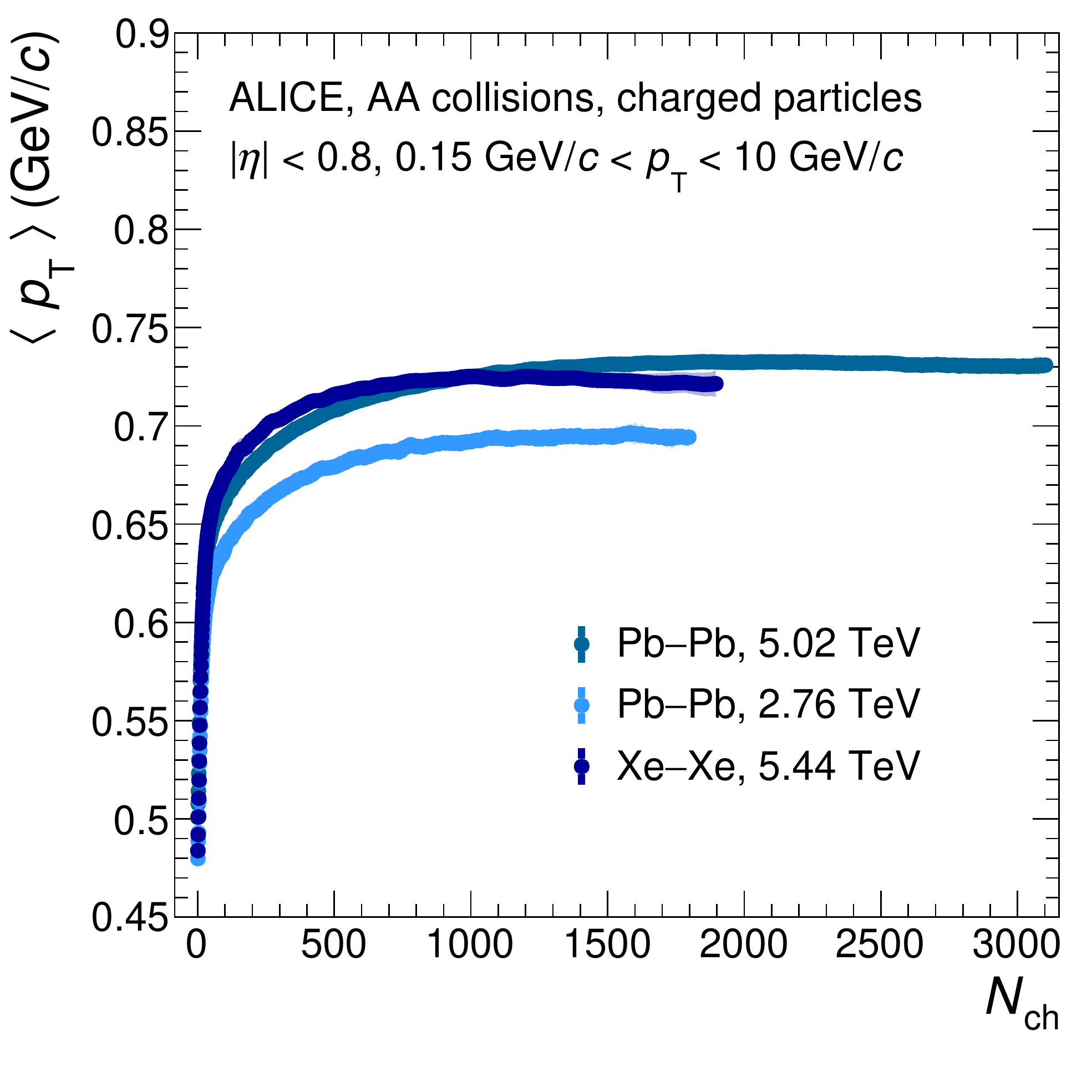}
	\hspace{\plotDistance}
	\includegraphics[width=\plotWidth]{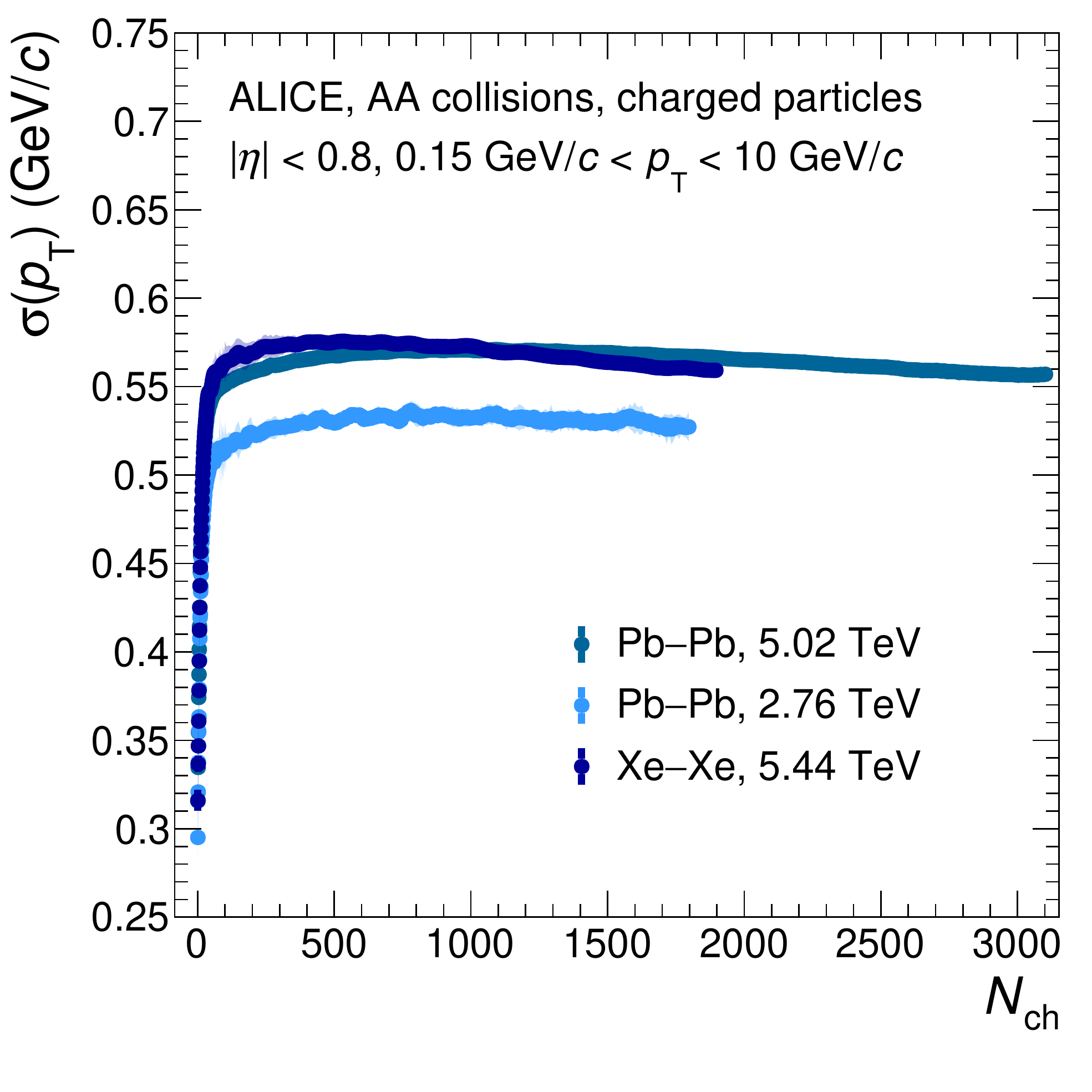}
	\caption{Mean (left) and standard deviation (right) of the charged-particle transverse momentum spectra as a function of the charged-particle multiplicity for \pp (top), \ppb (middle), and AA (bottom) collisions at different centre-of-mass energies per nucleon pair. Statistical and systematic uncertainties are shown as bars and semi-transparent bands, respectively.}
	\label{fig:mpt-sigmapt}
\end{figure}

\begin{figure}[htb]
	\center
	\includegraphics[width=\plotWidth]{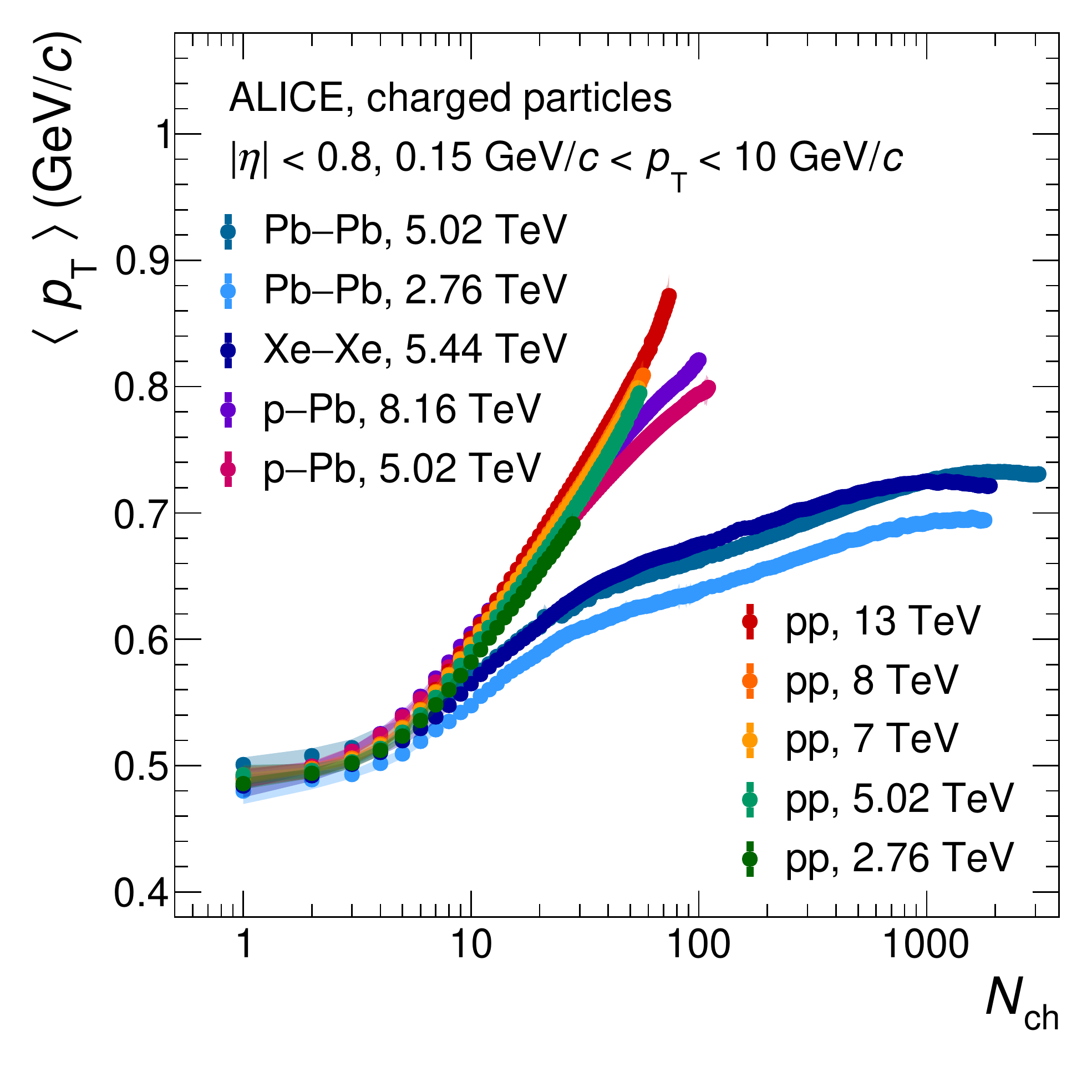}
	\hspace{\plotDistance}
	\includegraphics[width=\plotWidth]{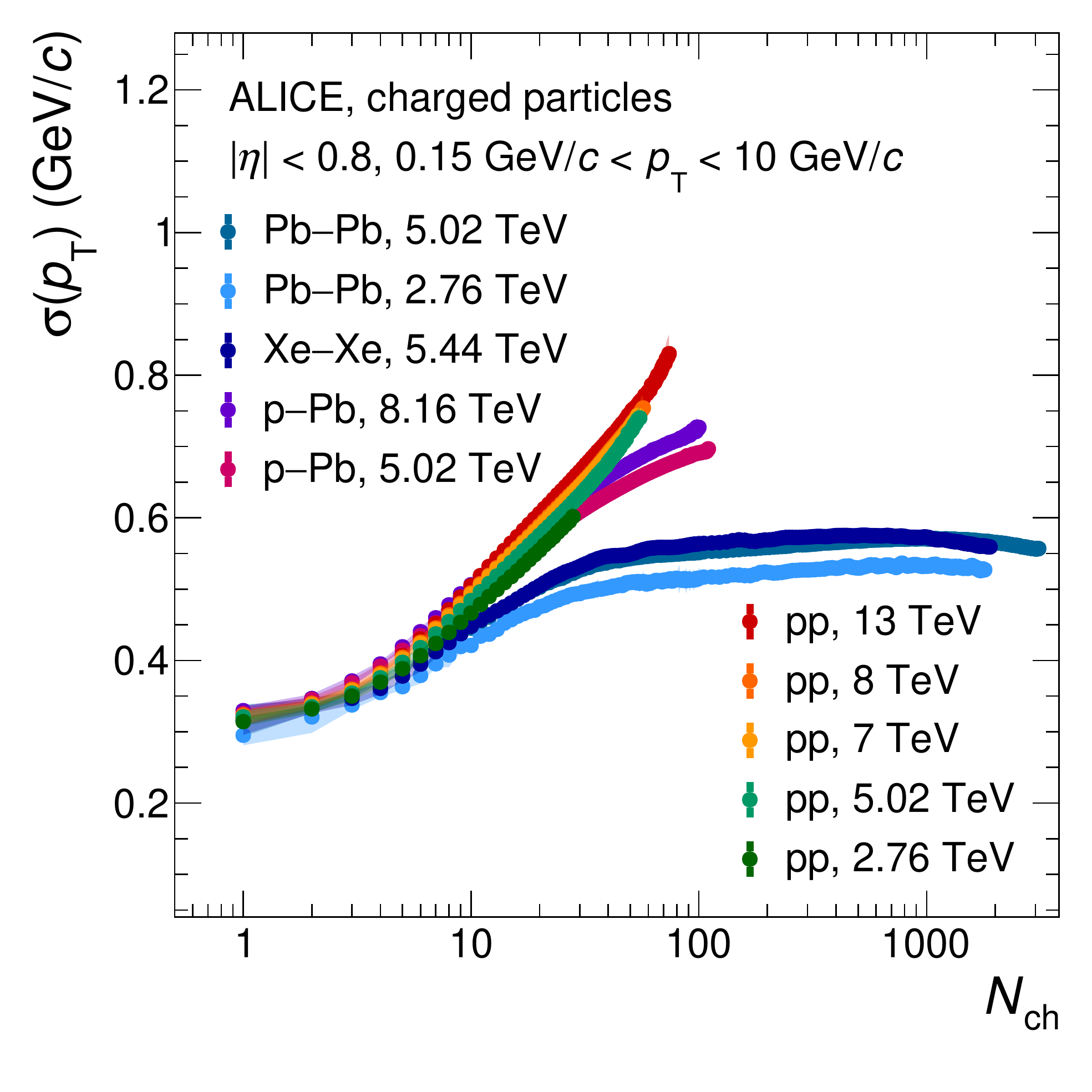}
    \\
	\includegraphics[width=\plotWidth]{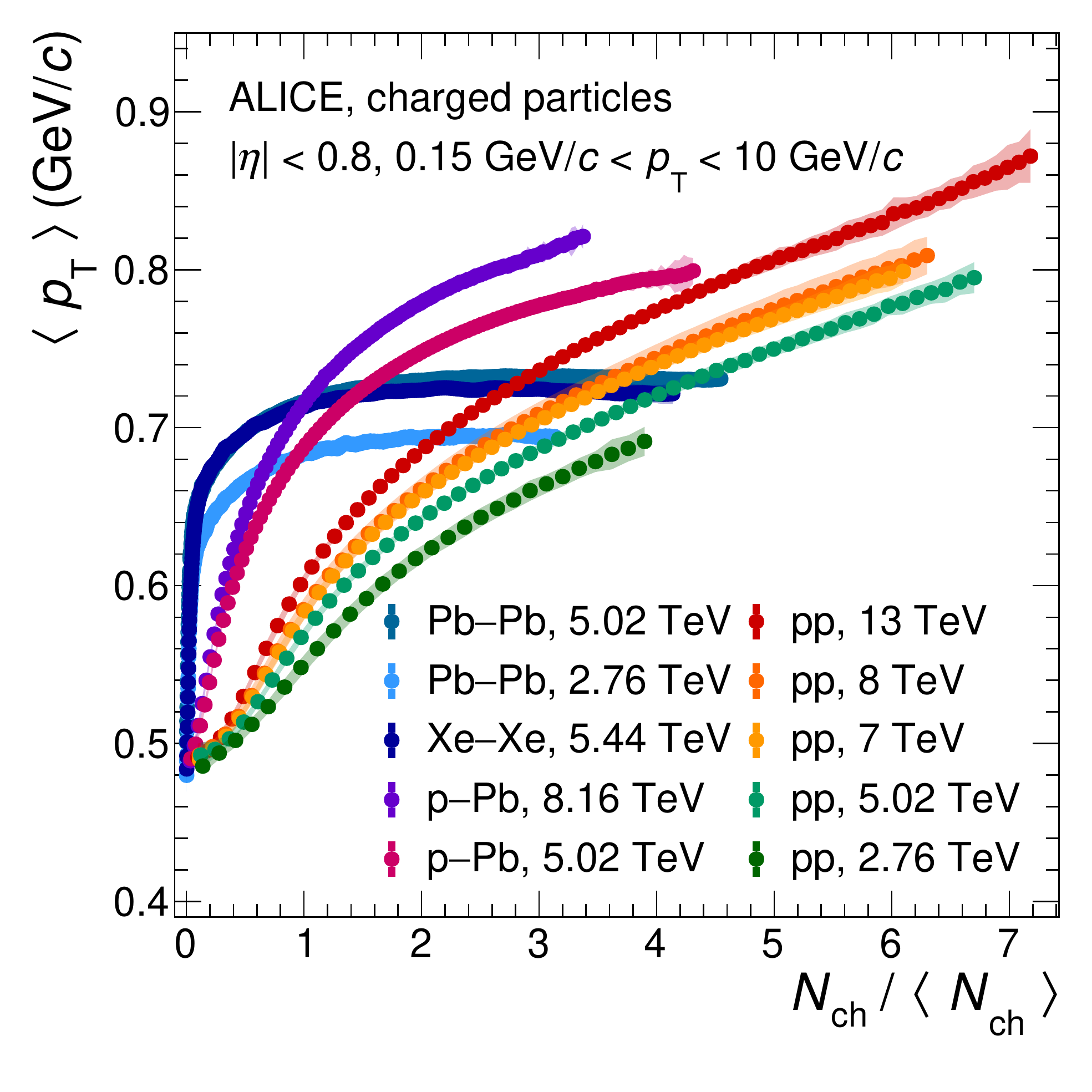}
	\hspace{\plotDistance}
	\includegraphics[width=\plotWidth]{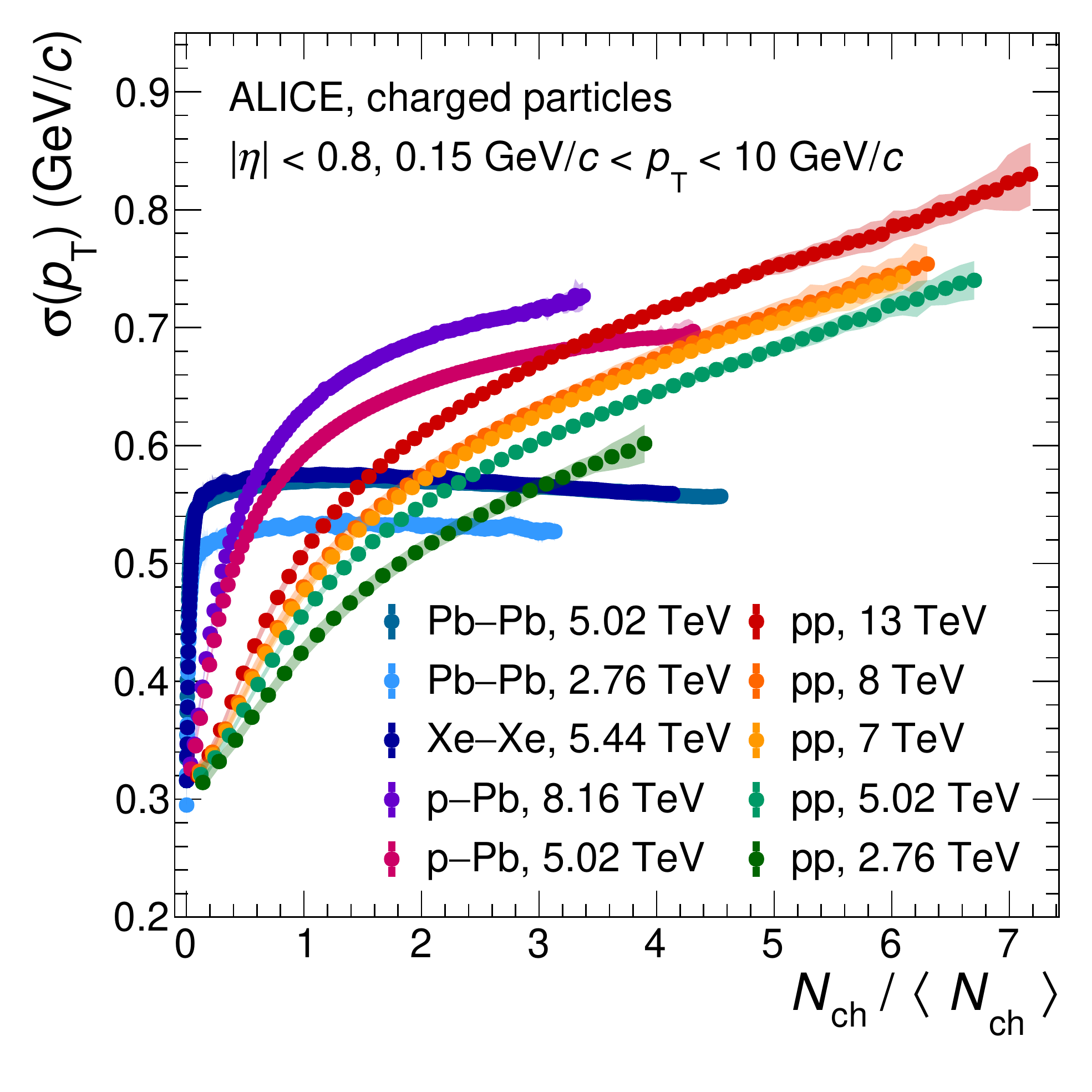}
	\\
	\includegraphics[width=\plotWidth]{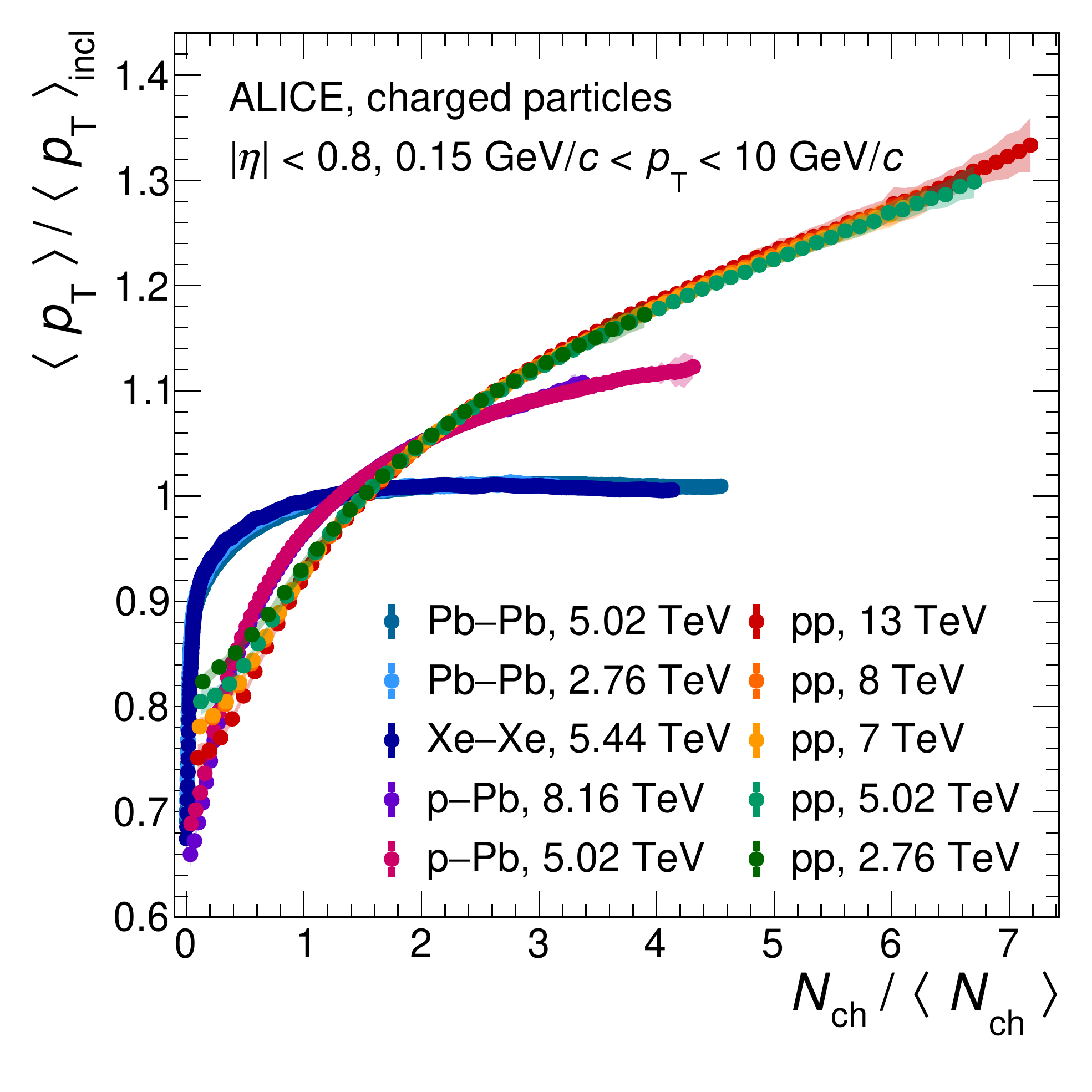}
	\hspace{\plotDistance}
	\includegraphics[width=\plotWidth]{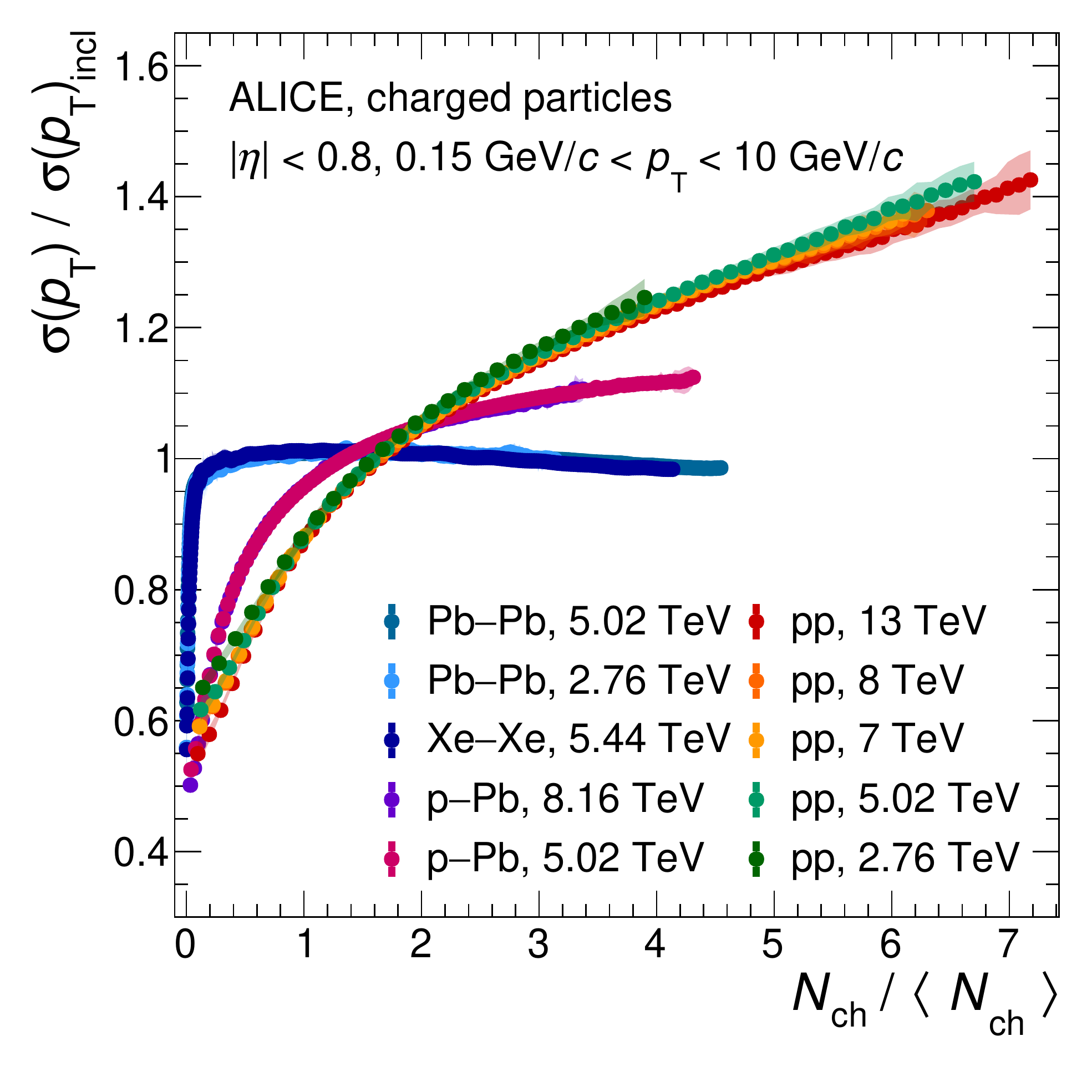}
	\caption{Mean (left) and standard deviation (right) of the charged-particle transverse momentum spectra as a function of the charged-particle multiplicity (top) and relative multiplicity $\nch / \mnch$ (middle, bottom) for \pp, \ppb, \xexe and \pbpb collisions at various centre-of-mass energies per nucleon pair. The bottom panels show both quantities relative to their multiplicity-inclusive value. 
	Statistical and systematic uncertainties are shown as bars and semi-transparent bands, respectively.}
	\label{fig:mpt-sigmapt-all}
\end{figure}

In Fig.~\ref{fig:mpt-sigmapt-5TeV} the mean and standard deviation of the \pt spectra are compared for \pp, \ppb and \pbpb collisions at the same centre-of-mass energy per nucleon pair of $\sqrsn = 5.02\tev$.
All three collision systems have similar values at $\nch = 1$ and then coincide until \pbpb deviates at $\nch \approx 5$ and \ppb deviates at $\nch \approx 25$ from the trend observed in \pp.
This observation is consistent with an earlier comparison of the \mpt--\nch correlation for the three systems at different centre-of-mass energies~\cite{Abelev_2013}.
Figure~\ref{fig:mpt-sigmapt} shows the mean (left) and standard deviation (right) of the transverse momentum spectra as a function of the charged-particle multiplicity \nch for \pp (top), \ppb (middle), and AA (bottom) collisions at different centre-of-mass energies per nucleon pair.
For all collision systems, a clear ordering of \mpt as well as \sigmapt with collision energy is observed, which can be attributed to the larger momentum transfers involved at higher \sqrsn.
For \pp collisions at all centre-of-mass energies, the average transverse momentum increases monotonically with an almost linear trend up to \nch $\approx 16$ and beyond that continues with an again almost linear dependence on \nch but reduced slope. 
In \ppb collisions, a similar multiplicity dependence is observed up to $\nch \approx 25$. 
At higher multiplicities, the increase in \mpt is slower than in \pp collisions.
In both \pp and \ppb, \sigmapt follows a similar trend as \mpt.
On the other hand, for AA collisions one observes an increase of \mpt with multiplicity up to about one third of the measured range, followed by a constant trend for the rest of the \nch range. 
The \sigmapt increases for $\nch \lesssim 100$ to a maximum and decreases afterwards. 
This is unique to large collision systems and is presumably a consequence of flow and jet quenching~\cite{ALICE:2018vuu}.
The high \nch resolution of this measurement makes it possible to spot differences between the spectral evolution with multiplicity in \xexe and \pbpb collisions at $\sqrsn = 5.44\tev$ and $\sqrsn = 5.02\tev$, respectively.
The observed difference in the trends might be a result of the slightly deformed Xe nuclei~\cite{Giacalone}.
In Fig.~\ref{fig:mpt-sigmapt-all} both the mean (left) and standard deviation (right) of the \pt spectra as a function of \nch are summarised for all data sets (top panels) and then shown as a function of relative multiplicity $\nch / \mnch$ (middle panels) as well as divided by their respective multiplicity-integrated values (bottom panels).
In the latter scaling, the overall energy-dependent increase of average kinematic energy and number of produced particles are accounted for.
As a result, the values for each collision system align almost perfectly for $\mpt / \mpt_{\mathrm{incl}}$ and $\sigmapt / \sigmapt_{\mathrm{incl}}$.

\clearpage

\begin{figure}[htb]
	\center
	\includegraphics[width=\plotWidth]{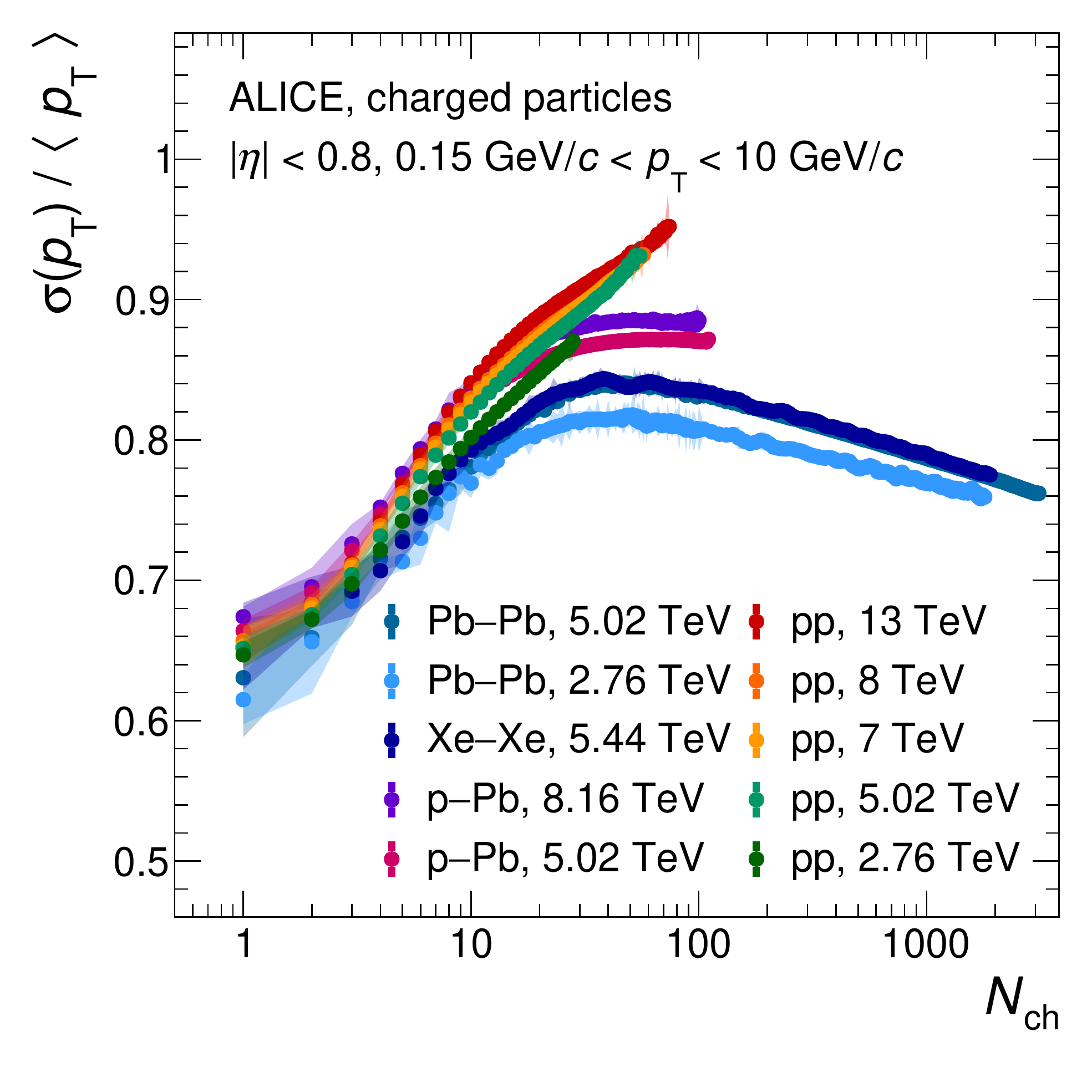}
	\hspace{\plotDistance}
	\includegraphics[width=\plotWidth]{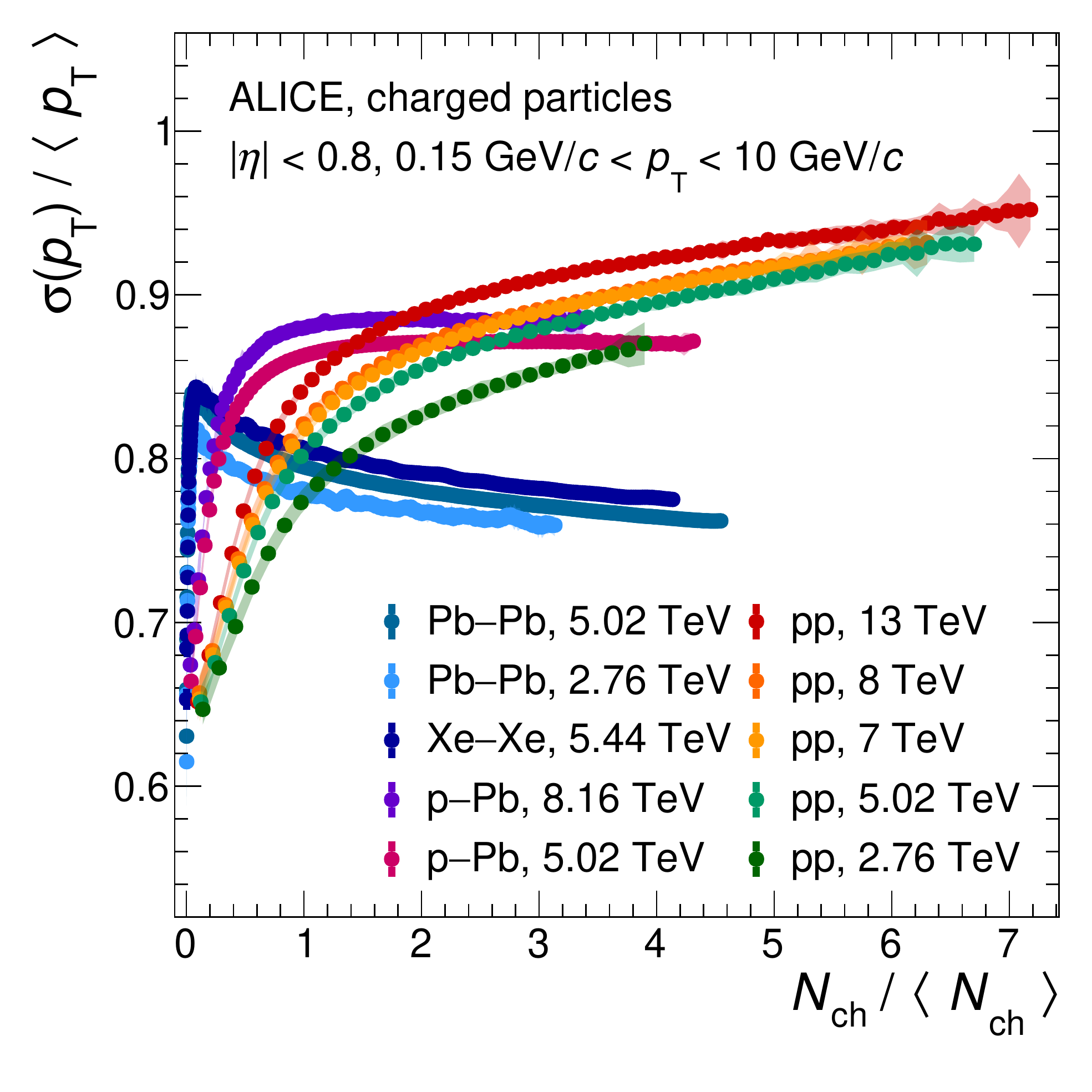}
	\caption{Relative standard deviation of the charged-particle transverse momentum spectra as a function of the absolute (left) and relative (right) charged-particle multiplicity for \pp, \ppb, \xexe and \pbpb collisions at various centre-of-mass energies per nucleon pair.
	Statistical and systematic uncertainties are shown as bars and semi-transparent bands, respectively.}
	\label{fig:relative-sigmapt}
\end{figure}

The left panel of Fig.~\ref{fig:relative-sigmapt} shows the relative standard deviation of the spectra $\sigmapt / \mpt$ as a function of \nch (left) and as a function of $\nch / \mnch$ (right).
For \pp collisions, this relative width of the \pt spectra increases with multiplicity.
The same trend is also observed for the larger collision systems.
However, after around $\nch \approx 20$ both for \ppb and AA collisions, the standard deviation rises at the same rate as the mean, resulting in a flattening of the $\sigmapt / \mpt$ ratio.
After this plateau, the spectra in AA collisions become narrower relative to their mean values. 
The right panel of Fig.~\ref{fig:relative-sigmapt} shows the relative standard deviation of the spectra $\sigmapt / \mpt$ as a function of the relative multiplicity $\nch / \mnch$. 
The plateau observed in \ppb collisions starts at the relative multiplicity $\nch / \mnch \approx 1$, while the decrease observed in AA collision already begins at lower relative multiplicities of around $\nch / \mnch \approx 0.2$.

\begin{figure}[htb]
	\center
	\includegraphics[width=\plotWidth]{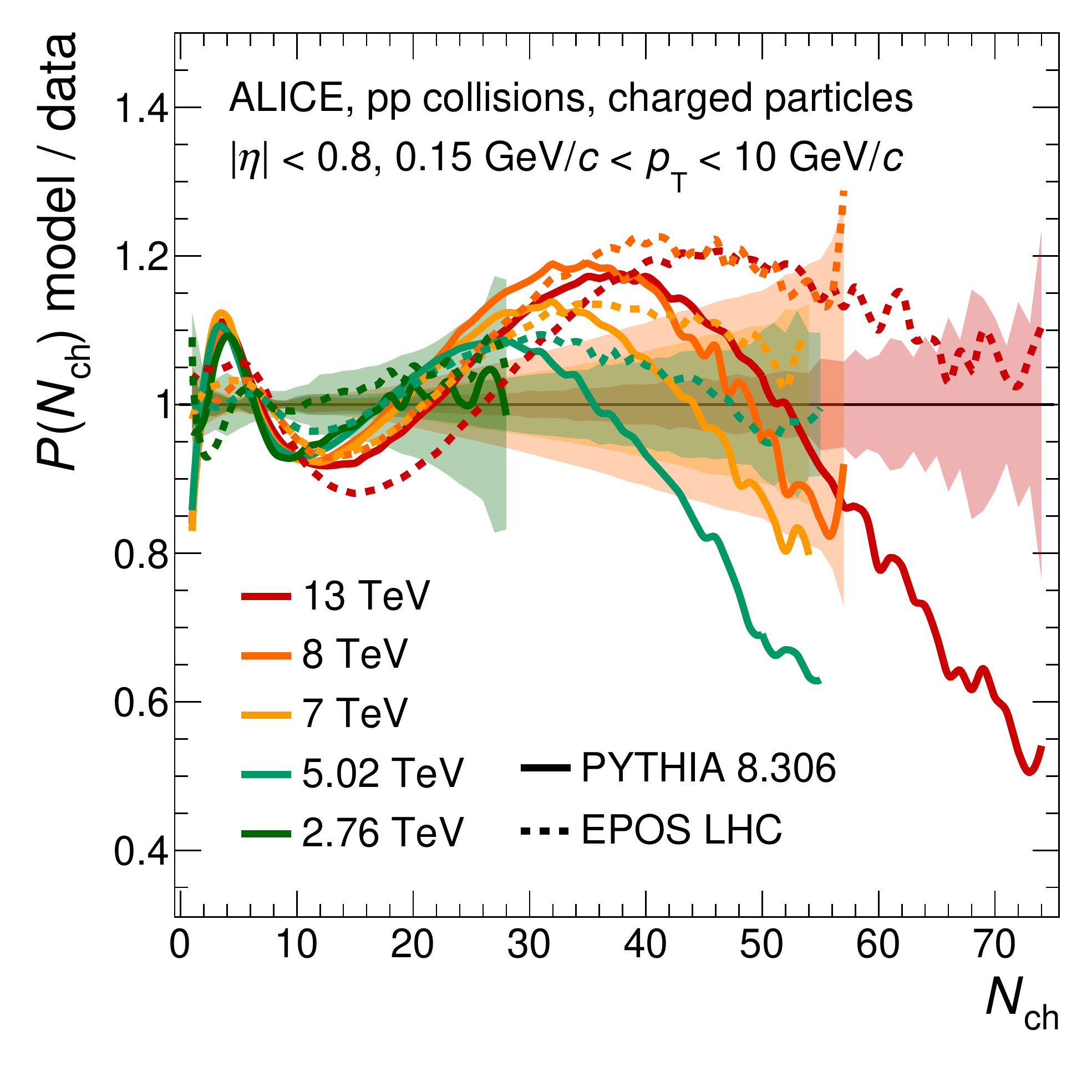}
	\hspace{\plotDistance}
	\includegraphics[width=\plotWidth]{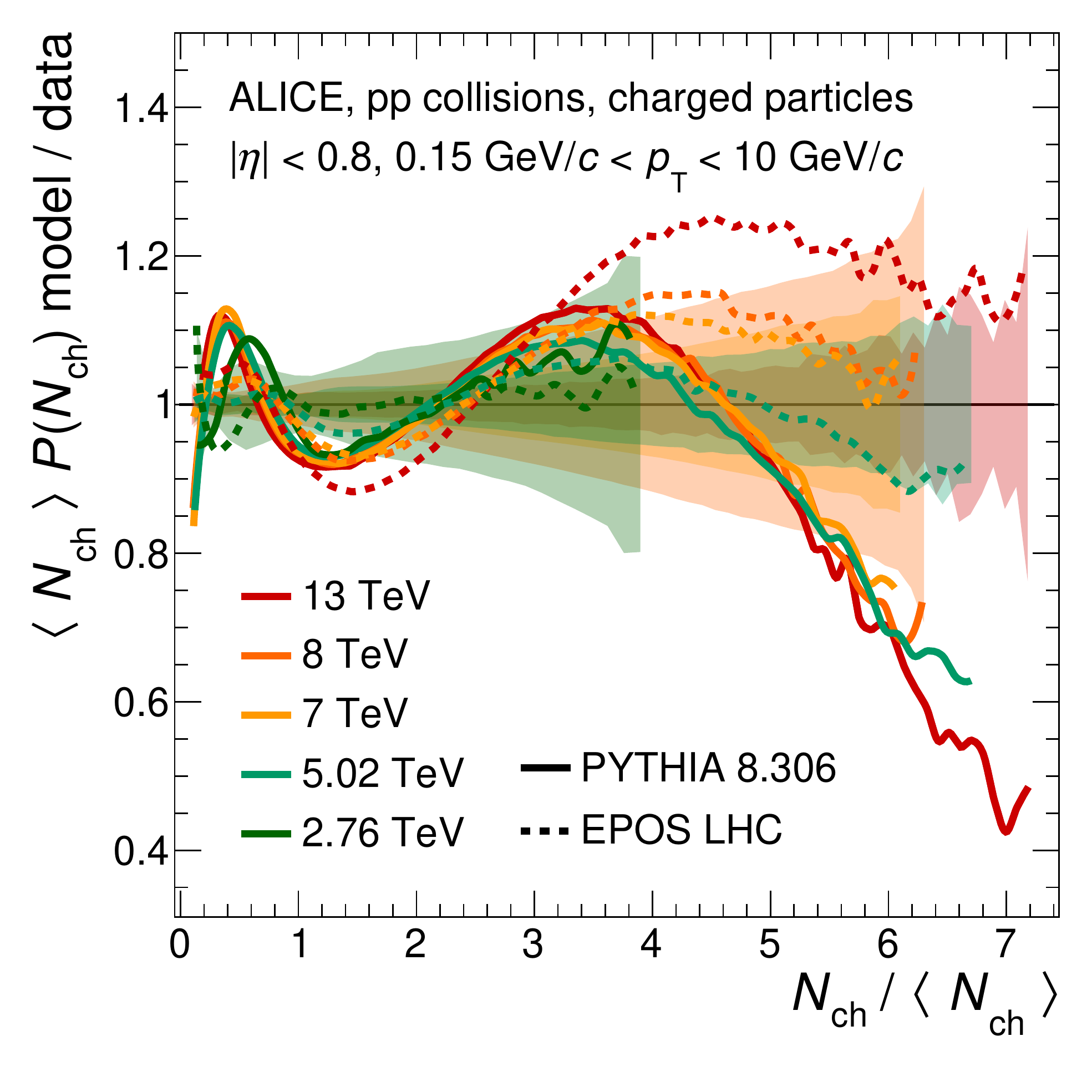}
	\\
	\includegraphics[width=\plotWidth]{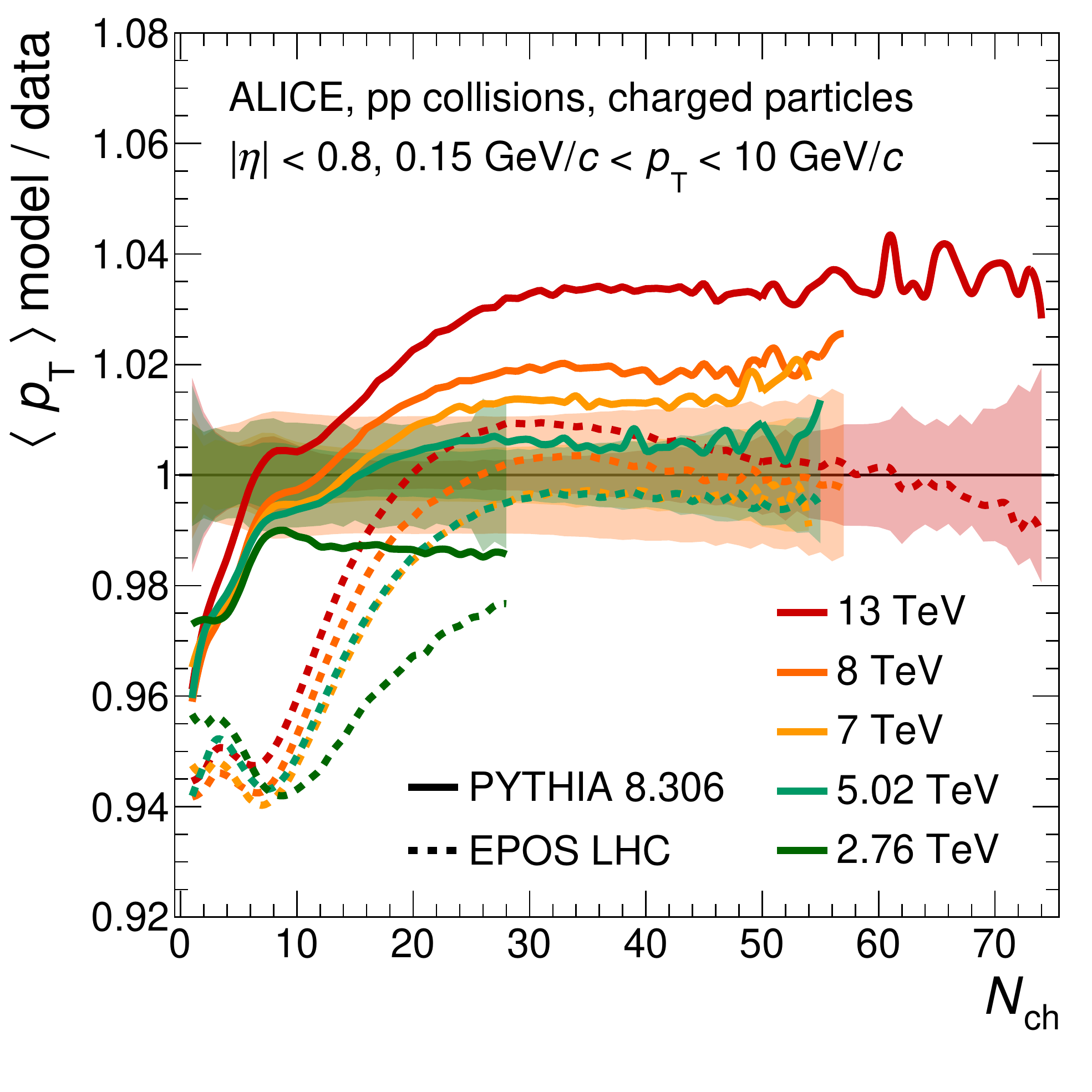}
	\hspace{\plotDistance}
    \includegraphics[width=\plotWidth]{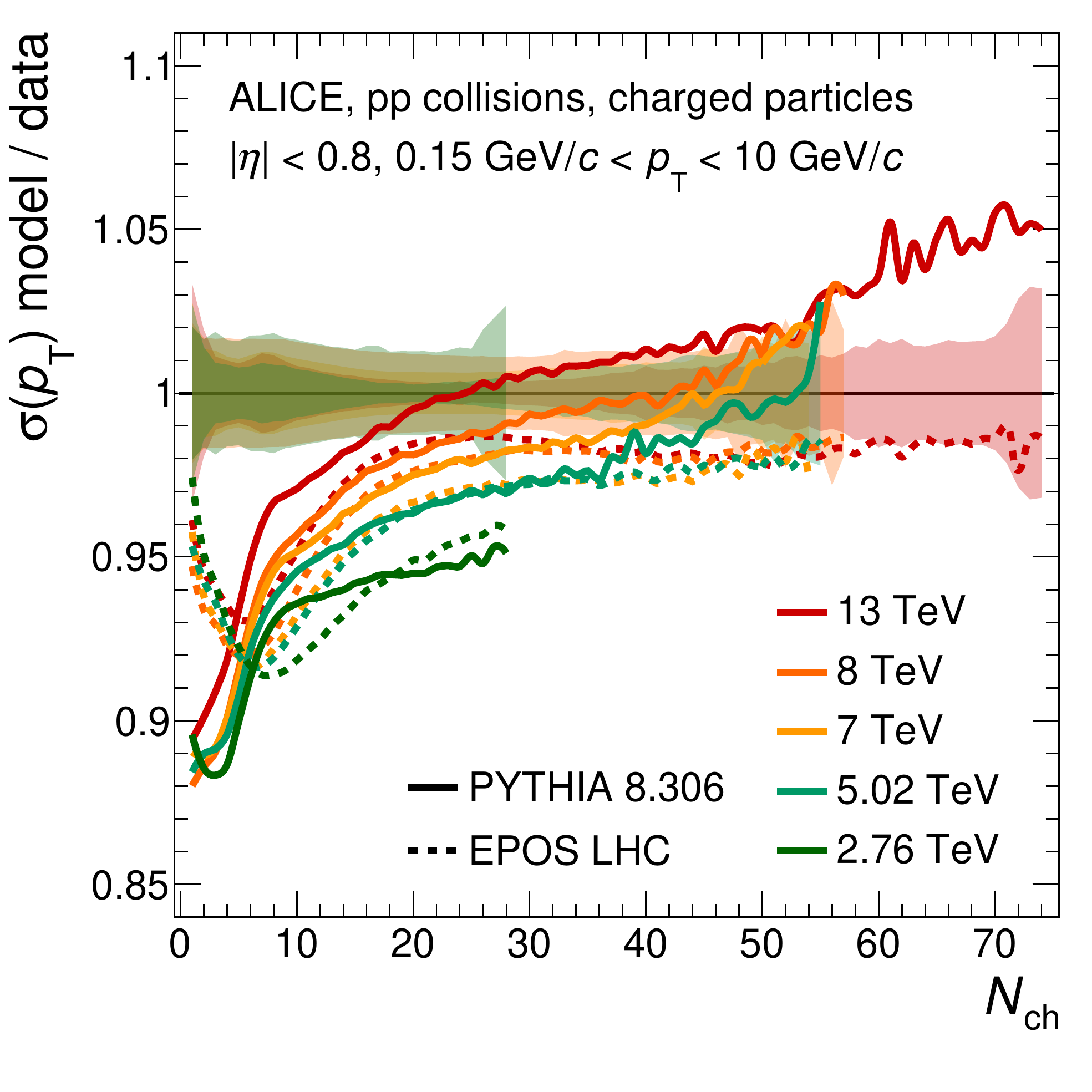}
	\caption{Ratio of model predictions to data for \pp collisions at various energies. The upper panels show it for the multiplicity distributions (left) and their KNO-scaling form (right), the bottom panels represent \mpt (left) and \sigmapt (right). The semi-transparent bands indicate the relative systematic uncertainties of the data.}
	\label{fig:cmp-pp}
\end{figure}

\begin{figure}[htb]
	\center
	\includegraphics[width=\plotWidth]{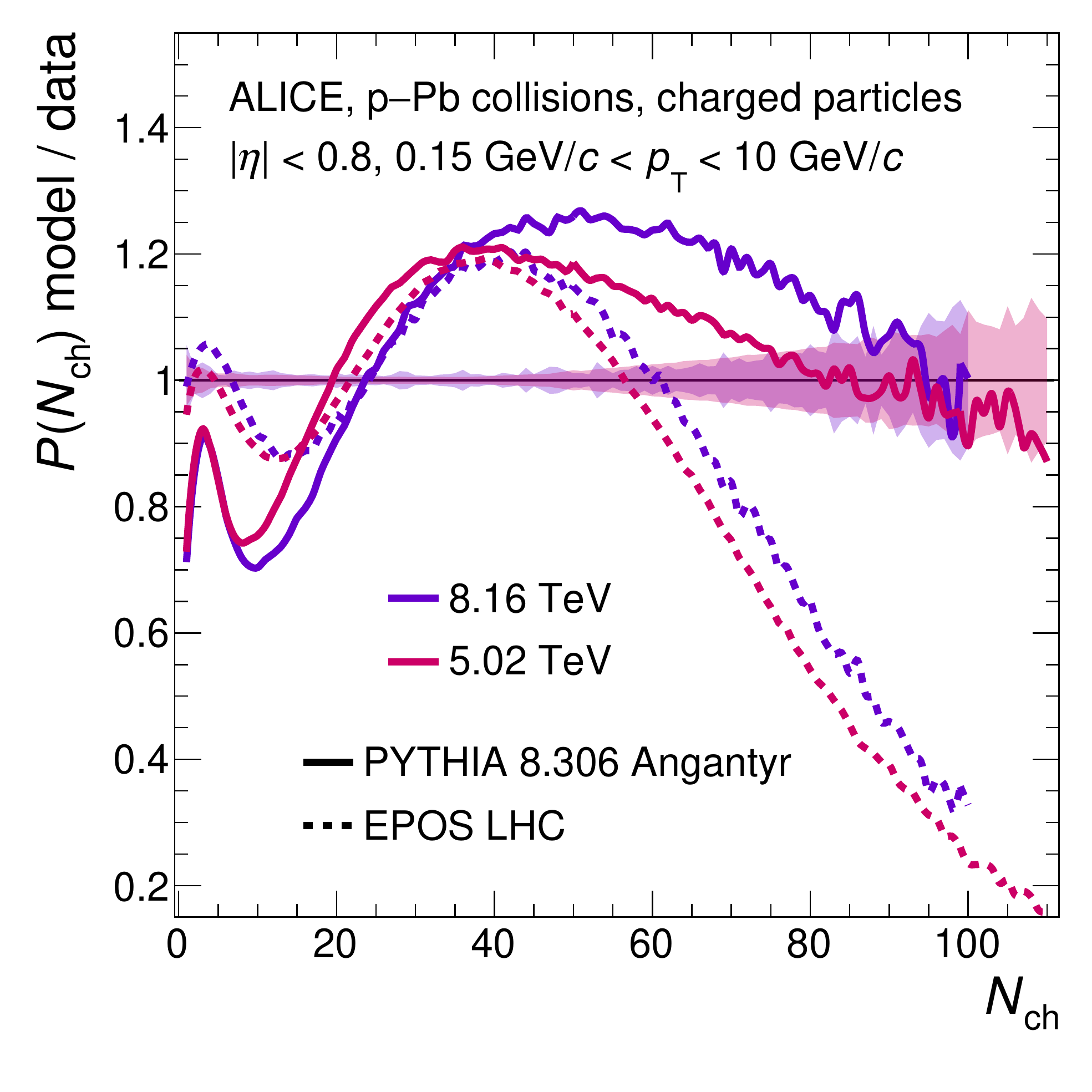}
	\hspace{\plotDistance}
	\includegraphics[width=\plotWidth]{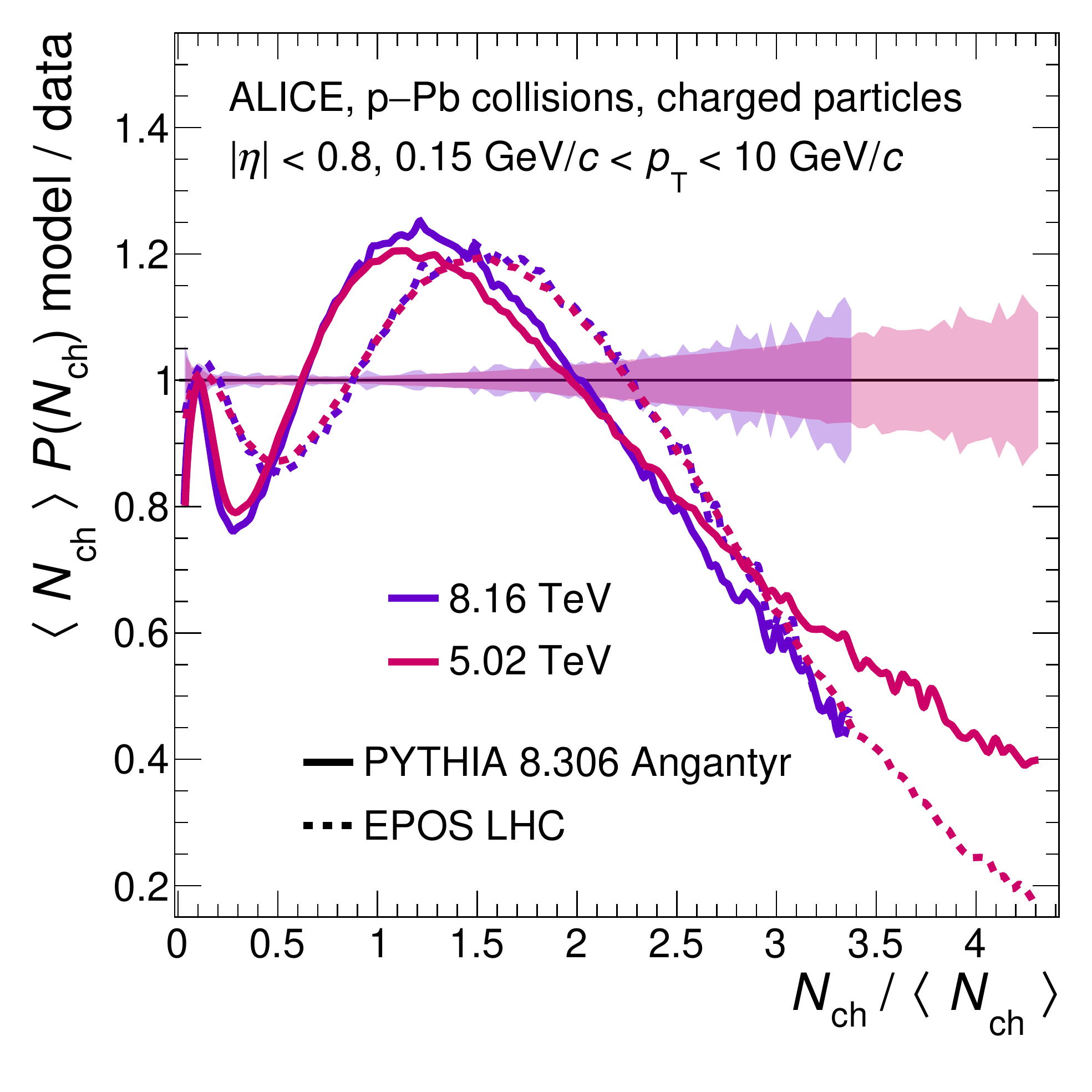}
	\\
	\includegraphics[width=\plotWidth]{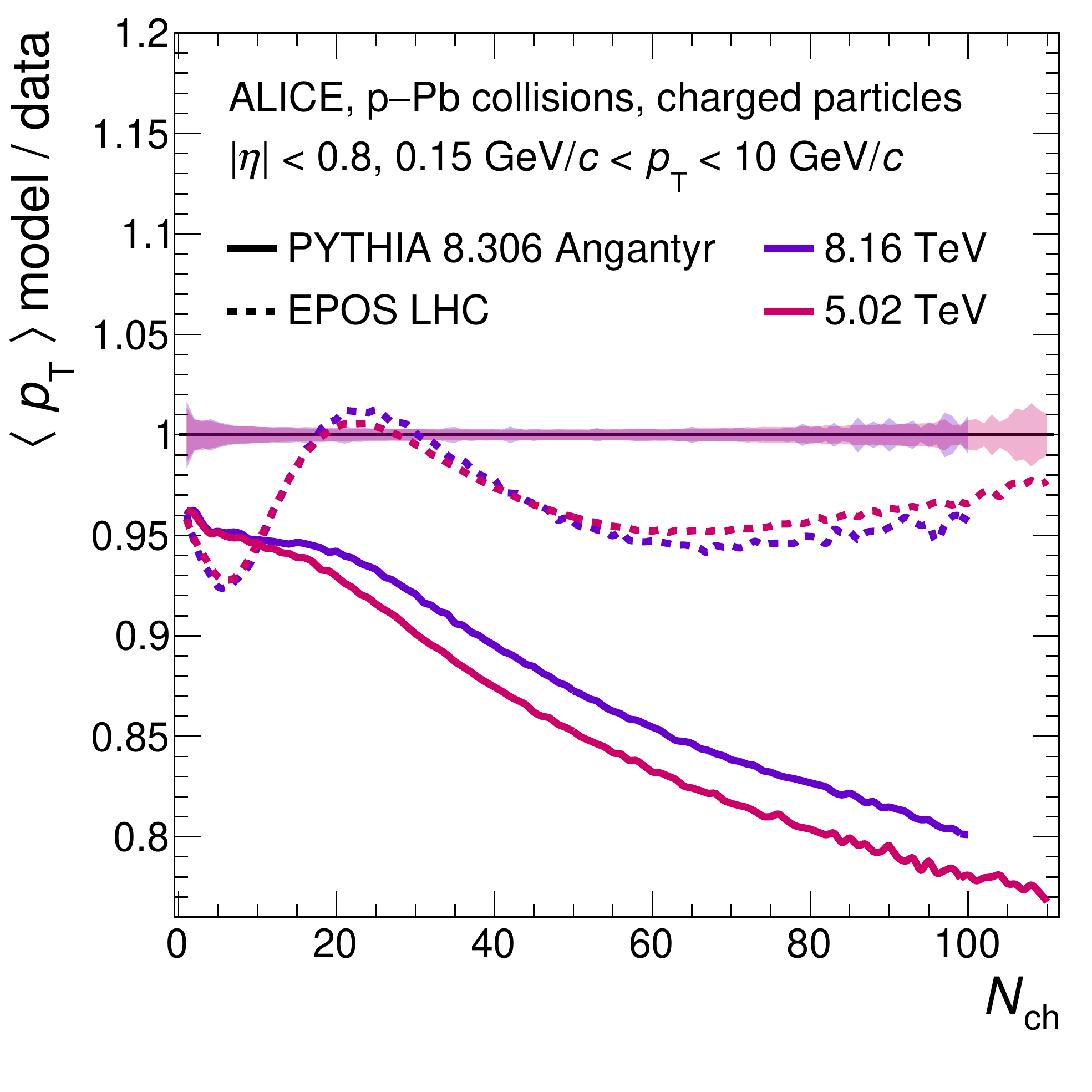}
	\hspace{\plotDistance}
    \includegraphics[width=\plotWidth]{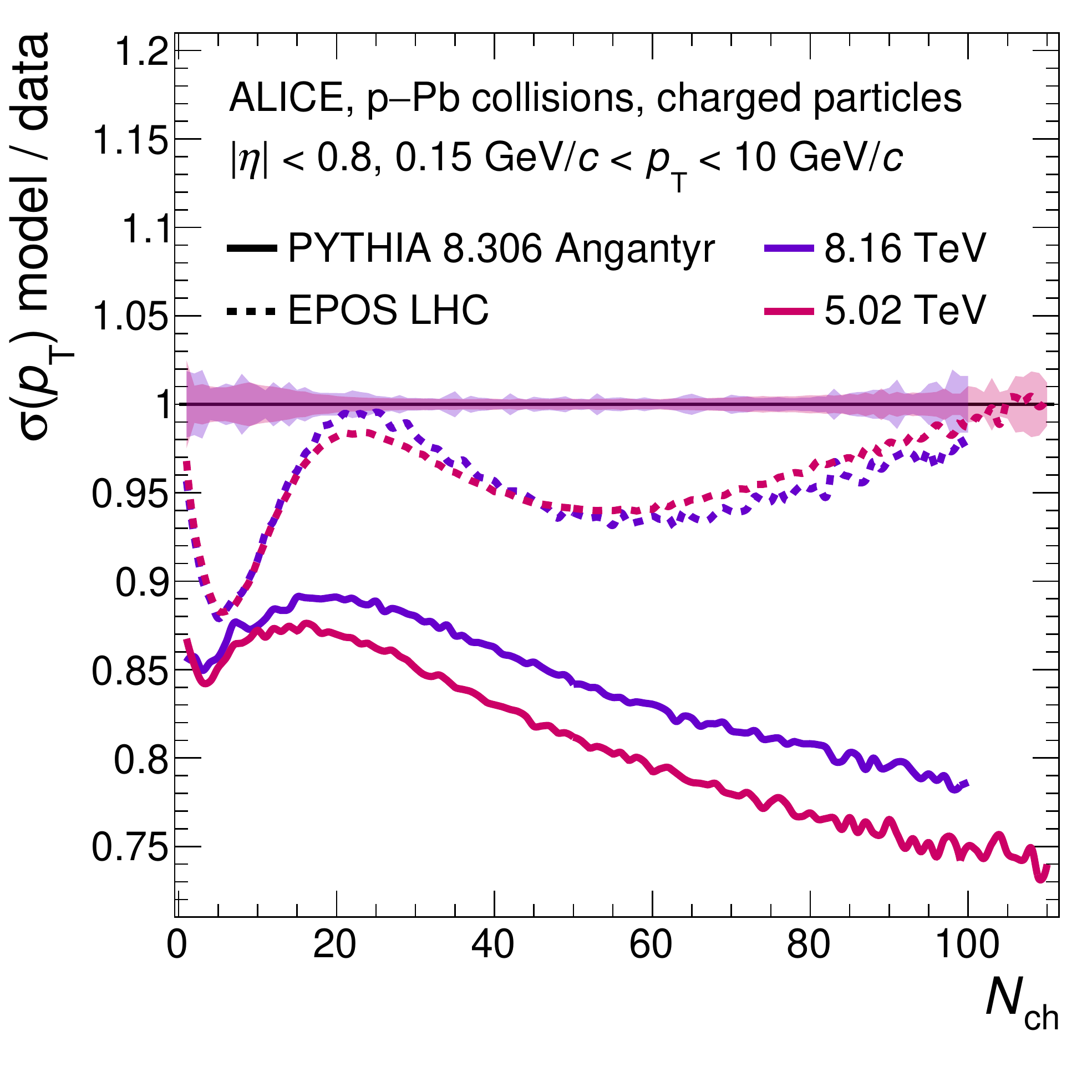}
	\caption{Ratio of model predictions to data for \ppb collisions at various energies. The upper panels show it for the multiplicity distributions (left) and their KNO-scaling form (right), the bottom panels represent \mpt (left) and \sigmapt (right). The semi-transparent bands indicate the relative systematic uncertainties of the data.}
	\label{fig:cmp-ppb}
\end{figure}

Figures~\ref{fig:cmp-pp} and~\ref{fig:cmp-ppb} compare measured results for \pp and \ppb collisions with predictions from PYTHIA8 (solid lines) and EPOS~LHC (dashed lines). Here, the PYTHIA8.306 event generator is used with the Monash-2013 tune~\cite{Skands:2014pea} for pp collisions and with the Angantyr model~\cite{Bierlich:2018xfw} for \ppb collisions.
The top left panel represents the ratio of models over measurements for the multiplicity distributions, the top right panel for the respective KNO-scaled multiplicity distributions and the bottom left and right panel show the ratios for \mpt and \sigmapt, respectively.

In pp collisions, the overall shapes of the multiplicity distribution and KNO-scaled distribution shown in the upper panels of Fig.~\ref{fig:cmp-pp} are better described by EPOS~LHC, while PYTHIA8 falls sharply off above $\nch/ \mnch \approx 4$.  
Both models agree with the experimental distributions within $25\%$ with larger deviations at highest multiplicities.
For \mpt and \sigmapt shown in the bottom panels of Fig.~\ref{fig:cmp-pp}, PYTHIA8 underpredicts the experimental data on \mpt at the lowest values of \nch by up to $4\%$. 
The \nch dependent \mpt values produced by PYTHIA8 increase faster than the measurements with an almost linear dependence up to $\nch \approx 20$, after which the ratio shows a flat multiplicity dependence with an offset from unity varying from $0.5\%$ at $\sqrs = 5.02\tev$ up to
$4\%$ at the highest centre-of-mass energy.
EPOS~LHC is further off at low multiplicities by up to $5\%$ and increases slower than the measurements, underestimating them by up to $6\%$ around $\nch \approx 9$. At higher multiplicities, the increase is faster with a linearly rising ratio up to $\nch \approx 20 -30$, reaching a plateau which
describes the measurements within $\pm2\%$.
The experimental data on \sigmapt is reproduced by both models within $10\%$ at charged-particle multiplicities \nch $ > 10$, with larger deviations at the lowest multiplicities.

Results from model calculations in comparison with measurements from \ppb collisions are shown in Fig.~\ref{fig:cmp-ppb}.
PYTHIA8/Angantyr predicts the charged-particle multiplicity distribution within $30\%$ (Fig.~\ref{fig:cmp-ppb}, top left) over the whole multiplicity range. EPOS~LHC agrees within $20\%$ for $\nch < 70$ but fails to describe the measurement at higher multiplicities.
The KNO-scaled multiplicity distributions shown in the top right panel of Fig.~\ref{fig:cmp-ppb} are described by both models within $20\%$ up to a relative multiplicity of 2.5. 
Beyond that, both models exhibit increasing deviations from the measurement.
PYTHIA8/Angantyr underpredicts \mpt by about $5\%$ at low multiplicities (Fig.~\ref{fig:cmp-ppb}, bottom left), $\nch < 20$, with the deviation increasing as a function of multiplicity, reaching about $25\%$ at $\nch = 110$.
This might result from the missing colour reconnection between the sub-collisions in the model. It is expected that high string density effects, as the recently-introduced shoving mechanism~\cite{Bierlich:2020naj}, will lead to an increase of \mpt as a function of the multiplicity. 
EPOS~LHC reproduces the \mpt and \sigmapt measurement within $10\%$.

\begin{figure}[htb]
	\center
    \includegraphics[width=1.\textwidth]{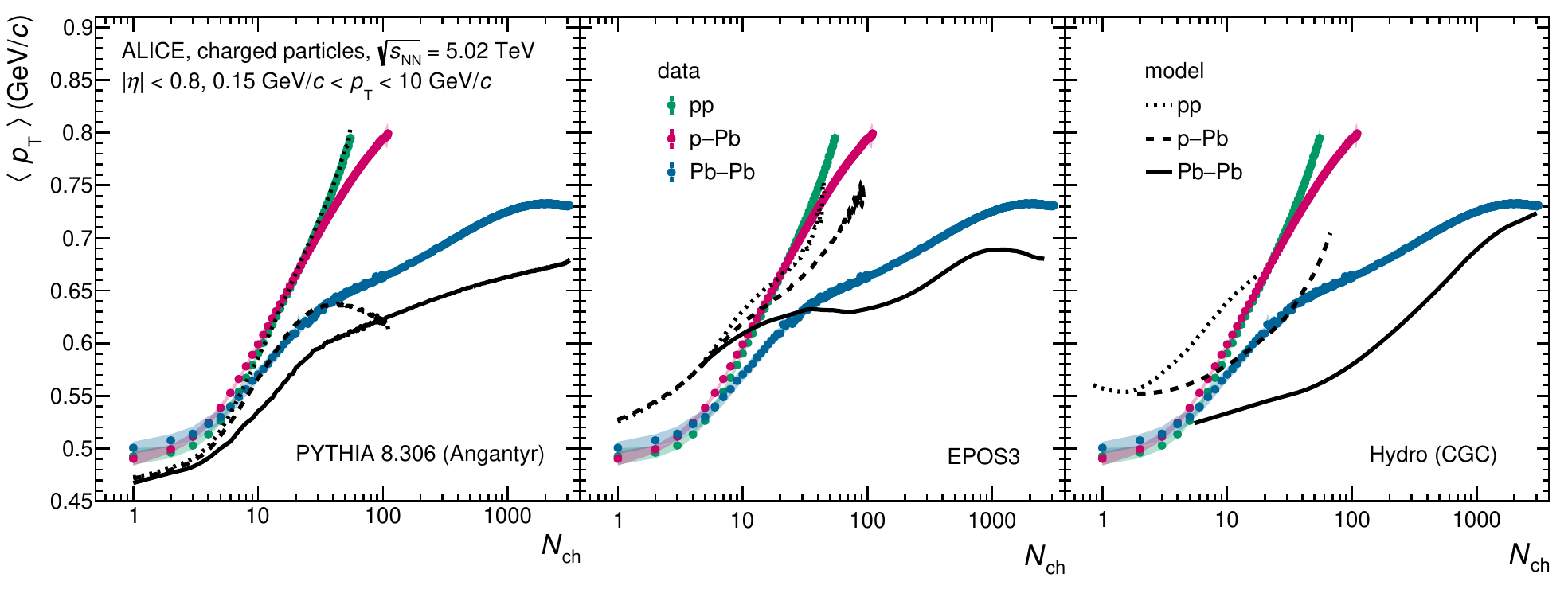}
	\caption{Comparisons of \mpt as a function of \nch at $\sqrsn = 5.02\tev$ for \pp, \ppb, and \pbpb collisions to three different model predictions. Statistical and systematic uncertainties are shown as bars and semi-transparent bands, respectively.}
    \label{fig:cmp-5tev}
\end{figure}

Figure~\ref{fig:cmp-5tev} shows a comparison of the measured \mpt as a function of \nch for \pp, \ppb and \pbpb collisions at the same centre-of-mass energy per nucleon pair, $\sqrsn = 5.02\tev$, with results from three different model calculations: PYTHIA8 (left panel; Angantyr for \ppb and \pbpb), EPOS3 (middle panel), and hydrodynamics with CGC initial conditions~\cite{Schenke:2020mbo} (right panel). 
As shown before, PYTHIA8 describes the \pp measurements very well over the entire multiplicity range, even at the highest multiplicities. However, it significantly underpredicts the \ppb measurements above $\nch > 10$, where the deviation increases with multiplicity. 
In \pbpb collisions, the \mpt is systematically underpredicted by PYTHIA8/Angantyr by around $8\%$ on average. 
Again, this points to the missing treatment of high string density effects which are not included in the PYTHIA8/Angantyr model, yet~\cite{Bierlich:2018xfw}. 
The EPOS3 model overpredicts the \mpt in all systems up to $\nch \approx 10-20$, and underpredicts \ppb and \pbpb measurements at higher multiplicities, less than PYTHIA8/Angantyr, but also cannot reproduce the \mpt evolution with \nch. 
The hydrodynamical model calculations do not describe the measurements well, except for \pbpb collisions at the highest multiplicities of $\nch > 1000$.

\begin{figure}[htb]
	\center
	\includegraphics[width=\plotWidth]{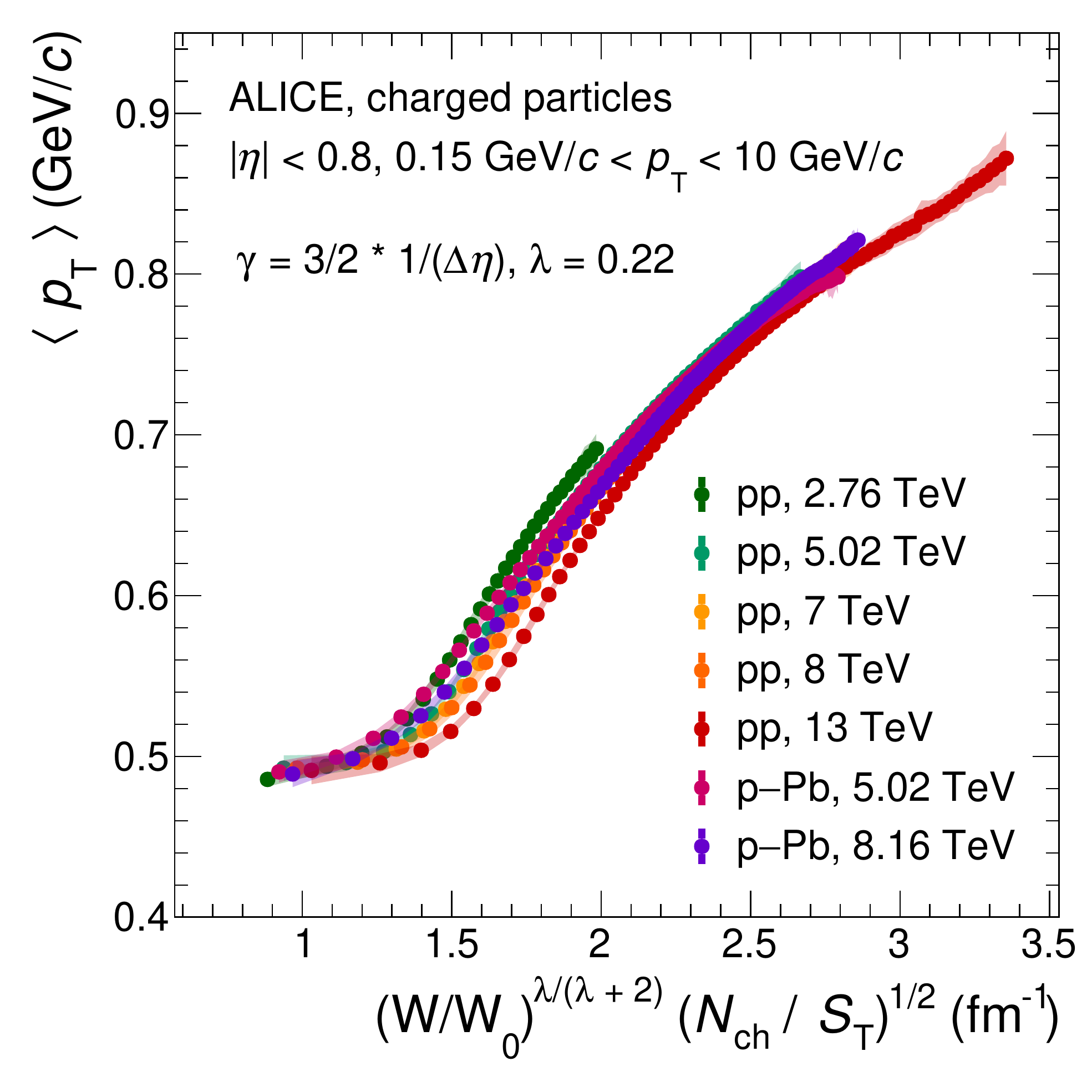}
	\hspace{\plotDistance}
	\includegraphics[width=\plotWidth]{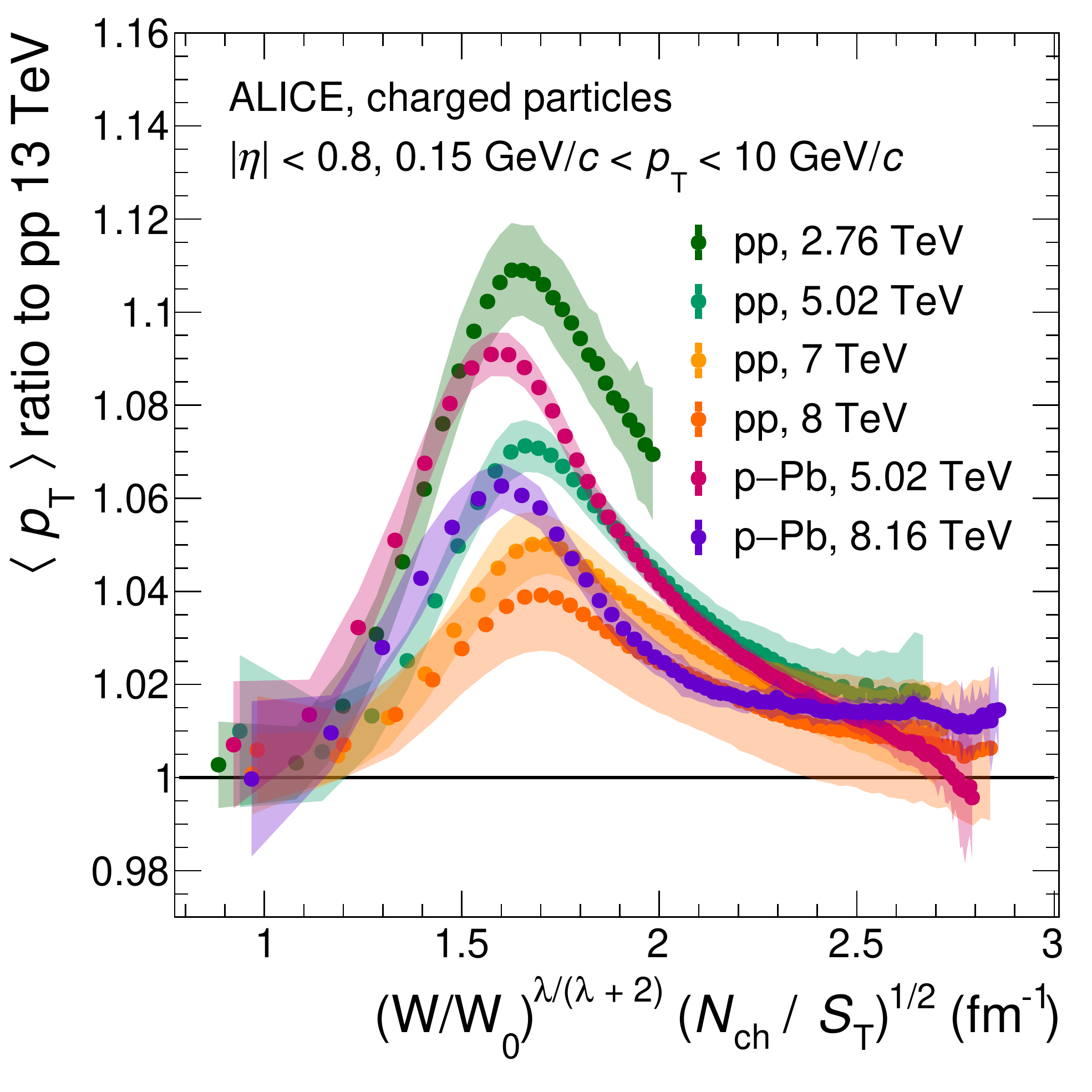}
	\caption{Average transverse momentum \mpt as a function of the scaling variable $(W/W_0)^{\lambda/(\lambda + 2)} \sqrt{\nch / S_\textrm{T}}$~\cite{McLerran:2014apa} for \pp and \ppb collisions at various energies (left) and the ratio of all data sets to that in \pp collisions at 13 TeV (right). The reference energy $W_0$ corresponds to $\sqrs = 7~\mathrm{TeV}$ for \pp and $\sqrsn = 5.02~\mathrm{TeV}$ for \ppb collisions. Statistical and systematic uncertainties are shown as bars and semi-transparent bands, respectively.}
	\label{fig:geom-sale}
\end{figure}

In the CGC approach, the average transverse momentum is a universal function of the ratio of charged-particle multiplicity and transverse area of a collision~\cite{McLerran:2014apa}.
The transverse area $S_\textrm{T} (\nch)$ is derived from the interaction radius $R(\sqrt[3]{\textrm{d}N_\textrm{g}/\textrm{d}y})$ as a function of gluon multiplicity $\textrm{d}N_\textrm{g}/\textrm{d}y$. 
This interaction radius was calculated within the CGC framework for pp collisions at a reference energy of $\sqrs = 7\ \textrm{TeV} = W_{0, \textrm{\pp}}$ and for \ppb collisions at $\sqrsn = 5.02\ \textrm{TeV} = W_{0, \textrm{\ppb}}$~\cite{PhysRevC.87.064906}. 
Parameterisations of these interaction radii proposed in  Ref.~\cite{McLerran:2013una} are used to calculate the interaction area.
Following the arguments in Ref.~\cite{McLerran:2014apa}, the \mpt vs. \nch measurements presented in this Letter are scaled for each collision energy $W = \sqrsn$  for the respective collision system (\pp, \ppb) with a factor of $(W/W_0)^{\lambda/(\lambda + 2)}$.
Here, the exponent $\lambda$ characterises the saturation scale 
and was determined in Ref.~\cite{McLerran:2014apa} to be $\lambda = 0.22$ as the best fit to the transverse momentum distributions measured with ALICE.
In order to approximate the gluon density corresponding to a measured final state multiplicity, a proportionality factor $\gamma$, defined by the equation $\textrm{d}N_\textrm{g}/\textrm{d}y = \gamma \nch$, is needed.
Here, the naive value $\gamma = 3/2\ \frac{1}{\Delta \eta}$ motivated by the ratio of the number of charged particles to all particles, as done in Ref.~\cite{McLerran:2013una}, is used. 
A weak dependence of the results on $\gamma$ was observed.
Figure~\ref{fig:geom-sale} shows the \mpt as a function of 
$(W/W_0)^{\lambda/(\lambda + 2)} \sqrt{\nch / S_\textrm{T}}$ for \pp and \ppb collisions at various collision energies (left), and the ratio of those curves to the 13 TeV result~(right), which has the highest reach in this scaling observable.
The disagreement from the scaling is significant given the measurement's uncertainties, but still within about $10\%$ over the entire range.
At comparable values of the scaling variable, the ratio shows a distinct energy ordering, and all ratios exhibit a noticeable peak.
In a more recent study~\cite{Prasza_owicz_2015}, the determination of the exponent $\lambda$ was revisited and it was found that the differential cross sections are better described when using $\lambda \approx 0.32$ instead of $\lambda = 0.22$.
However, with this updated value for the characteristic exponent $\lambda$, the geometrical scaling of the data presented in this Letter agrees only within $\pm 15\%$.
In general, the scaling results are found to be very sensitive to the value of $\lambda$ and the best agreement is actually found for $\lambda = 0$, which effectively removes the energy scaling term $(W/W_0)^{\lambda/(\lambda + 2)}$ proposed in Ref.~\cite{McLerran:2014apa}.
An approximate geometrical scaling of \mpt was also observed in AA collisions as discussed in Ref.~\cite{Petrovici:2018mpq}.

%% file: Content/5--Summary.tex
\section{Summary and conclusions}\label{sec:summary}

A comprehensive study of inclusive charged-particle production at the LHC is presented, spanning a wide range of collision energies in \pp, \ppb, \xexe and \pbpb collisions.
Multiplicity distributions are compared across centre-of-mass energies for all collision systems, and, in addition, shown in the KNO-scaling form.
The KNO scaling is observed to hold within about $20\%$, not only for \pp collisions, but  also for \ppb and AA collisions.
Transverse momentum spectra are measured as a function of charged-particle multiplicity in narrow \nch intervals.
The mean and standard deviation of these \pt spectra as a function of \nch are compared with PYTHIA8~(Angantyr), EPOS LHC, EPOS3, and hydrodynamical model predictions.
For pp collisions, the spectral shape evolution with multiplicity is described fairly well by PYTHIA8 and EPOS LHC for all centre-of-mass energies, while EPOS3 and the hydrodynamical model fail to predict this observable.
In general, for \ppb and AA collisions there is a large tension with the data for all considered models,
except for EPOS LHC which offers the best model prediction for \ppb collisions.
The geometric scaling of \mpt proposed within the colour glass condensate framework is found to hold in first order, with deviations at the level of $10\%$.

Since the study of charged-particle production as a function of multiplicity plays a key role in understanding the properties of strongly-interacting matter created in collision systems of different sizes and energy densities, in the future, this rich high-precision set of multidimensional measurements can help to improve the theoretical modelling of the complex interplay of hard and soft QCD processes that govern particle production at LHC energies.

%% file: fa_2022-11-03_Opt_C.tex

The ALICE Collaboration would like to thank all its engineers and technicians for their invaluable contributions to the construction of the experiment and the CERN accelerator teams for the outstanding performance of the LHC complex.
The ALICE Collaboration gratefully acknowledges the resources and support provided by all Grid centres and the Worldwide LHC Computing Grid (WLCG) collaboration.
The ALICE Collaboration acknowledges the following funding agencies for their support in building and running the ALICE detector:
A. I. Alikhanyan National Science Laboratory (Yerevan Physics Institute) Foundation (ANSL), State Committee of Science and World Federation of Scientists (WFS), Armenia;
Austrian Academy of Sciences, Austrian Science Fund (FWF): [M 2467-N36] and Nationalstiftung f\"{u}r Forschung, Technologie und Entwicklung, Austria;
Ministry of Communications and High Technologies, National Nuclear Research Center, Azerbaijan;
Conselho Nacional de Desenvolvimento Cient\'{\i}fico e Tecnol\'{o}gico (CNPq), Financiadora de Estudos e Projetos (Finep), Funda\c{c}\~{a}o de Amparo \`{a} Pesquisa do Estado de S\~{a}o Paulo (FAPESP) and Universidade Federal do Rio Grande do Sul (UFRGS), Brazil;
Bulgarian Ministry of Education and Science, within the National Roadmap for Research Infrastructures 2020-2027 (object CERN), Bulgaria;
Ministry of Education of China (MOEC) , Ministry of Science \& Technology of China (MSTC) and National Natural Science Foundation of China (NSFC), China;
Ministry of Science and Education and Croatian Science Foundation, Croatia;
Centro de Aplicaciones Tecnol\'{o}gicas y Desarrollo Nuclear (CEADEN), Cubaenerg\'{\i}a, Cuba;
Ministry of Education, Youth and Sports of the Czech Republic, Czech Republic;
The Danish Council for Independent Research | Natural Sciences, the VILLUM FONDEN and Danish National Research Foundation (DNRF), Denmark;
Helsinki Institute of Physics (HIP), Finland;
Commissariat \`{a} l'Energie Atomique (CEA) and Institut National de Physique Nucl\'{e}aire et de Physique des Particules (IN2P3) and Centre National de la Recherche Scientifique (CNRS), France;
Bundesministerium f\"{u}r Bildung und Forschung (BMBF) and GSI Helmholtzzentrum f\"{u}r Schwerionenforschung GmbH, Germany;
General Secretariat for Research and Technology, Ministry of Education, Research and Religions, Greece;
National Research, Development and Innovation Office, Hungary;
Department of Atomic Energy Government of India (DAE), Department of Science and Technology, Government of India (DST), University Grants Commission, Government of India (UGC) and Council of Scientific and Industrial Research (CSIR), India;
National Research and Innovation Agency - BRIN, Indonesia;
Istituto Nazionale di Fisica Nucleare (INFN), Italy;
Japanese Ministry of Education, Culture, Sports, Science and Technology (MEXT) and Japan Society for the Promotion of Science (JSPS) KAKENHI, Japan;
Consejo Nacional de Ciencia (CONACYT) y Tecnolog\'{i}a, through Fondo de Cooperaci\'{o}n Internacional en Ciencia y Tecnolog\'{i}a (FONCICYT) and Direcci\'{o}n General de Asuntos del Personal Academico (DGAPA), Mexico;
Nederlandse Organisatie voor Wetenschappelijk Onderzoek (NWO), Netherlands;
The Research Council of Norway, Norway;
Commission on Science and Technology for Sustainable Development in the South (COMSATS), Pakistan;
Pontificia Universidad Cat\'{o}lica del Per\'{u}, Peru;
Ministry of Education and Science, National Science Centre and WUT ID-UB, Poland;
Korea Institute of Science and Technology Information and National Research Foundation of Korea (NRF), Republic of Korea;
Ministry of Education and Scientific Research, Institute of Atomic Physics, Ministry of Research and Innovation and Institute of Atomic Physics and University Politehnica of Bucharest, Romania;
Ministry of Education, Science, Research and Sport of the Slovak Republic, Slovakia;
National Research Foundation of South Africa, South Africa;
Swedish Research Council (VR) and Knut \& Alice Wallenberg Foundation (KAW), Sweden;
European Organization for Nuclear Research, Switzerland;
Suranaree University of Technology (SUT), National Science and Technology Development Agency (NSTDA), Thailand Science Research and Innovation (TSRI) and National Science, Research and Innovation Fund (NSRF), Thailand;
Turkish Energy, Nuclear and Mineral Research Agency (TENMAK), Turkey;
National Academy of  Sciences of Ukraine, Ukraine;
Science and Technology Facilities Council (STFC), United Kingdom;
National Science Foundation of the United States of America (NSF) and United States Department of Energy, Office of Nuclear Physics (DOE NP), United States of America.
In addition, individual groups or members have received support from:
Marie Sk\l{}odowska Curie, European Research Council, Strong 2020 - Horizon 2020 (grant nos. 950692, 824093, 896850), European Union;
Academy of Finland (Center of Excellence in Quark Matter) (grant nos. 346327, 346328), Finland;
Programa de Apoyos para la Superaci\'{o}n del Personal Acad\'{e}mico, UNAM, Mexico.

%% file: 2022-11-03-Alice_Authorlist_2022-11-03_0_Opt_C.tex
\begin{flushleft} 
\small

S.~Acharya\,\orcidlink{0000-0002-9213-5329}\,$^{\rm 124}$, 
D.~Adamov\'{a}\,\orcidlink{0000-0002-0504-7428}\,$^{\rm 85}$, 
A.~Adler$^{\rm 69}$, 
G.~Aglieri Rinella\,\orcidlink{0000-0002-9611-3696}\,$^{\rm 32}$, 
M.~Agnello\,\orcidlink{0000-0002-0760-5075}\,$^{\rm 29}$, 
N.~Agrawal\,\orcidlink{0000-0003-0348-9836}\,$^{\rm 50}$, 
Z.~Ahammed\,\orcidlink{0000-0001-5241-7412}\,$^{\rm 132}$, 
S.~Ahmad\,\orcidlink{0000-0003-0497-5705}\,$^{\rm 15}$, 
S.U.~Ahn\,\orcidlink{0000-0001-8847-489X}\,$^{\rm 70}$, 
I.~Ahuja\,\orcidlink{0000-0002-4417-1392}\,$^{\rm 37}$
% ...

\section*{Affiliation Notes}

$^{\rm I}$ Deceased\\
$^{\rm II}$ Also at: Max-Planck-Institut f\"{u}r Physik, Munich, Germany\\
$^{\rm III}$ Also at: Italian National Agency for New Technologies, Energy and Sustainable Economic Development (ENEA), Bologna, Italy\\
$^{\rm IV}$ Also at: Dipartimento DET del Politecnico di Torino, Turin, Italy\\
$^{\rm V}$ Also at: Department of Applied Physics, Aligarh Muslim University, Aligarh, India\\
$^{\rm VI}$ Also at: Institute of Theoretical Physics, University of Wroclaw, Poland\\
$^{\rm VII}$ Also at: An institution covered by a cooperation agreement with CERN\\

\section*{Collaboration Institutes}

$^{1}$ A.I. Alikhanyan National Science Laboratory (Yerevan Physics Institute) Foundation, Yerevan, Armenia\\
$^{2}$ AGH University of Krakow, Cracow, Poland\\
$^{3}$ Bogolyubov Institute for Theoretical Physics, National Academy of Sciences of Ukraine, Kiev, Ukraine\\
$^{4}$ Bose Institute, Department of Physics  and Centre for Astroparticle Physics and Space Science (CAPSS), Kolkata, India\\
$^{5}$ California Polytechnic State University, San Luis Obispo, California, United States\\
$^{6}$ Central China Normal University, Wuhan, China\\
$^{7}$ Centro de Aplicaciones Tecnol\'{o}gicas y Desarrollo Nuclear (CEADEN), Havana, Cuba\\
$^{8}$ Centro de Investigaci\'{o}n y de Estudios Avanzados (CINVESTAV), Mexico City and M\'{e}rida, Mexico\\
$^{9}$ Chicago State University, Chicago, Illinois, United States\\
$^{10}$ China Institute of Atomic Energy, Beijing, China\\
$^{11}$ Chungbuk National University, Cheongju, Republic of Korea\\
$^{12}$ Comenius University Bratislava, Faculty of Mathematics, Physics and Informatics, Bratislava, Slovak Republic\\
$^{13}$ COMSATS University Islamabad, Islamabad, Pakistan\\
$^{14}$ Creighton University, Omaha, Nebraska, United States\\
$^{15}$ Department of Physics, Aligarh Muslim University, Aligarh, India\\
$^{16}$ Department of Physics, Pusan National University, Pusan, Republic of Korea\\
$^{17}$ Department of Physics, Sejong University, Seoul, Republic of Korea\\
$^{18}$ Department of Physics, University of California, Berkeley, California, United States\\
$^{19}$ Department of Physics, University of Oslo, Oslo, Norway\\
$^{20}$ Department of Physics and Technology, University of Bergen, Bergen, Norway\\
$^{21}$ Dipartimento di Fisica, Universit\`{a} di Pavia, Pavia, Italy\\
$^{22}$ Dipartimento di Fisica dell'Universit\`{a} and Sezione INFN, Cagliari, Italy\\
$^{23}$ Dipartimento di Fisica dell'Universit\`{a} and Sezione INFN, Trieste, Italy\\
$^{24}$ Dipartimento di Fisica dell'Universit\`{a} and Sezione INFN, Turin, Italy\\
$^{25}$ Dipartimento di Fisica e Astronomia dell'Universit\`{a} and Sezione INFN, Bologna, Italy\\
$^{26}$ Dipartimento di Fisica e Astronomia dell'Universit\`{a} and Sezione INFN, Catania, Italy\\
$^{27}$ Dipartimento di Fisica e Astronomia dell'Universit\`{a} and Sezione INFN, Padova, Italy\\
$^{28}$ Dipartimento di Fisica `E.R.~Caianiello' dell'Universit\`{a} and Gruppo Collegato INFN, Salerno, Italy\\
$^{29}$ Dipartimento DISAT del Politecnico and Sezione INFN, Turin, Italy\\
$^{30}$ Dipartimento di Scienze MIFT, Universit\`{a} di Messina, Messina, Italy\\
$^{31}$ Dipartimento Interateneo di Fisica `M.~Merlin' and Sezione INFN, Bari, Italy\\
$^{32}$ European Organization for Nuclear Research (CERN), Geneva, Switzerland\\
$^{33}$ Faculty of Electrical Engineering, Mechanical Engineering and Naval Architecture, University of Split, Split, Croatia\\
$^{34}$ Faculty of Engineering and Science, Western Norway University of Applied Sciences, Bergen, Norway\\
$^{35}$ Faculty of Nuclear Sciences and Physical Engineering, Czech Technical University in Prague, Prague, Czech Republic\\
$^{36}$ Faculty of Physics, Sofia University, Sofia, Bulgaria\\
$^{37}$ Faculty of Science, P.J.~\v{S}af\'{a}rik University, Ko\v{s}ice, Slovak Republic\\
$^{38}$ Frankfurt Institute for Advanced Studies, Johann Wolfgang Goethe-Universit\"{a}t Frankfurt, Frankfurt, Germany\\
$^{39}$ Fudan University, Shanghai, China\\
$^{40}$ Gangneung-Wonju National University, Gangneung, Republic of Korea\\
$^{41}$ Gauhati University, Department of Physics, Guwahati, India\\
$^{42}$ Helmholtz-Institut f\"{u}r Strahlen- und Kernphysik, Rheinische Friedrich-Wilhelms-Universit\"{a}t Bonn, Bonn, Germany\\
$^{43}$ Helsinki Institute of Physics (HIP), Helsinki, Finland\\
$^{44}$ High Energy Physics Group,  Universidad Aut\'{o}noma de Puebla, Puebla, Mexico\\
$^{45}$ Horia Hulubei National Institute of Physics and Nuclear Engineering, Bucharest, Romania\\
$^{46}$ Indian Institute of Technology Bombay (IIT), Mumbai, India\\
$^{47}$ Indian Institute of Technology Indore, Indore, India\\
$^{48}$ INFN, Laboratori Nazionali di Frascati, Frascati, Italy\\
$^{49}$ INFN, Sezione di Bari, Bari, Italy\\
$^{50}$ INFN, Sezione di Bologna, Bologna, Italy\\
$^{51}$ INFN, Sezione di Cagliari, Cagliari, Italy\\
$^{52}$ INFN, Sezione di Catania, Catania, Italy\\
$^{53}$ INFN, Sezione di Padova, Padova, Italy\\
$^{54}$ INFN, Sezione di Pavia, Pavia, Italy\\
$^{55}$ INFN, Sezione di Torino, Turin, Italy\\
$^{56}$ INFN, Sezione di Trieste, Trieste, Italy\\
$^{57}$ Inha University, Incheon, Republic of Korea\\
$^{58}$ Institute for Gravitational and Subatomic Physics (GRASP), Utrecht University/Nikhef, Utrecht, Netherlands\\
$^{59}$ Institute of Experimental Physics, Slovak Academy of Sciences, Ko\v{s}ice, Slovak Republic\\
$^{60}$ Institute of Physics, Homi Bhabha National Institute, Bhubaneswar, India\\
$^{61}$ Institute of Physics of the Czech Academy of Sciences, Prague, Czech Republic\\
$^{62}$ Institute of Space Science (ISS), Bucharest, Romania\\
$^{63}$ Institut f\"{u}r Kernphysik, Johann Wolfgang Goethe-Universit\"{a}t Frankfurt, Frankfurt, Germany\\
$^{64}$ Instituto de Ciencias Nucleares, Universidad Nacional Aut\'{o}noma de M\'{e}xico, Mexico City, Mexico\\
$^{65}$ Instituto de F\'{i}sica, Universidade Federal do Rio Grande do Sul (UFRGS), Porto Alegre, Brazil\\
$^{66}$ Instituto de F\'{\i}sica, Universidad Nacional Aut\'{o}noma de M\'{e}xico, Mexico City, Mexico\\
$^{67}$ iThemba LABS, National Research Foundation, Somerset West, South Africa\\
$^{68}$ Jeonbuk National University, Jeonju, Republic of Korea\\
$^{69}$ Johann-Wolfgang-Goethe Universit\"{a}t Frankfurt Institut f\"{u}r Informatik, Fachbereich Informatik und Mathematik, Frankfurt, Germany\\
$^{70}$ Korea Institute of Science and Technology Information, Daejeon, Republic of Korea\\
$^{71}$ KTO Karatay University, Konya, Turkey\\
$^{72}$ Laboratoire de Physique Subatomique et de Cosmologie, Universit\'{e} Grenoble-Alpes, CNRS-IN2P3, Grenoble, France\\
$^{73}$ Lawrence Berkeley National Laboratory, Berkeley, California, United States\\
$^{74}$ Lund University Department of Physics, Division of Particle Physics, Lund, Sweden\\
$^{75}$ Nagasaki Institute of Applied Science, Nagasaki, Japan\\
$^{76}$ Nara Women{'}s University (NWU), Nara, Japan\\
$^{77}$ National and Kapodistrian University of Athens, School of Science, Department of Physics , Athens, Greece\\
$^{78}$ National Centre for Nuclear Research, Warsaw, Poland\\
$^{79}$ National Institute of Science Education and Research, Homi Bhabha National Institute, Jatni, India\\
$^{80}$ National Nuclear Research Center, Baku, Azerbaijan\\
$^{81}$ National Research and Innovation Agency - BRIN, Jakarta, Indonesia\\
$^{82}$ Niels Bohr Institute, University of Copenhagen, Copenhagen, Denmark\\
$^{83}$ Nikhef, National institute for subatomic physics, Amsterdam, Netherlands\\
$^{84}$ Nuclear Physics Group, STFC Daresbury Laboratory, Daresbury, United Kingdom\\
$^{85}$ Nuclear Physics Institute of the Czech Academy of Sciences, Husinec-\v{R}e\v{z}, Czech Republic\\
$^{86}$ Oak Ridge National Laboratory, Oak Ridge, Tennessee, United States\\
$^{87}$ Ohio State University, Columbus, Ohio, United States\\
$^{88}$ Physics department, Faculty of science, University of Zagreb, Zagreb, Croatia\\
$^{89}$ Physics Department, Panjab University, Chandigarh, India\\
$^{90}$ Physics Department, University of Jammu, Jammu, India\\
$^{91}$ Physics Program and International Institute for Sustainability with Knotted Chiral Meta Matter (SKCM2), Hiroshima University, Hiroshima, Japan\\
$^{92}$ Physikalisches Institut, Eberhard-Karls-Universit\"{a}t T\"{u}bingen, T\"{u}bingen, Germany\\
$^{93}$ Physikalisches Institut, Ruprecht-Karls-Universit\"{a}t Heidelberg, Heidelberg, Germany\\
$^{94}$ Physik Department, Technische Universit\"{a}t M\"{u}nchen, Munich, Germany\\
$^{95}$ Politecnico di Bari and Sezione INFN, Bari, Italy\\
$^{96}$ Research Division and ExtreMe Matter Institute EMMI, GSI Helmholtzzentrum f\"ur Schwerionenforschung GmbH, Darmstadt, Germany\\
$^{97}$ Saga University, Saga, Japan\\
$^{98}$ Saha Institute of Nuclear Physics, Homi Bhabha National Institute, Kolkata, India\\
$^{99}$ School of Physics and Astronomy, University of Birmingham, Birmingham, United Kingdom\\
$^{100}$ Secci\'{o}n F\'{\i}sica, Departamento de Ciencias, Pontificia Universidad Cat\'{o}lica del Per\'{u}, Lima, Peru\\
$^{101}$ Stefan Meyer Institut f\"{u}r Subatomare Physik (SMI), Vienna, Austria\\
$^{102}$ SUBATECH, IMT Atlantique, Nantes Universit\'{e}, CNRS-IN2P3, Nantes, France\\
$^{103}$ Sungkyunkwan University, Suwon City, Republic of Korea\\
$^{104}$ Suranaree University of Technology, Nakhon Ratchasima, Thailand\\
$^{105}$ Technical University of Ko\v{s}ice, Ko\v{s}ice, Slovak Republic\\
$^{106}$ The Henryk Niewodniczanski Institute of Nuclear Physics, Polish Academy of Sciences, Cracow, Poland\\
$^{107}$ The University of Texas at Austin, Austin, Texas, United States\\
$^{108}$ Universidad Aut\'{o}noma de Sinaloa, Culiac\'{a}n, Mexico\\
$^{109}$ Universidade de S\~{a}o Paulo (USP), S\~{a}o Paulo, Brazil\\
$^{110}$ Universidade Estadual de Campinas (UNICAMP), Campinas, Brazil\\
$^{111}$ Universidade Federal do ABC, Santo Andre, Brazil\\
$^{112}$ University of Cape Town, Cape Town, South Africa\\
$^{113}$ University of Houston, Houston, Texas, United States\\
$^{114}$ University of Jyv\"{a}skyl\"{a}, Jyv\"{a}skyl\"{a}, Finland\\
$^{115}$ University of Kansas, Lawrence, Kansas, United States\\
$^{116}$ University of Liverpool, Liverpool, United Kingdom\\
$^{117}$ University of Science and Technology of China, Hefei, China\\
$^{118}$ University of South-Eastern Norway, Kongsberg, Norway\\
$^{119}$ University of Tennessee, Knoxville, Tennessee, United States\\
$^{120}$ University of the Witwatersrand, Johannesburg, South Africa\\
$^{121}$ University of Tokyo, Tokyo, Japan\\
$^{122}$ University of Tsukuba, Tsukuba, Japan\\
$^{123}$ University Politehnica of Bucharest, Bucharest, Romania\\
$^{124}$ Universit\'{e} Clermont Auvergne, CNRS/IN2P3, LPC, Clermont-Ferrand, France\\
$^{125}$ Universit\'{e} de Lyon, CNRS/IN2P3, Institut de Physique des 2 Infinis de Lyon, Lyon, France\\
$^{126}$ Universit\'{e} de Strasbourg, CNRS, IPHC UMR 7178, F-67000 Strasbourg, France, Strasbourg, France\\
$^{127}$ Universit\'{e} Paris-Saclay, Centre d'Etudes de Saclay (CEA), IRFU, D\'{e}partment de Physique Nucl\'{e}aire (DPhN), Saclay, France\\
$^{128}$ Universit\'{e}  Paris-Saclay, CNRS/IN2P3, IJCLab, Orsay, France\\
$^{129}$ Universit\`{a} degli Studi di Foggia, Foggia, Italy\\
$^{130}$ Universit\`{a} del Piemonte Orientale, Vercelli, Italy\\
$^{131}$ Universit\`{a} di Brescia, Brescia, Italy\\
$^{132}$ Variable Energy Cyclotron Centre, Homi Bhabha National Institute, Kolkata, India\\
$^{133}$ Warsaw University of Technology, Warsaw, Poland\\
$^{134}$ Wayne State University, Detroit, Michigan, United States\\
$^{135}$ Westf\"{a}lische Wilhelms-Universit\"{a}t M\"{u}nster, Institut f\"{u}r Kernphysik, M\"{u}nster, Germany\\
$^{136}$ Wigner Research Centre for Physics, Budapest, Hungary\\
$^{137}$ Yale University, New Haven, Connecticut, United States\\
$^{138}$ Yonsei University, Seoul, Republic of Korea\\
$^{139}$  Zentrum  f\"{u}r Technologie und Transfer (ZTT), Worms, Germany\\
$^{140}$ Affiliated with an institute covered by a cooperation agreement with CERN\\
$^{141}$ Affiliated with an international laboratory covered by a cooperation agreement with CERN.\\

\end{flushleft} 